# Eigenwaves: Finite-frequency Eigenrays in Elastic Anisotropic Media

Zvi Koren[1,2]

[1]Emerson-AspenTech, Subsurface Science & Engineering
[2]Department of Geophysics, Tel Aviv University
(draft for arXiv submission)

**Abstract**

The Eigenray method (Koren and Ravve, 2021; Ravve and Koren, 2021) solves two-point boundary-value ray-tracing problems in heterogeneous anisotropic elastic media through a variational formulation based on the eikonal equation. As a result, it inherits the high-frequency limitations of classical ray theory. In this paper, we extend the Eigenray framework to finite frequencies using the trajectory-mechanics formulation of Vasco and Nihei (2019), in which finite-frequency effects are represented by a frequency-dependent potential that couples phase and amplitude. We show that finite-frequency wave propagation can be described within the same variational framework as the original Eigenray method through an equivalent effective medium. This preserves the structure of the method while introducing finite-frequency corrections to wave trajectories, phase, amplitudes, geometric spreading, wavefront curvature, and caustic behavior. We derive the corresponding variational and Hamiltonian formulations and develop a weak-coupling approximation that relates finite-frequency trajectories and phase shifts to perturbations of the underlying eikonal rays. Because the finite-element Eigenray solver requires only local velocity information and its derivatives, the extension to finite frequencies can be implemented with minimal modifications to the existing framework. The finite-frequency corrections vanish in the high-frequency limit, recovering classical ray theory. Numerical examples involving multipathing and cusp caustics demonstrate that the proposed Eigenwave formulation accurately reproduces finite-frequency propagation effects. While ray-based amplitudes become inadequate near caustics, a phase-space (Maslov) treatment successfully reproduces finite-difference wavefield solutions. These results establish Eigenwave theory as a unified variational framework connecting high-frequency ray theory and finite-frequency wave propagation. The method also provides a foundation for phase-space Local Angle Domain Full-Waveform Inversion (LAD-FWI) in anisotropic elastic media.

Key words: Anisotropy; Rays; Finite-frequency; Trajectory mechanics; Variational calculus; Dynamic ray tracing.

## 1 Introduction

Two-point (boundary-value) ray tracing is a cornerstone of seismic modelling, imaging and tomography. In a series of papers we introduced the Eigenray method (Koren and Ravve, 2020a,b,c; Ravve and Koren, 2020a,b,c,d; Koren and Ravve, 2021; Ravve and Koren, 2021), a ray-bending finite-element approach for finding stationary-traveltime paths between two fixed endpoints in 3D smooth heterogeneous general anisotropic elastic media (kinematics), and for computing the paraxial rays, the ray Jacobian, the relative geometric spreading and the caustics along the resolved stationary path (dynamics).

The kinematic problem is governed by the nonlinear, second-order Euler-Lagrange equation applied to an arclength-related Lagrangian equal to the reciprocal ray (group) velocity, and the dynamic problem by the linear, second-order Jacobi equation obtained from the second traveltime variation (Bliss, 1916). A distinctive feature of the method is that the resolving matrix of the (linear) dynamic finite-element problem coincides with the global traveltime Hessian already computed in the (nonlinear) kinematic stage.

Like all methods based on the eikonal equation, the Eigenray method describes the high-frequency asymptotic limit of the elastodynamic equations. Its predictions become questionable when the medium varies on scales comparable to the wavelength, e.g. across thin layers or narrow transition zones (e.g., Červený, 2000; Chapman, 2004). Finite-frequency effects on traveltimes are traditionally introduced through sensitivity kernels (Marquering et al., 1999; Dahlen et al., 2000), or through the summation of paraxial Gaussian beams (Červený et al., 1982; Popov, 1982). A different route, developed by Foreman (1989) for the Helmholtz equation and, in the elastic anisotropic case, by Vasco and Nihei (2019a) (see also Vasco, 2018; Vasco and Nihei, 2019b; Protasov and Gadylshin, 2017), is to keep the exact polar decomposition of the monochromatic wavefield, $\mathbf{u} = \mathbf{a}e^{i\varphi}$, and to derive *exact* (i.e. not asymptotic) trajectory equations. The price is a real potential $W(\mathbf{x}, \omega)$, proportional to $\omega^{-2}$ and to second spatial derivatives of the amplitude field, that enters the Hamiltonian and couples the trajectories to the amplitudes; the trajectories become frequency dependent, and a suite of such trajectories for a band of frequencies delineates the region of sensitivity of the band-limited arrival. This is the trajectory-mechanics analogue of the Bohmian formulation of quantum mechanics (Bohm, 1952; Wyatt, 2005), where $W$ plays the role of the quantum potential.

We refer to the resulting stationary finite-frequency paths as **Eigenwaves**. The answer turns out to be remarkably simple. For a fixed wave mode and a given amplitude field, we show that the finite-frequency Hamiltonian is dynamically equivalent to the classical Christoffel Hamiltonian of an effective medium with conformally rescaled stiffness tensor, $\tilde{\mathbf{C}}^{\omega} = \tilde{\mathbf{C}}/(1 + W)$. The finite-frequency Lagrangian is therefore obtained by multiplying the classical Lagrangian by an effective refractive index, $n = \sqrt{1 + W}$. As a result, the key properties underlying the Eigenray formulation, including first-degree homogeneity, the Bliss identities, the Jacobi equation, its tangential degeneracy, the standard initial conditions, and the relation between the ray Jacobian and geometric spreading, carry over directly, requiring only explicit finite-frequency corrections to the ray-velocity derivatives.

The present development is also motivated by our recent work on phase-space Local Angle Domain Full-Waveform Inversion (LAD-FWI) in anisotropic elastic media (Koren, 2026, https://arxiv.org/abs/2608.04901). In that framework, seismic imaging and inversion are formulated in terms of localized incident-scattered wave-pair interactions in phase space, represented by angle-domain gathers parameterized by propagation directions, opening angles, wave modes, frequencies, and spatial location. These gathers provide a physically interpretable decomposition of the wavefield and establish a direct link between measured data residuals and the model parameters that control them. However, the current formulation is largely based on high-frequency asymptotic propagation and therefore inherits the limitations of ray theory. The finite-frequency trajectories developed in the present paper provide a natural extension of this framework. By replacing frequency-independent rays with frequency-dependent stationary paths, Eigenwaves introduce finite-frequency sensitivity directly into the phase-space representation and offer a physically consistent description of wave propagation, focusing, multipathing, and parameter sensitivity. More broadly, the Eigenwave formulation suggests a pathway toward a finite-frequency phase-space inversion methodology, in which residuals measured on local angle-domain gathers are associated with the finite-frequency trajectories that generate them. Such a framework could bridge the gap between asymptotic angle-domain tomography and full-waveform inversion, combining the interpretability of local angle-domain analysis with the physics of finite-frequency wave propagation.

The paper is organized as follows. Section 2 recalls the Eigenray formulation. Section 3 summarizes the trajectory-mechanics equations of Vasco and Nihei (2019a) and establishes the effective-medium equivalence. Section 4 derives the finite-frequency Euler-Lagrange equation, the matching Hamiltonian and the weak (finite-element) form. Section 5 derives the finite-frequency Hessians of the Lagrangian and the Jacobi dynamic ray-tracing equation, together with its initial conditions, ray Jacobian, geometric spreading, and caustic criteria, and establishes the equivalence with Hamiltonian dynamic ray tracing. Section 6 develops the weak-coupling theory (first order in $W$), leading to an inhomogeneous Jacobi equation for the finite-frequency path correction and to a closed-form phase-time correction. Section 7 discusses the self-consistent coupling between trajectories and amplitudes and the resulting workflow, while Section 8 analyzes the behavior of the closure near caustics and admissible regularizations of the amplitude term. Section 9 presents a two-dimensional numerical example in which a focusing low-velocity inclusion produces multipathing and a cusp caustic at the receiver and compares the resulting Eigenwave and Eigenray seismograms with finite-difference solutions. The example demonstrates that the Eigenwave equations are exact when the finite-frequency potential is known exactly, but that their ray-closed form alone does not accurately reproduce amplitudes at caustics. It further shows that a phase-space (Maslov) closure field constructed from the ray family accurately reconstructs the wavefield across the caustic region, whereas the finite-frequency phase correction remains reliable primarily away from strong caustic singularities. Section 10 concludes.

## 2 The Eigenray method: a summary

We use the notation of Koren and Ravve (2021); Ravve and Koren (2021). Let $\mathbf{x}(s)$ be a trajectory between the source $S$ and the receiver $R$, parametrized by the arclength $s$, and let $\mathbf{r}(s) = \dot{\mathbf{x}} = \mathrm{d}\mathbf{x}/\mathrm{d}s$ be the unit ray-direction (ray-velocity direction) vector, $\mathbf{r} \cdot \mathbf{r} = 1$. Fermat's principle (Bona and Slawinski, 2003) states that the traveltime

$$t = \int_S^R L\,(\mathbf{x}, \mathbf{r})\mathrm{d}s \to \text{stationary}, \quad L(\mathbf{x}, \mathbf{r}) = \frac{\mathrm{d}t}{\mathrm{d}s} = \frac{\sqrt{\mathbf{r} \cdot \mathbf{r}}}{v_{\mathrm{ray}}(\mathbf{x}, \mathbf{r})} = \frac{\sqrt{\mathbf{r} \cdot \mathbf{r}}}{v_{\mathrm{ray}}\left[\tilde{\mathbf{C}}(\mathbf{x}), \mathbf{r}\right]} \tag{1}$$

is stationary along the ray, where $v_{\mathrm{ray}}$ is the ray (group) velocity magnitude and $\tilde{\mathbf{C}}(\mathbf{x})$ the density-normalized stiffness tensor. The square root is kept because it affects the directional derivatives. The Lagrangian is a first-degree homogeneous function of $\mathbf{r}$; by Euler's theorem and its consequences (Bliss, 1916),

$$L_{\mathbf{r}} \cdot \mathbf{r} = L, \qquad L_{\mathbf{xr}} \cdot \mathbf{r} = \mathbf{r} \cdot L_{\mathbf{rx}} = L_{\mathbf{x}}, \qquad L_{\mathbf{rr}} \cdot \mathbf{r} = \mathbf{0}. \tag{2}$$

The stationarity condition is the Euler-Lagrange equation

$$\frac{\mathrm{d}}{\mathrm{d}s} L_{\mathbf{r}} = L_{\mathbf{x}}, \qquad L_{\mathbf{r}} = \frac{\mathbf{r}}{v_{\mathrm{ray}}} - \frac{\nabla_{\mathbf{r}} v_{\mathrm{ray}}}{v_{\mathrm{ray}}^2} = \mathbf{p}, \qquad L_{\mathbf{x}} = -\frac{\nabla_{\mathbf{x}} v_{\mathrm{ray}}}{v_{\mathrm{ray}}^2}, \tag{3}$$

where $\mathbf{p}$ is the slowness vector (the momentum conjugate to $\mathbf{x}$), decomposed into a (radial) component $\mathbf{r}/v_{\mathrm{ray}}$ along the ray and a (traverse) component $-\nabla_{\mathbf{r}} v_{\mathrm{ray}}/v_{\mathrm{ray}}^2 = \mathbf{r} \times \mathbf{p} \times \mathbf{r}$ in the plane normal to the ray. The gradients and Hessians of the Lagrangian are (eq. 18 of Koren and Ravve, 2021)

$$
\begin{aligned}
L_{\mathbf{xx}} &= 2\frac{\nabla_{\mathbf{x}} v_{\mathrm{ray}} \otimes \nabla_{\mathbf{x}} v_{\mathrm{ray}}}{v_{\mathrm{ray}}^3} - \frac{\nabla_{\mathbf{x}}\nabla_{\mathbf{x}} v_{\mathrm{ray}}}{v_{\mathrm{ray}}^2}, \\
L_{\mathbf{xr}} &= (L_{\mathbf{rx}})^{\mathsf{T}} = -\frac{\nabla_{\mathbf{x}} v_{\mathrm{ray}} \otimes \mathbf{r}}{v_{\mathrm{ray}}^2} + 2\frac{\nabla_{\mathbf{x}} v_{\mathrm{ray}} \otimes \nabla_{\mathbf{r}} v_{\mathrm{ray}}}{v_{\mathrm{ray}}^3} - \frac{\nabla_{\mathbf{x}}\nabla_{\mathbf{r}} v_{\mathrm{ray}}}{v_{\mathrm{ray}}^2}, \\
L_{\mathbf{rr}} &= \frac{\mathbf{I} - \mathbf{r}\otimes\mathbf{r}}{v_{\mathrm{ray}}} - \frac{\mathbf{r}\otimes\nabla_{\mathbf{r}} v_{\mathrm{ray}} + \nabla_{\mathbf{r}} v_{\mathrm{ray}} \otimes \mathbf{r}}{v_{\mathrm{ray}}^2} + 2\frac{\nabla_{\mathbf{r}} v_{\mathrm{ray}} \otimes \nabla_{\mathbf{r}} v_{\mathrm{ray}}}{v_{\mathrm{ray}}^3} - \frac{\nabla_{\mathbf{r}}\nabla_{\mathbf{r}} v_{\mathrm{ray}}}{v_{\mathrm{ray}}^2}
\end{aligned}
\tag{4}
$$

where the spatial and directional gradients and Hessians of the ray velocity are computed as in Ravve and Koren (2019). The arclength-related Hamiltonian $H(\mathbf{x},\mathbf{p})$ that matches $L$ through the Legendre transformation $L = \mathbf{p}\cdot\mathbf{r} - H$ is obtained from the (vanishing) reference Christoffel Hamiltonian,

$$
\begin{gathered}
H^{\bar{\tau}}(\mathbf{x},\mathbf{p}) = \det[\boldsymbol{\Gamma}(\mathbf{x},\mathbf{p}) - \mathbf{I}] = 0, \\
\boldsymbol{\Gamma}(\mathbf{x},\mathbf{p}) = \mathbf{p}\cdot\tilde{\mathbf{C}}(\mathbf{x})\cdot\mathbf{p} \quad , \quad H(\mathbf{x},\mathbf{p}) = \frac{H^{\bar{\tau}}(\mathbf{x},\mathbf{p})}{\sqrt{H^{\bar{\tau}}_{\mathbf{p}}\cdot H^{\bar{\tau}}_{\mathbf{p}}}},
\end{gathered}
\tag{5}
$$

with the kinematic equations $\mathrm{d}\mathbf{x}/\mathrm{d}s = H_{\mathbf{p}} = \mathbf{r}$, $\mathrm{d}\mathbf{p}/\mathrm{d}s = -H_{\mathbf{x}}$. In the Eigenray finite-element solver (Koren and Ravve, 2021) the weak form of (3) is minimized with respect to nodal locations and directions of Hermite elements; the required inputs at any point are $v_{\mathrm{ray}}$, its gradients and Hessians.

Along the resolved stationary path (the central ray) the dynamics is governed by the Jacobi equation, obtained by applying the Euler-Lagrange equation to the second traveltime variation (Bliss, 1916; Ravve and Koren, 2021),

$$
\frac{\mathrm{d}}{\mathrm{d}s}(L_{\mathbf{rx}}\cdot\mathbf{y} + L_{\mathbf{rr}}\cdot\dot{\mathbf{y}}) = L_{\mathbf{xx}}\cdot\mathbf{y} + L_{\mathbf{xr}}\cdot\dot{\mathbf{y}}, \qquad \mathbf{y}\cdot\mathbf{r} = 0,
\tag{6}
$$

where $\mathbf{y}(s) = \delta\mathbf{x}(s)$ is the normal shift of a paraxial ray (note that, to keep consistent notation, the vector $\mathbf{y}$ in this paper replaces its equivalent vector $\mathbf{u}$ used in Ravve and Koren, 2021). Because of (2) only two of the three scalar equations are independent, any tangential shift $\rho_t(s)\mathbf{r}(s)$ is a (trivial) solution, and the fundamental solution is a combination of four basic solutions with point-source ($\mathbf{y}_{1,2}$) and plane-wave ($\mathbf{y}_{3,4}$) initial conditions. The ray Jacobian $J = \mathbf{y}_1 \times \mathbf{y}_2 \cdot \mathbf{r}$, its zeros (caustics) and the relative geometric spreading $L_{\mathrm{GS}}$ follow (Ravve and Koren, 2021).

## 3 Trajectory mechanics and the effective-medium equivalence

### 3.1 The finite-frequency Hamiltonian

Following Vasco and Nihei (2019a), consider the frequency-domain elastodynamic equation away from the source, $-\omega^2\rho\mathbf{u} = \nabla_\mathbf{x}\cdot\left(\rho\,\tilde{\mathbf{C}}:\nabla_\mathbf{x}\mathbf{u}\right)$, where $\tilde{\mathbf{C}}(\mathbf{x})$ is the density-normalized fourth-order stiffness tensor of equation (1), $\rho(\mathbf{x})$ is the density, and the double contraction $\rho\,\tilde{\mathbf{C}}:\nabla_\mathbf{x}\mathbf{u}$ is the stress tensor of the displacement field (in components, $\left(\tilde{\mathbf{C}}:\nabla_\mathbf{x}\mathbf{u}\right)_{ij} = \tilde{c}_{ijkl}\ \partial u_k/\ \partial x_l$). The coherent body wave is written in exact polar form

$$\mathbf{u}(\mathbf{x},\omega) = \mathbf{a}(\mathbf{x},\omega)e^{i\varphi(\mathbf{x},\omega)}, \tag{7}$$

with real amplitude vector $\mathbf{a} = a\hat{\mathbf{a}}$ and real phase $\varphi$. Substituting (7) and separating real and imaginary parts yields two coupled equations. With the slowness vector $\mathbf{p} = \nabla_\mathbf{x}\varphi/\omega$, the real part is

$$[\boldsymbol{\Gamma}(\mathbf{x},\mathbf{p}) - \mathbf{I}]\mathbf{a} = \mathbf{w}, \qquad \mathbf{w}(\mathbf{x},\omega) = \frac{\nabla_\mathbf{x}\cdot\left(\rho\,\tilde{\mathbf{C}}:\nabla_\mathbf{x}\mathbf{a}\right)}{\rho\,\omega^2}, \tag{8}$$

and the imaginary part is the transport equation $\nabla_\mathbf{x}\cdot\left[\rho\,\tilde{\mathbf{C}}:(\mathbf{a}\otimes\mathbf{p})\right] + \rho\left(\tilde{\mathbf{C}}:\nabla_\mathbf{x}\mathbf{a}\right)\mathbf{p} = \mathbf{0}$, with $\mathbf{a}\otimes\mathbf{p}$ the dyadic product. Equation (8) differs from the Christoffel (eikonal) equation only by the right-hand side, which is $O(\omega^{-2})$ and contains the second spatial derivatives of the amplitude field: this is exactly the term neglected in high-frequency asymptotic ray theory. Projecting on $\hat{\mathbf{a}}$ and dividing by $a$, Vasco and Nihei (2019a) define the Hamiltonian

$$\begin{aligned} H_{VN}(\mathbf{x},\mathbf{p},\omega) &= \hat{\mathbf{a}}\cdot\boldsymbol{\Gamma}(\mathbf{x},\mathbf{p})\,\hat{\mathbf{a}} - W(\mathbf{x},\omega) = 1, \\ W(\mathbf{x},\omega) &= \frac{\hat{\mathbf{a}}\cdot\mathbf{w}}{a} = \frac{\hat{\mathbf{a}}\cdot\nabla_\mathbf{x}\cdot\left(\rho\,\tilde{\mathbf{C}}:\nabla_\mathbf{x}\mathbf{a}\right)}{\rho\,\omega^2\,a}. \end{aligned} \tag{9}$$

For a wave that propagates as a single mode - in this paper we focus on the quasi-compressional (qP) wave, but the derivation holds for any mode - the amplitude vector is aligned with the corresponding polarization (eigenvector) $\hat{\mathbf{g}}$ of the Christoffel tensor $\boldsymbol{\Gamma}(\mathbf{x},\mathbf{p}) = \mathbf{p}\cdot\tilde{\mathbf{C}}(\mathbf{x})\cdot\mathbf{p}$ of equation (5), $\boldsymbol{\Gamma}\hat{\mathbf{g}} = \lambda\hat{\mathbf{g}}$, so that

$$H_{VN}(\mathbf{x},\mathbf{p},\omega) = \lambda(\mathbf{x},\mathbf{p}) - W(\mathbf{x},\omega) = 1 \quad , \quad \lambda(\mathbf{x},\mathbf{p}) = \hat{\mathbf{g}}\cdot\boldsymbol{\Gamma}(\mathbf{x},\mathbf{p})\,\hat{\mathbf{g}}, \tag{10}$$

where $\lambda(\mathbf{x},\mathbf{p})$ is the qP eigenvalue of $\boldsymbol{\Gamma}$ (a second-degree homogeneous function of $\mathbf{p}$) and $W$ is evaluated with the qP amplitude field.

It is convenient to introduce the companion sandwich of the Christoffel tensor: the *ray-velocity tensor* $\widetilde{\mathbf{A}}(\hat{\mathbf{g}})$, the symmetric second-order tensor obtained by contracting $\tilde{\mathbf{C}}$ with the polarization over its outer index pair, defined coordinate-free by the bilinear identity $\mathbf{b}\cdot$ $\widetilde{\mathbf{A}}(\hat{\mathbf{g}})\,\mathbf{c} = (\hat{\mathbf{g}} \otimes \mathbf{b}) : \tilde{\mathbf{C}} : (\hat{\mathbf{g}} \otimes \mathbf{c})$ for all vectors $\mathbf{b}$, $\mathbf{c}$, and linked to $\boldsymbol{\Gamma}$ by the reciprocity identity $\hat{\mathbf{g}} \cdot \boldsymbol{\Gamma}(\mathbf{p})\,\hat{\mathbf{g}} = \mathbf{p} \cdot \widetilde{\mathbf{A}}(\hat{\mathbf{g}})\,\mathbf{p}$, so that $\lambda = \mathbf{p} \cdot \widetilde{\mathbf{A}}(\hat{\mathbf{g}})\,\mathbf{p}$ and the ray (group) velocity is $\widetilde{\mathbf{A}}(\hat{\mathbf{g}})\,\mathbf{p}$. The trajectory equations in Vasco and Nihei (2019a) are Hamilton's equations for (9),

$$\frac{d\mathbf{x}}{dT} = \frac{1}{2}\frac{\partial H_{VN}}{\partial \mathbf{p}} = \widetilde{\mathbf{A}}(\hat{\mathbf{g}})\,\mathbf{p},$$
$$\frac{d\mathbf{p}}{dT} = -\frac{1}{2}\frac{\partial H_{VN}}{\partial \mathbf{x}} = -\frac{1}{2}\nabla_{\mathbf{x}}\left[\mathbf{p} \cdot \widetilde{\mathbf{A}}(\hat{\mathbf{g}})\,\mathbf{p}\right] + \frac{1}{2}\nabla_{\mathbf{x}} W, \tag{11}$$

where $T$ is the flow parameter, and the spatial gradient in the second equation is taken at fixed $\mathbf{p}$ and $\hat{\mathbf{g}}$. In the derivation, the amplitude field and hence $W$, is treated as a known function of $\mathbf{x}$ and $\omega$ that does not depend on $\mathbf{p}$; the trajectories then depend on the frequency through $W$. When $|W| \ll 1$ the equations reduce to the classical anisotropic ray equations, and the magnitude of $W$ measures the validity of the eikonal approximation.

**Remark 1** (Amplitude direction versus polarization). The unit amplitude vector $\hat{\mathbf{a}}$ of (7) and the polarization $\hat{\mathbf{g}}$ of (10) are distinct objects. $\hat{\mathbf{a}}$ is an exact, frequency-dependent property of the wavefield: it is the direction of the displacement amplitude $\mathbf{a}$ at $(\mathbf{x}, \omega)$, whatever that field is, and it carries near-field, mode-mixing and interference effects. It is in general not an eigenvector of $\boldsymbol{\Gamma}$, since the right-hand side $\mathbf{w}$ of (8) tilts $\mathbf{a}$ away from the Christoffel eigenvectors by $O(\omega^{-2})$. The polarization $\hat{\mathbf{g}}$, in contrast, is a local algebraic quantity: the eigenvector of $\boldsymbol{\Gamma}(\mathbf{x}, \mathbf{p})$ for the selected mode, which depends only on the medium and the slowness, not on ω or on the wavefield, and coincides with the polarization of high-frequency ray theory. The replacement $\hat{\mathbf{a}} \to \hat{\mathbf{g}}$ that turns (9) into (10) is therefore an approximation, but a second-order one: the Rayleigh quotient $\hat{\mathbf{a}} \cdot \boldsymbol{\Gamma}\,\hat{\mathbf{a}}$ is stationary at the eigenvector $\hat{\mathbf{g}}$, so a tilt $\delta$ of $\hat{\mathbf{a}}$ off $\hat{\mathbf{g}}$ changes it only by $O(\delta^2) = O(\omega^{-4})$, beyond the $O(\omega^{-2})$ accuracy of W itself. What survives at first order is the scalar projection $W = \hat{\mathbf{a}} \cdot \mathbf{w}/a$ of (9), which is evaluated with the actual (qP) amplitude field. The distinction becomes material precisely where a single, well-separated mode cannot be assumed - near caustics and in regions of mode coupling - which are the situations addressed in Sections 8–10.

### 3.2 Effective-medium equivalence

Throughout, we assume that $W(\mathbf{x}, \omega)$ is a given, smooth field with

$$1 + W(\mathbf{x}, \omega) > 0, \qquad n(\mathbf{x}, \omega) \equiv \sqrt{1 + W(\mathbf{x}, \omega)}, \tag{12}$$

which is the case in the propagating regime (Vasco and Nihei's numerical examples show $|W|$ well below 1 except at sharp layer edges and near the source at low frequency). Since $\lambda(\mathbf{x}, \mathbf{p})$ is homogeneous of degree two in $\mathbf{p}$ and its eigenvector $\hat{\mathbf{g}}$ is homogeneous of degree zero, the constraint (10) can be written as

$$\lambda(\mathbf{x}, \mathbf{p}) = n^2(\mathbf{x}, \omega) \Leftrightarrow \lambda\left(\mathbf{x}, \frac{\mathbf{p}}{n(\mathbf{x}, \omega)}\right) = 1 \Leftrightarrow \mathbf{p} \cdot \tilde{\mathbf{C}}^{\omega}(\mathbf{x}) \cdot \mathbf{p} \text{ has qP eigenvalue } 1, \tag{13}$$

where we have introduced the *effective, frequency-dependent* density-normalized stiffness tensor

$$\tilde{\mathbf{C}}^{\omega}(\mathbf{x}) = \frac{\tilde{\mathbf{C}}(\mathbf{x})}{1 + W(\mathbf{x},\omega)} = \frac{\tilde{\mathbf{C}}(\mathbf{x})}{n^2(\mathbf{x},\omega)} \tag{14}$$

Equation (13) is precisely the classical Christoffel (eikonal) equation of a medium with stiffness $\tilde{\mathbf{C}}^{\omega}$. We now show that not only the constraint surface but also the trajectories coincide.

**Proposition 1** (Effective-medium equivalence). For a fixed wave mode and a given field $\mathrm{W}(\mathbf{x},\omega)$ satisfying (12), the finite-frequency trajectories of Vasco and Nihei (2019a), i.e. the integral curves of (11) on the level set $\mathrm{H}_{\mathrm{VN}} = 1$, coincide, up to reparametrization, with the classical (eikonal) rays of the effective medium $\tilde{\mathbf{C}}^{\omega}(\mathbf{x})$ of (14).

*Proof.* Consider the two vanishing Hamiltonians $H_1(\mathbf{x},\mathbf{p}) = \lambda(\mathbf{x},\mathbf{p}) - 1 - W(\mathbf{x},\omega)$ and $H_2(\mathbf{x},\mathbf{p}) = \lambda(\mathbf{x},\mathbf{p})/n^2(\mathbf{x},\omega) - 1$, so that $H_2 = fH_1$ with $f = n^{-2} > 0$. The Hamiltonian vector fields are related by $X_{H_2} = fX_{H_1} + H_1X_f$; on the common zero level set $H_1 = H_2 = 0$ the second term vanishes and $X_{H_2} = fX_{H_1}$, i.e. the two flows have the same trajectories with flow parameters related by $\mathrm{d}\sigma_2 = n^{-2}\mathrm{d}\sigma_1$. Since $H_2$ is the classical Christoffel Hamiltonian of $\tilde{\mathbf{C}}^{\omega}$ (its zero level set is the qP slowness surface of $\tilde{\mathbf{C}}^{\omega}$, and Hamilton's equations for $H_2$ are the classical anisotropic ray equations of that medium), the claim follows. □

The proposition is the key to the whole extension. It says that, once the amplitude field (hence $W$) is fixed, the finite-frequency kinematics and dynamics are those of a classical anisotropic medium in which *all* stiffness components are locally rescaled by the same factor $n^{-2}(\mathbf{x},\omega)$. Since a uniform rescaling of the stiffness tensor rescales all velocities by the square root of the factor and leaves polarizations and the shape of the slowness and ray-velocity surfaces unchanged, we obtain immediately the finite-frequency phase and ray velocities, slowness vector and polarization:

$$\begin{gathered} v_{\mathrm{phs}}^{\omega}(\mathbf{x},\mathbf{n}) = \frac{v_{\mathrm{phs}}(\mathbf{x},\mathbf{n})}{n(\mathbf{x},\omega)}, \qquad v_{\mathrm{ray}}^{\omega}(\mathbf{x},\mathbf{r}) = \frac{v_{\mathrm{ray}}(\mathbf{x},\mathbf{r})}{n(\mathbf{x},\omega)}, \\ \mathbf{p}^{\omega}(\mathbf{x},\mathbf{r}) = n(\mathbf{x},\omega)\mathbf{p}(\mathbf{x},\mathbf{r}), \\ \hat{\mathbf{g}}^{\omega} = \hat{\mathbf{g}}, \end{gathered} \tag{15}$$

where $\mathbf{n} = \mathbf{p}/|\mathbf{p}|$ is the phase direction and $\mathbf{p}(\mathbf{x},\mathbf{r})$ is the classical slowness vector for the ray direction $\mathbf{r}$ (computed by the Newton solver described in Ravve and Koren, 2019). The map $\mathbf{r} \leftrightarrow \mathbf{n}$ between ray and phase directions is unaffected, and so are the angles between phase and ray velocities. The finite-frequency effects therefore enter only through the scalar field $n(\mathbf{x},\omega)$ and its first and second spatial derivatives,

$$\nabla_{\mathbf{x}} n = \frac{\nabla_{\mathbf{x}} W}{2n}, \qquad \nabla_{\mathbf{x}}\nabla_{\mathbf{x}} n = \frac{\nabla_{\mathbf{x}}\nabla_{\mathbf{x}} W}{2n} - \frac{\nabla_{\mathbf{x}} W \otimes \nabla_{\mathbf{x}} W}{4n^3}. \tag{16}$$

The equivalence is illustrated in Fig. 1 for a tilted transversely isotropic medium: the slowness and ray-velocity surfaces are rescaled uniformly by $n$ and $1/n$, while their shape, the polarizations and the map between ray and phase directions are unchanged.

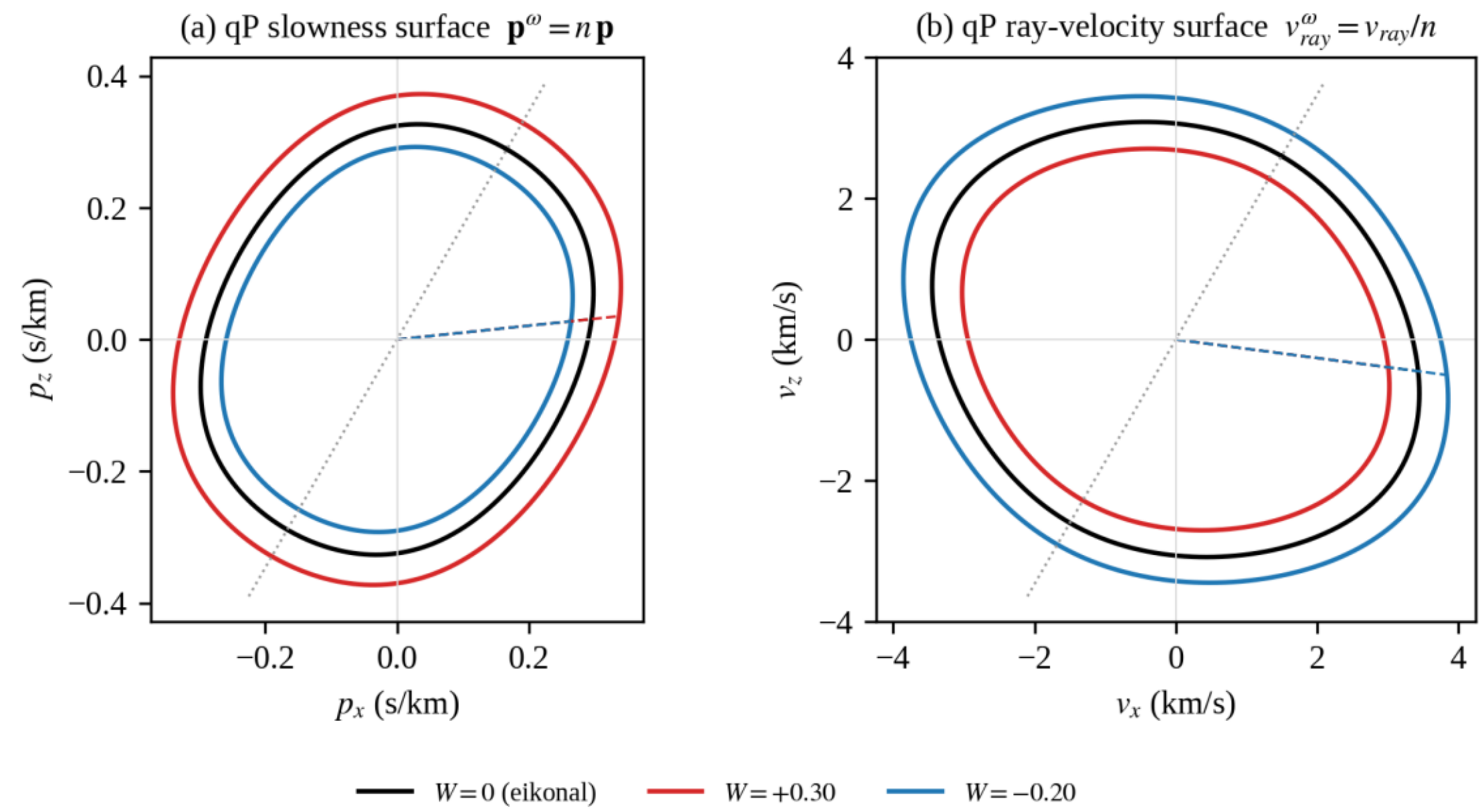

**Figure 1.** Effective-medium equivalence (Proposition 1) for a TTI medium ($v_{P0} = 3$ km/s, $v_{S0} = 1.5$ km/s, $\varepsilon = 0.25$, $\delta = 0.05$, symmetry axis tilted by 30°, dotted line), in the plane containing the axis. (a) qP slowness surface for $W = 0$ (eikonal), $W = +0.30$ and $W = -0.20$; the finite-frequency slowness is $\mathbf{p}^{\omega} = n\,\mathbf{p}$, $n = \sqrt{1 + W}$. (b) The corresponding qP ray-velocity surfaces, $v^{\omega}_{\mathrm{ray}} = v_{\mathrm{ray}}/n$. The surfaces are rescaled uniformly, so that their shape, the polarizations and the map between phase and ray directions are unchanged (dashed: the slowness vectors of one phase direction in (a) and the corresponding ray-velocity vectors in (b), collinear for the three values of $W$); the finite-frequency effects enter only through the scalar field $n(\mathbf{x}, \omega)$.

**Remark 2** (Optical analogy). In the isotropic acoustic case $\lambda = v^2 \mathbf{p} \cdot \mathbf{p}$ and (13) reads $|\mathbf{p}| = n/v$: the medium acquires the frequency-dependent index of refraction $n = \sqrt{1 + W}$ relative to the eikonal velocity $v$, exactly as in the "exact ray theory" of Foreman (1989) for the Helmholtz equation. Proposition 1 extends this to general anisotropy: the index is a scalar and multiplies the whole stiffness tensor.

**Remark 3** (The flow parameter of Vasco and Nihei versus the phase time). Along a trajectory of (11), $\mathbf{p} \cdot d\mathbf{x}/dT = \mathbf{p} \cdot 1/2\, \nabla_{\mathbf{p}}\lambda = \lambda = n^2$ by Euler's theorem, hence $\mathbf{p} \cdot d\mathbf{x} = n^2\, dT$. The phase of the wave (7) accumulates as $d\varphi = \omega\, \mathbf{p} \cdot d\mathbf{x}$, so the phase time $\tau = \varphi/\omega$ satisfies $d\tau = \mathbf{p} \cdot d\mathbf{x} = (1 + W)dT$, which is eq. (46) of Vasco and Nihei (2019a). In this paper the traveltime functional of the Eigenwave is the phase time $\tau$: it is the quantity whose stationarity defines the trajectory (Section 4), and it reduces to the eikonal traveltime as $W \to 0$.

## 4 The Eigenwave Lagrangian and the finite-frequency Euler-Lagrange equation

### 4.1 The Lagrangian

By Proposition 1 and (15), the arclength-related Eigenwave Lagrangian is the classical Eigenray Lagrangian (1) evaluated in the effective medium:

$$L^{\omega}(\mathbf{x},\mathbf{r};\omega)=\frac{\mathrm{d}\tau}{\mathrm{d}s}=\frac{\sqrt{\mathbf{r}\cdot\mathbf{r}}}{v_{\mathrm{ray}}^{\omega}(\mathbf{x},\mathbf{r})}=n(\mathbf{x},\omega)\frac{\sqrt{\mathbf{r}\cdot\mathbf{r}}}{v_{\mathrm{ray}}(\mathbf{x},\mathbf{r})}=n(\mathbf{x},\omega)L(\mathbf{x},\mathbf{r}) \tag{17}$$

and Fermat's principle for the Eigenwave reads

$$\tau=\int_S^R L^{\omega}\,(\mathbf{x},\mathbf{r};\omega)\mathrm{d}s=\int_S^R n\,(\mathbf{x},\omega)L(\mathbf{x},\mathbf{r})\mathrm{d}s\to\text{stationary},\qquad \mathbf{x}(S),\mathbf{x}(R)\,\text{fixed.} \tag{18}$$

Since $n$ does not depend on $\mathbf{r}$, $L^{\omega}$ inherits the first-degree homogeneity in $\mathbf{r}$, and the Bliss identities (2) hold for $L^{\omega}$:

$$L^{\omega}_{\mathbf{r}}\cdot\mathbf{r}=L^{\omega},\qquad L^{\omega}_{\mathbf{xr}}\cdot\mathbf{r}=\mathbf{r}\cdot L^{\omega}_{\mathbf{rx}}=L^{\omega}_{\mathbf{x}},\qquad L^{\omega}_{\mathbf{rr}}\cdot\mathbf{r}=\mathbf{0}. \tag{19}$$

The stationary value of (18) is the finite-frequency phase time $\tau=\int\mathbf{p}^{\omega}\cdot\mathrm{d}\mathbf{x}$ (Remark 3), and the classical traveltime is recovered as $W\to 0$ (equivalently $\omega\to\infty$).

### 4.2 Gradients and the Euler-Lagrange equation

Differentiating (17),

$$L^{\omega}_{\mathbf{r}}=nL_{\mathbf{r}}=n\mathbf{p}=\mathbf{p}^{\omega},\qquad L^{\omega}_{\mathbf{x}}=nL_{\mathbf{x}}+L\nabla_{\mathbf{x}}n=-n\frac{\nabla_{\mathbf{x}}v_{\mathrm{ray}}}{v_{\mathrm{ray}}^2}+\frac{\nabla_{\mathbf{x}}W}{2nv_{\mathrm{ray}}}, \tag{20}$$

where we used $L=1/v_{\mathrm{ray}}$ (for $\mathbf{r}\cdot\mathbf{r}=1$) and (16). The first relation is the finite-frequency momentum equation: the momentum conjugate to $\mathbf{x}$ is the finite-frequency slowness $\mathbf{p}^{\omega}=n\mathbf{p}$, in agreement with (15). Its decomposition into ray-tangent and ray-normal components is

$$\mathbf{p}^{\omega}=n\left(\frac{\mathbf{r}}{v_{\mathrm{ray}}}-\frac{\nabla_{\mathbf{r}}v_{\mathrm{ray}}}{v_{\mathrm{ray}}^2}\right)=\frac{\mathbf{r}}{v_{\mathrm{ray}}^{\omega}}-\frac{\nabla_{\mathbf{r}}v_{\mathrm{ray}}^{\omega}}{\left(v_{\mathrm{ray}}^{\omega}\right)^2},\qquad \mathbf{p}^{\omega}\cdot\mathbf{r}=\frac{1}{v_{\mathrm{ray}}^{\omega}}=\frac{n}{v_{\mathrm{ray}}}. \tag{21}$$

The finite-frequency Euler-Lagrange (kinematic Eigenwave) equation $\mathrm{d}L^{\omega}_{\mathbf{r}}/\mathrm{d}s=L^{\omega}_{\mathbf{x}}$ becomes

$$\frac{\mathrm{d}}{\mathrm{d}s}\left[n\left(\frac{\mathbf{r}}{v_{\mathrm{ray}}}-\frac{\nabla_{\mathbf{r}}v_{\mathrm{ray}}}{v_{\mathrm{ray}}^2}\right)\right]=-n\frac{\nabla_{\mathbf{x}}v_{\mathrm{ray}}}{v_{\mathrm{ray}}^2}+\frac{\nabla_{\mathbf{x}}W}{2nv_{\mathrm{ray}}} \tag{22}$$

or, in terms of the finite-frequency slowness and equivalently in terms of the effective ray velocity,

$$\frac{\mathrm{d}\mathbf{p}^{\omega}}{\mathrm{d}s}=nL_{\mathbf{x}}+L\nabla_{\mathbf{x}}n=-\frac{\nabla_{\mathbf{x}}v_{\mathrm{ray}}^{\omega}}{\left(v_{\mathrm{ray}}^{\omega}\right)^2},\qquad \mathbf{p}^{\omega}=\frac{\mathbf{r}}{v_{\mathrm{ray}}^{\omega}}-\frac{\nabla_{\mathbf{r}}v_{\mathrm{ray}}^{\omega}}{\left(v_{\mathrm{ray}}^{\omega}\right)^2}. \tag{23}$$

Equation (22) is a nonlinear, second-order, vector-form ODE for $\mathbf{x}(s)$; the term $\nabla_{\mathbf{x}}W/\left(2nv_{\mathrm{ray}}\right)$ is the finite-frequency "force" that bends the trajectory towards regions of larger $W$ (larger effective slowness). Expanding the left-hand side, $\mathrm{d}(n\mathbf{p})/\mathrm{d}s=n\dot{\mathbf{p}}+(\nabla_{\mathbf{x}}n\cdot\mathbf{r})\mathbf{p}$, and using $L=\mathbf{p}\cdot\mathbf{r}$, (22) can also be written as

$$n[\dot{\mathbf{p}}-L_{\mathbf{x}}]=(\mathbf{p}\cdot\mathbf{r})\nabla_{\mathbf{x}}n-(\nabla_{\mathbf{x}}n\cdot\mathbf{r})\mathbf{p}=\mathbf{r}\times(\nabla_{\mathbf{x}}n\times\mathbf{p}), \tag{24}$$

which shows explicitly that the departure from the classical ray equation $\dot{\mathbf{p}}=L_{\mathbf{x}}$ is driven by the component of $\nabla_{\mathbf{x}}n$ that is not aligned with the ray. In the isotropic case, $\mathbf{p}=\mathbf{r}/v$, $L_{\mathbf{x}}=-\nabla_{\mathbf{x}}v/v^2$, and (24) reduces to the ray equation of a medium with velocity $v/n$, as it must.

Equation (22) is consistent with the Hamiltonian trajectory equations (11): eliminating the flow parameter through $\mathrm{d}s=|\mathrm{d}\mathbf{x}/\mathrm{d}T|\mathrm{d}T$ and using $\lambda=n^2$ one recovers (11) from (22) and vice versa (Appendix D). Its advantage is that it is expressed in the degrees of freedom of the Eigenray method - locations and ray directions - and is directly amenable to the weak formulation.

### 4.3 The matching Hamiltonian and Legendre transformation

Following (5), the reference (vanishing) finite-frequency Christoffel Hamiltonian and its arclength-related counterpart are

$$H^{\bar{\tau},\omega}(\mathbf{x},\mathbf{p})=\det\left[\frac{\mathbf{\Gamma}(\mathbf{x},\mathbf{p})}{n^2(\mathbf{x},\omega)}-\mathbf{I}\right]=0,\qquad H^{\omega}(\mathbf{x},\mathbf{p})=\frac{H^{\bar{\tau},\omega}(\mathbf{x},\mathbf{p})}{\sqrt{H_{\mathbf{p}}^{\bar{\tau},\omega}\cdot H_{\mathbf{p}}^{\bar{\tau},\omega}}}, \tag{25}$$

with kinematic equations $\mathrm{d}\mathbf{x}/\mathrm{d}s=H_{\mathbf{p}}^{\omega}=\mathbf{r}$, $\mathrm{d}\mathbf{p}/\mathrm{d}s=-H_{\mathbf{x}}^{\omega}$, and Legendre transformation $L^{\omega}(\mathbf{x},\mathbf{r})=\mathbf{p}^{\omega}\cdot\mathbf{r}-H^{\omega}(\mathbf{x},\mathbf{p}^{\omega})$ (with $H^{\omega}=0$ on the trajectory, $L^{\omega}=\mathbf{p}^{\omega}\cdot\mathbf{r}=1/v_{\mathrm{ray}}^{\omega}$). For a single mode one may equivalently use $H^{\bar{\tau},\omega}=\lambda(\mathbf{x},\mathbf{p})-n^2(\mathbf{x},\omega)$, which is $H_{\mathrm{VN}}-1$ of Vasco and Nihei (2019a); by Proposition 1 both reference Hamiltonians generate the same arclength flow. The inverse problem of finding the slowness given the ray direction, needed at every node of the finite-element solver, is solved by the classical Newton procedure of Ravve and Koren (2019) on $\tilde{\mathbf{C}}$ (or on $\tilde{\mathbf{C}}^{\omega}$; both give the same $\mathbf{n}$), followed by the scaling $\mathbf{p}^{\omega}=n\mathbf{p}$.

## 4.4 Weak formulation and finite-element implementation

The weak Galerkin form of (22) within a Hermite element with internal parameter $\xi \in [-1,1]$ and metric $l(\xi)/2 = \mathrm{d}s/\mathrm{d}\xi$ is identical to eq. (33) of Koren and Ravve (2021) with $L \to L^{\omega}$:

$$
\begin{aligned}
&\int_{-1}^{+1} [1/2\, l(\xi) h_I(\xi) L^{\omega}_{\mathbf{x}} + h_I{}'(\xi) L^{\omega}_{\mathbf{r}}]\, \mathrm{d}\xi = \mathbf{0}, \\
&\int_{-1}^{+1} [1/2\, l(\xi) d_I(\xi) L^{\omega}_{\mathbf{x}} + d_I{}'(\xi) L^{\omega}_{\mathbf{r}}]\, \mathrm{d}\xi = \mathbf{0},
\end{aligned}
\tag{26}
$$

where $h_I, d_I$ are the location and direction Hermite shape functions of node $I$. The nodal degrees of freedom, the node-distribution and direction-normalization soft constraints, the target functions for minimum and saddle-point stationary solutions, the Newton/gradient minimization and the assembly of the global gradient and Hessian are unchanged. *The only modification is in the ray-velocity kernel:* wherever the Eigenray code evaluates $v_{\mathrm{ray}}$, $\nabla_{\mathbf{x}} v_{\mathrm{ray}}$, $\nabla_{\mathbf{r}} v_{\mathrm{ray}}$, $\nabla_{\mathbf{x}}\nabla_{\mathbf{x}} v_{\mathrm{ray}}$, $\nabla_{\mathbf{x}}\nabla_{\mathbf{r}} v_{\mathrm{ray}}$, $\nabla_{\mathbf{r}}\nabla_{\mathbf{r}} v_{\mathrm{ray}}$, the Eigenwave code evaluates the effective quantities

$$
\begin{aligned}
v^{\omega}_{\mathrm{ray}} &= \frac{v_{\mathrm{ray}}}{n}, \nabla_{\mathbf{x}} v^{\omega}_{\mathrm{ray}} = \frac{\nabla_{\mathbf{x}} v_{\mathrm{ray}}}{n} - \frac{v_{\mathrm{ray}} \nabla_{\mathbf{x}} n}{n^2}, \nabla_{\mathbf{r}} v^{\omega}_{\mathrm{ray}} = \frac{\nabla_{\mathbf{r}} v_{\mathrm{ray}}}{n}, \\
\nabla_{\mathbf{x}}\nabla_{\mathbf{x}} v^{\omega}_{\mathrm{ray}} &= \frac{\nabla_{\mathbf{x}}\nabla_{\mathbf{x}} v_{\mathrm{ray}}}{n} - \frac{\nabla_{\mathbf{x}} v_{\mathrm{ray}} \otimes \nabla_{\mathbf{x}} n + \nabla_{\mathbf{x}} n \otimes \nabla_{\mathbf{x}} v_{\mathrm{ray}}}{n^2} - \frac{v_{\mathrm{ray}} \nabla_{\mathbf{x}}\nabla_{\mathbf{x}} n}{n^2} + \frac{2 v_{\mathrm{ray}}}{} \\
\nabla_{\mathbf{x}}\nabla_{\mathbf{r}} v^{\omega}_{\mathrm{ray}} &= \frac{\nabla_{\mathbf{x}}\nabla_{\mathbf{r}} v_{\mathrm{ray}}}{n} - \frac{\nabla_{\mathbf{x}} n \otimes \nabla_{\mathbf{r}} v_{\mathrm{ray}}}{n^2}, \nabla_{\mathbf{r}}\nabla_{\mathbf{r}} v^{\omega}_{\mathrm{ray}} = \frac{\nabla_{\mathbf{r}}\nabla_{\mathbf{r}} v_{\mathrm{ray}}}{n},
\end{aligned}
\tag{27}
$$

with $\nabla_{\mathbf{x}} n$, $\nabla_{\mathbf{x}}\nabla_{\mathbf{x}} n$ from (16). Thus, the finite-frequency extension of the kinematic Eigenray solver requires the field $W(\mathbf{x}, \omega)$ and its first and second spatial derivatives, and nothing else. As in the classical case, $W$ has to be preconditioned (smoothed) so that these derivatives are stable; this is the same requirement that Vasco and Nihei (2019a) impose on the amplitude field.

# 5 The finite-frequency Jacobi equation: dynamic Eigenwaves

## 5.1 Hessians of the Eigenwave Lagrangian

Differentiating (20) once more (Appendix C) we obtain the four second-order tensors of the Eigenwave Lagrangian in terms of those of the Eigenray Lagrangian (4):

$$
\begin{aligned}
L^{\omega}_{\mathbf{xx}} &= n L_{\mathbf{xx}} + \nabla_{\mathbf{x}} n \otimes L_{\mathbf{x}} + L_{\mathbf{x}} \otimes \nabla_{\mathbf{x}} n + L \nabla_{\mathbf{x}}\nabla_{\mathbf{x}} n, \\
L^{\omega}_{\mathbf{xr}} &= n L_{\mathbf{xr}} + \nabla_{\mathbf{x}} n \otimes \mathbf{p}, L^{\omega}_{\mathbf{rx}} = (L^{\omega}_{\mathbf{xr}})^{\mathsf{T}} = n L_{\mathbf{rx}} + \mathbf{p} \otimes \nabla_{\mathbf{x}} n, \\
L^{\omega}_{\mathbf{rr}} &= n L_{\mathbf{rr}}.
\end{aligned}
\tag{28}
$$

Equivalently, (28) is (4) with $v_{\mathrm{ray}} \to v^{\omega}_{\mathrm{ray}}$ and the derivatives (27). Two properties are worth noting. First, $L^{\omega}_{\mathbf{rr}} = n L_{\mathbf{rr}}$: the directional Hessian is simply rescaled, so it remains positive semidefinite with $\mathbf{r}$ as the eigenvector of the zero eigenvalue and with the two nonzero eigenvalues $\lambda_{1,2}^{\omega} = n\lambda_{1,2}$ (their eigenvectors are unchanged). Second, the mixed Hessian acquires the rank-one term $\nabla_{\mathbf{x}} n \otimes \mathbf{p}$; the symmetric combination $L_{\mathbf{xr}}^{\omega} \cdot \mathbf{r} = nL_{\mathbf{x}} + (\nabla_{\mathbf{x}} n)(\mathbf{p} \cdot \mathbf{r}) = L_{\mathbf{x}}^{\omega}$ verifies (19).

### 5.2 The Jacobi equation

Applying the Euler-Lagrange equation to the second variation of the phase-time functional (18) along the resolved Eigenwave $\mathbf{x}(s)$ (Appendix C) yields the linear, second-order, vector-form Jacobi equation for the normal paraxial shift $\mathbf{y}(s) = \delta\mathbf{x}(s)$:

$$\frac{\mathrm{d}}{\mathrm{d}s}\left(L_{\mathbf{rx}}^{\omega} \cdot \mathbf{y} + L_{\mathbf{rr}}^{\omega} \cdot \dot{\mathbf{y}}\right) = L_{\mathbf{xx}}^{\omega} \cdot \mathbf{y} + L_{\mathbf{xr}}^{\omega} \cdot \dot{\mathbf{y}}, \qquad \mathbf{y} \cdot \mathbf{r} = 0, \dot{\mathbf{y}} \cdot \mathbf{r} + \mathbf{y} \cdot \dot{\mathbf{r}} = 0. \tag{29}$$

Substituting (28) and grouping the finite-frequency terms,

$$\begin{aligned}\frac{\mathrm{d}}{\mathrm{d}s}\left[n(L_{\mathbf{rx}} \cdot \mathbf{y} + L_{\mathbf{rr}} \cdot \dot{\mathbf{y}}) + (\nabla_{\mathbf{x}} n \cdot \mathbf{y})\mathbf{p}\right] = \; & n(L_{\mathbf{xx}} \cdot \mathbf{y} + L_{\mathbf{xr}} \cdot \dot{\mathbf{y}}) + (L_{\mathbf{x}} \cdot \mathbf{y} + \mathbf{p} \cdot \dot{\mathbf{y}})' \\ & + (\nabla_{\mathbf{x}} n \cdot \mathbf{y}) L_{\mathbf{x}} + L(\nabla_{\mathbf{x}} \nabla_{\mathbf{x}} n) \cdot \mathbf{y}.\end{aligned} \tag{30}$$

The scalar $L_{\mathbf{x}} \cdot \mathbf{y} + \mathbf{p} \cdot \dot{\mathbf{y}} = L_{\mathbf{x}} \cdot \mathbf{y} + L_{\mathbf{r}} \cdot \dot{\mathbf{y}} = \delta L$ is the first variation of the classical Lagrangian induced by the paraxial shift, so that (30) may be written compactly as

$$\frac{\mathrm{d}}{\mathrm{d}s}\left[n\mathbf{m} + (\nabla_{\mathbf{x}} n \cdot \mathbf{y})\mathbf{p}\right] = n\mathbf{k} + \delta L \nabla_{\mathbf{x}} n + (\nabla_{\mathbf{x}} n \cdot \mathbf{y}) L_{\mathbf{x}} + L(\nabla_{\mathbf{x}} \nabla_{\mathbf{x}} n) \cdot \mathbf{y}, \tag{31}$$

where $\mathbf{m} = L_{\mathbf{rx}}\mathbf{y} + L_{\mathbf{rr}}\dot{\mathbf{y}}$ and $\mathbf{k} = L_{\mathbf{xx}}\mathbf{y} + L_{\mathbf{xr}}\dot{\mathbf{y}}$ are the classical Jacobi "momentum" and "force". The classical Jacobi equation (6) is $\dot{\mathbf{m}} = \mathbf{k}$; the finite-frequency corrections are $O(\nabla_{\mathbf{x}} n, \nabla_{\mathbf{x}} \nabla_{\mathbf{x}} n)$ and vanish where $W$ is constant. It is important to note, however, that even where $\nabla_{\mathbf{x}} n = 0$ locally, the coefficients in (29) are evaluated along the *Eigenwave* $\mathbf{x}(s)$, not along the eikonal ray, and the paraxial shifts differ globally.

The structural properties of the classical Jacobi equation established by Bliss (1916) and Ravve and Koren (2021) rest solely on the first-degree homogeneity of the Lagrangian; by (19) they hold for (29):

1. only two of the three scalar equations are independent; any tangential shift $\rho_t(s)\mathbf{r}(s)$ solves (29), and the normal shifts are chosen as the physical solutions;
2. the fundamental solution is a linear combination of four basic normal-shift solutions, $\mathbf{y} = \sum_{i=1}^{4} \gamma_i \mathbf{y}_i$, with $\gamma_i$ the ray coordinates;
3. the finite-element solution of (29) by the weak formulation with the same Hermite elements as in the kinematic stage has a resolving matrix equal to the global (all-node) Hessian of the phase-time functional (18), already assembled during the kinematic Eigenwave iterations.

### 5.3 Initial conditions, ray Jacobian, geometric spreading and caustics

The point-source and plane-wave initial conditions of Ravve and Koren (2021) are defined through the eigenstructure of $L_{\mathbf{rr}}$ at the source. Since $L^{\omega}_{\mathbf{rr}} = nL_{\mathbf{rr}}$ has the same eigenvectors, the finite-frequency initial conditions are literally the same:

$$L^{\omega}_{\mathbf{rr},S}\dot{\mathbf{y}}_{1,S} = n_S\lambda_{1,S}\dot{\mathbf{y}}_{1,S}, L^{\omega}_{\mathbf{rr},S}\dot{\mathbf{y}}_{2,S} = n_S\lambda_{2,S}\dot{\mathbf{y}}_{2,S}, L^{\omega}_{\mathbf{rr},S}\mathbf{r}_S = \mathbf{0},$$
$$\mathbf{y}_{1,S} = \mathbf{y}_{2,S} = \mathbf{0}, |\dot{\mathbf{y}}_{1,S}| = |\dot{\mathbf{y}}_{2,S}| = 1, \dot{\mathbf{y}}_{1,S} \times \dot{\mathbf{y}}_{2,S} \cdot \mathbf{r}_S = 1, \tag{32}$$

for the point source, and, for the plane wave, $\mathbf{y}_{3,S}, \mathbf{y}_{4,S}$ equal to the normalized eigenvectors of $L^{\omega}_{\mathbf{rr},S}$ (i.e. of $L_{\mathbf{rr},S}$) with $\dot{\mathbf{y}}_{3,S} = -(\dot{\mathbf{r}}_S \cdot \mathbf{y}_{3,S})\mathbf{r}_S$, $\dot{\mathbf{y}}_{4,S} = -(\dot{\mathbf{r}}_S \cdot \mathbf{y}_{4,S})\mathbf{r}_S$. (The isotropic degenerate case is handled as in Ravve and Koren, 2021.) The point-source ray Jacobian, the caustic criteria and the direction of a line caustic are unchanged in form,

$$J^{\omega}(s) = \mathbf{y}_1(s) \times \mathbf{y}_2(s) \cdot \mathbf{r}(s), c_P = y_1^2 + y_2^2 = 0 \text{ (point caustic)}, \mathbf{r}_L = \mathbf{z}_o \times \mathbf{r}, \tag{33}$$

where $\mathbf{y}_{1,2}$ now solve (29) and $\mathbf{z}_o$ is the null eigenvector of $[\mathbf{y}_1\ \mathbf{y}_2\ \mathbf{r}]^{\mathrm{T}}$. The relative geometric spreading follows from eq. (24) of Ravve and Koren (2021) applied to the effective medium; using (15) and $\lambda^{\omega}_{i,S} = n_S\lambda_{i,S}$,

$$L^{\omega}_{\mathrm{GS}}(s) = \sqrt{\frac{v^{\omega}_{\mathrm{ray},S}}{v^{\omega}_{\mathrm{phs},S}}\frac{v^{\omega}_{\mathrm{ray}}(s)}{v^{\omega}_{\mathrm{phs}}(s)}\frac{\sqrt{|J^{\omega}(s)|}}{\sqrt{\lambda^{\omega}_{1,S}\lambda^{\omega}_{2,S}}}} = \sqrt{\frac{v_{\mathrm{ray},S}}{v_{\mathrm{phs},S}}\frac{v_{\mathrm{ray}}(s)}{v_{\mathrm{phs}}(s)}\frac{\sqrt{|J^{\omega}(s)|}}{n_S\sqrt{\lambda_{1,S}\lambda_{2,S}}}}, \tag{34}$$

because the ratio $v_{\mathrm{ray}}/v_{\mathrm{phs}}$ is invariant under the conformal rescaling. The normalizing parameter of the complexity criterion becomes $\sigma^{\omega} = \int_S^R v^{\omega}_{\mathrm{ray}}\, \mathrm{d}s = \int_S^R v_{\mathrm{ray}}/n\, \mathrm{d}s$, and the source-point limit of the normalized geometric spreading, eq. (27) of Ravve and Koren (2021), is

$$\left(\frac{L^{\omega}_{\mathrm{GS}}}{\sigma^{\omega}}\right)_S = \frac{1}{v^{\omega}_{\mathrm{phs},S}\sqrt{\lambda^{\omega}_{1,S}\lambda^{\omega}_{2,S}}} = \frac{1}{v_{\mathrm{phs},S}\sqrt{\lambda_{1,S}\lambda_{2,S}}} = \left(\frac{L_{\mathrm{GS}}}{\sigma}\right)_S, \tag{35}$$

i.e. *frequency independent*: a purely local rescaling of the medium cannot change the near-source spreading rate, and finite-frequency effects on the normalized spreading accumulate only along the path through $\nabla_{\mathbf{x}}n$ and $\nabla_{\mathbf{x}}\nabla_{\mathbf{x}}n$. The KMAH index is accumulated from the zeros of $J^{\omega}$ exactly as in the classical case, and the weighted propagation complexity $c_r = s_{\max}^{-1}\int (L^{\omega}_{\mathrm{GS}}/\sigma^{\omega} - 1)^2 \mathrm{d}s$ can be used to qualify the Eigenwave solutions.

### 5.4 Equivalence with the Hamiltonian dynamic ray tracing

The Hamiltonian dynamic ray tracing of the Eigenwave is the linearization of the arclength Hamilton system of $H^{\omega}$ (Appendix D),

$$\dot{\mathbf{z}} = H^{\omega}_{\mathbf{px}}\mathbf{z} + H^{\omega}_{\mathbf{pp}}\mathbf{q}, \quad \dot{\mathbf{q}} = -H^{\omega}_{\mathbf{xx}}\mathbf{z} - H^{\omega}_{\mathbf{xp}}\mathbf{q}, H^{\omega}_{\mathbf{x}} \cdot \mathbf{z} + H^{\omega}_{\mathbf{p}} \cdot \mathbf{q} = 0, \tag{36}$$

where $\mathbf{z} = \partial \mathbf{x}_{\text{prx}}/\partial\gamma$ and $\mathbf{q} = \partial \mathbf{p}^{\omega}_{\text{prx}}/\partial\gamma$. Since $L^{\omega}$ and $H^{\omega}$ are related by the Legendre transformation and $L^{\omega}$ is first-degree homogeneous, the Lagrangian-Hamiltonian Hessian relations of Ravve and Koren (2021); Ravve and Koren (2020c) hold verbatim,

$$\begin{aligned} L^{\omega}_{\mathbf{rr}} &= \left(H^{\omega}_{\mathbf{pp}}\right)^{-1} - \lambda_{\mathbf{r}}\mathbf{r}\otimes\mathbf{r}, &\quad L^{\omega}_{\mathbf{rx}} &= -\left(H^{\omega}_{\mathbf{pp}}\right)^{-1} H^{\omega}_{\mathbf{px}}, \\ L^{\omega}_{\mathbf{xx}} &= H^{\omega}_{\mathbf{xp}}\left(H^{\omega}_{\mathbf{pp}}\right)^{-1} H^{\omega}_{\mathbf{px}} - H^{\omega}_{\mathbf{xx}}, &\quad L^{\omega}_{\mathbf{xr}} &= -H^{\omega}_{\mathbf{xp}}\left(H^{\omega}_{\mathbf{pp}}\right)^{-1}, \end{aligned} \tag{37}$$

the Hamiltonian shift differs from the Lagrangian normal shift by an inessential tangential component, $\mathbf{z} = \mathbf{y} + \rho_t\mathbf{r}$, and both approaches yield the same ray Jacobian $J^{\omega} = \mathbf{z}_1 \times \mathbf{z}_2 \cdot \mathbf{r} = \mathbf{y}_1 \times \mathbf{y}_2 \cdot \mathbf{r}$ and geometric spreading. In particular, the propagator-matrix formalism (Červený, 2000; Farra, 1992) for the finite-frequency trajectories of Vasco and Nihei (2019a) can be obtained from (36); the Lagrangian form (29) has the advantage of requiring only the normal shifts and of reusing the kinematic Hessian. Appendix B translates the finite-frequency shift-vector solutions into the classical ray-centred paraxial matrices $\mathbf{Q}^{\omega}, \mathbf{P}^{\omega}$, the propagator and the wavefront-curvature matrix $\mathbf{M}^{\omega} = \mathbf{P}^{\omega}(\mathbf{Q}^{\omega})^{-1}$, with the explicit finite-frequency corrections to the dynamic ray tracing coefficients.

## 6 Weak-coupling theory: the Eigenwave as a perturbation of the Eigenray

In many practical situations $|W| \ll 1$ except in narrow zones, and the Eigenwave is close to the Eigenray. It is then natural to seek the finite-frequency path and phase time as first-order corrections to the classical ones. This yields a compact and instructive result: the correction to the path solves the *classical Jacobi equation with a forcing term*, and the correction to the phase time is a path integral of $W$ along the classical ray.

Let $\mathbf{x}_0(s)$ be the Eigenray (classical stationary path) between $S$ and $R$, with $\mathbf{p}_0$, $\mathbf{r}_0$, $v_{\text{ray}}$ etc. evaluated along it, and let $\mathcal{E}[\mathbf{x}] = \mathrm{d}L_{\mathbf{r}}/\mathrm{d}s - L_{\mathbf{x}}$ and $\mathcal{E}^{\omega}[\mathbf{x}] = \mathrm{d}L^{\omega}_{\mathbf{r}}/\mathrm{d}s - L^{\omega}_{\mathbf{x}}$ denote the classical and finite-frequency Euler-Lagrange operators, so that $\mathcal{E}[\mathbf{x}_0] = \mathbf{0}$. To first order in $W$, $n \simeq 1 + W/2$ and $\nabla_{\mathbf{x}} n \simeq \nabla_{\mathbf{x}} W/2$. Writing the Eigenwave as $\mathbf{x} = \mathbf{x}_0 + \delta\mathbf{x}$ with $\delta\mathbf{x}(S) = \delta\mathbf{x}(R) = \mathbf{0}$ and $\delta\mathbf{x} \cdot \mathbf{r}_0 = 0$, and linearizing $\mathcal{E}^{\omega}$ around $\mathbf{x}_0$ (Appendix E) we obtain

$$\frac{\mathrm{d}}{\mathrm{d}s}\left(L_{\mathbf{rx}} \cdot \delta\mathbf{x} + L_{\mathbf{rr}} \cdot \delta\dot{\mathbf{x}}\right) - L_{\mathbf{xx}} \cdot \delta\mathbf{x} - L_{\mathbf{xr}} \cdot \delta\dot{\mathbf{x}} = \frac{\nabla_{\mathbf{x}} W}{2v_{\text{ray}}} - \frac{1}{2}(\nabla_{\mathbf{x}} W \cdot \mathbf{r}_0)\mathbf{p}_0 \equiv \mathbf{f}(s), \tag{38}$$

where all coefficients are the classical ones along $\mathbf{x}_0$. Thus, the first-order finite-frequency path correction is the solution of the classical Jacobi equation (6) with the forcing $\mathbf{f}$ and homogeneous Dirichlet (two-point) boundary conditions.

Note that $\mathbf{f} \cdot \mathbf{r}_0 = 1/2\,(\nabla_{\mathbf{x}} W \cdot \mathbf{r}_0)\left(1/v_{\text{ray}} - \mathbf{p}_0 \cdot \mathbf{r}_0\right) = 0$: the forcing lies in the plane normal to the ray, consistently with the tangential degeneracy of the Jacobi operator, and indeed $\mathbf{f} =$ $1/2\ \mathbf{r}_0 \times (\nabla_\mathbf{x} W \times \mathbf{p}_0)$, which is the first-order form of the right-hand side of (24). In the finite-element implementation, (38) is a linear system whose matrix is the classical global traveltime Hessian of the Eigenray, and whose right-hand side is the assembled forcing; the Eigenray solution can therefore be "upgraded" to a first-order Eigenwave at the cost of one back-substitution per frequency. This is the finite-frequency analogue of the ray-perturbation approach of Snieder and Spencer (1993), with the role of the medium perturbation played by $W$.

The first-order correction to the phase time follows from Fermat's principle: since $\mathbf{x}_0$ is stationary for $\int L\, \mathrm{d}s$, the $O(\delta\mathbf{x})$ path change does not contribute, and

$$\tau(\omega) = \int_S^R n\, L \mathrm{d}s \simeq t_0 + \delta\tau, \qquad \delta\tau = \frac{1}{2}\int_{\mathbf{x}_0} \frac{W(\mathbf{x},\omega)}{v_{\mathrm{ray}}(\mathbf{x},\mathbf{r}_0)} \mathrm{d}s, \tag{39}$$

where $t_0$ is the eikonal traveltime. Equation (39) is a ray-based finite-frequency traveltime correction that is positive where $W > 0$ (the effective medium is slower) and vanishes as $\omega \to \infty$. Because the correction is quadratic in $\delta\mathbf{x}$, (39) remains second-order accurate even if the path correction (38) is not computed at all.

*6.0.0.1 Frequency scaling.*

By (9), $W(\mathbf{x},\omega) = \kappa(\mathbf{x},\omega)/\omega^2$ with $\kappa = \hat{\mathbf{a}} \cdot \nabla_\mathbf{x} \cdot \left(\rho\, \tilde{\mathbf{C}} : \nabla_\mathbf{x} \mathbf{a}\right)/(\rho a)$. To the extent that the shape of the amplitude field varies slowly with frequency (e.g. when $\mathbf{a}$ is taken from the zeroth-order transport equation), $\kappa(\mathbf{x})$ is a frequency-independent *coupling scalar* of the medium and of the wavefield geometry, and the finite-frequency corrections have an explicit $\omega^{-2}$ signature:

*6.0.0.2 Validity.*

The weak-coupling theory requires $|W| \ll 1$ along the classical ray and that the classical Jacobi operator with Dirichlet conditions be invertible, i.e. that $R$ be not conjugate to $S$ along $\mathbf{x}_0$ (no caustic exactly at the receiver). Near a caustic of the eikonal ray the amplitude $a$ grows and $W$ computed from it is not small; there the full nonlinear Eigenwave problem (22) should be solved, with $W$ obtained from a regularized (finite-frequency) amplitude field, see Section 7. In this respect it is worth recalling that trajectories obeying (11) with a self-consistent $W$ do not cross (Foreman, 1989; Wyatt, 2005): the finite-frequency potential is precisely the mechanism that regularizes the caustics of ray theory.

## 7 Coupling with the amplitude field and workflow

The Eigenwave equations of Sections 4-5 are exact for a given $W(\mathbf{x},\omega)$. The potential itself is a functional of the amplitude field, which in turn is governed by the transport (imaginary-part) equation. For the qP mode, Vasco and Nihei (2019a) reduce it to $\nabla_\mathbf{x} \cdot (\rho a^2 \mathbf{v}) = 0$, where $\mathbf{v} = \mathrm{d}\mathbf{x}/\mathrm{d}T = 1/2\ \nabla_\mathbf{p}\lambda$ is the group-velocity vector of the trajectory equations. In terms of the arclength quantities of this paper, $\mathbf{v} = n v_{\mathrm{ray}}\, \mathbf{r}$ (Remark 3), and the conservation of $\rho a^2 |\mathbf{v}|$ times the normal cross-section of the ray tube gives, along an Eigenwave,

$$a(s) \propto (\rho n v_{\mathrm{ray}} |J^{\omega}|)^{-1/2}, \tag{40}$$

where $J^{\omega}$ is the point-source ray Jacobian obtained from the finite-frequency Jacobi equation (29). Equation (40) closes the system: the dynamic Eigenwave delivers the amplitude field of a family of Eigenwaves, from which $W$ can be recomputed and the trajectories updated. We stress two caveats already emphasized by Vasco and Nihei (2019a): (i) $W$ contains second spatial derivatives of $a$ *across* neighbouring trajectories, so a family of Eigenwaves (or a hybrid grid/trajectory representation) is required, and (ii) the ray-theoretical amplitude (40) is singular at the caustics of the current iterate, whereas the converged finite-frequency amplitude is not; near caustics $W$ must be regularized (e.g. by evaluating it from a phase-space (Maslov) representation of the ray family, Sections 8.3 and 9.3, or from a numerical wavefield as done by Vasco and Nihei, 2019a).

A practical workflow consistent with the philosophy of the Eigenray method proceeds from the Eigenray towards lower frequencies. In it, the potential $W$ is assembled from ray-based quantities: a numerical solution of the elastic, or even acoustic, wave equation would supply $W$ directly (as implemented in Vasco and Nihei, 2019a), but it would make the method practically infeasible in applications where millions of point source wavefields and the corresponding Eigenwaves must be computed, and it is therefore reserved for verification along selected paths.

*The workflow consists of four steps.*

1. Solve the kinematic and dynamic Eigenray problems (eikonal limit, $W = 0$) between $S$ and $R$; this provides the initial-guess trajectory, the global traveltime Hessian, $J$ and $a$.

2. Assemble $W(\mathbf{x}, \omega)$ at the upper end of the band from step 1 through (9), from the closure field of each branch built on a fan of neighbouring Eigenrays: the ray amplitude (40) away from caustics and its phase-space (Maslov) representation near them (Sections 8.3 and 9.3); smooth it so that $\nabla_{\mathbf{x}} W$ and $\nabla_{\mathbf{x}} \nabla_{\mathbf{x}} W$ are stable, as described in Section 8.

3. Weak coupling: for each frequency solve the forced Jacobi equation (38) (one back-substitution with the classical Hessian) and evaluate $\delta\tau$ from (39).

4. Strong coupling: run the kinematic Eigenwave solver (weak form (26), kernel (27)) starting from the corrected path of step 3, then the dynamic Eigenwave solver (29) for $J^{\omega}$, $L_{\mathrm{GS}}^{\omega}$ and caustics; optionally iterate steps 2-4 to self-consistency, then step downward in frequency, each converged path serving as the initial guess and background for the next (Section 8.4).

The workflow retains every stationary path of the eikonal problem, including multiple arrivals, and is the natural choice where $|W|$ is small over most of the band. It terminates where $|W|$ approaches unity, below which no single-ray background is adequate.

The workflow is summarized in Fig. 2.

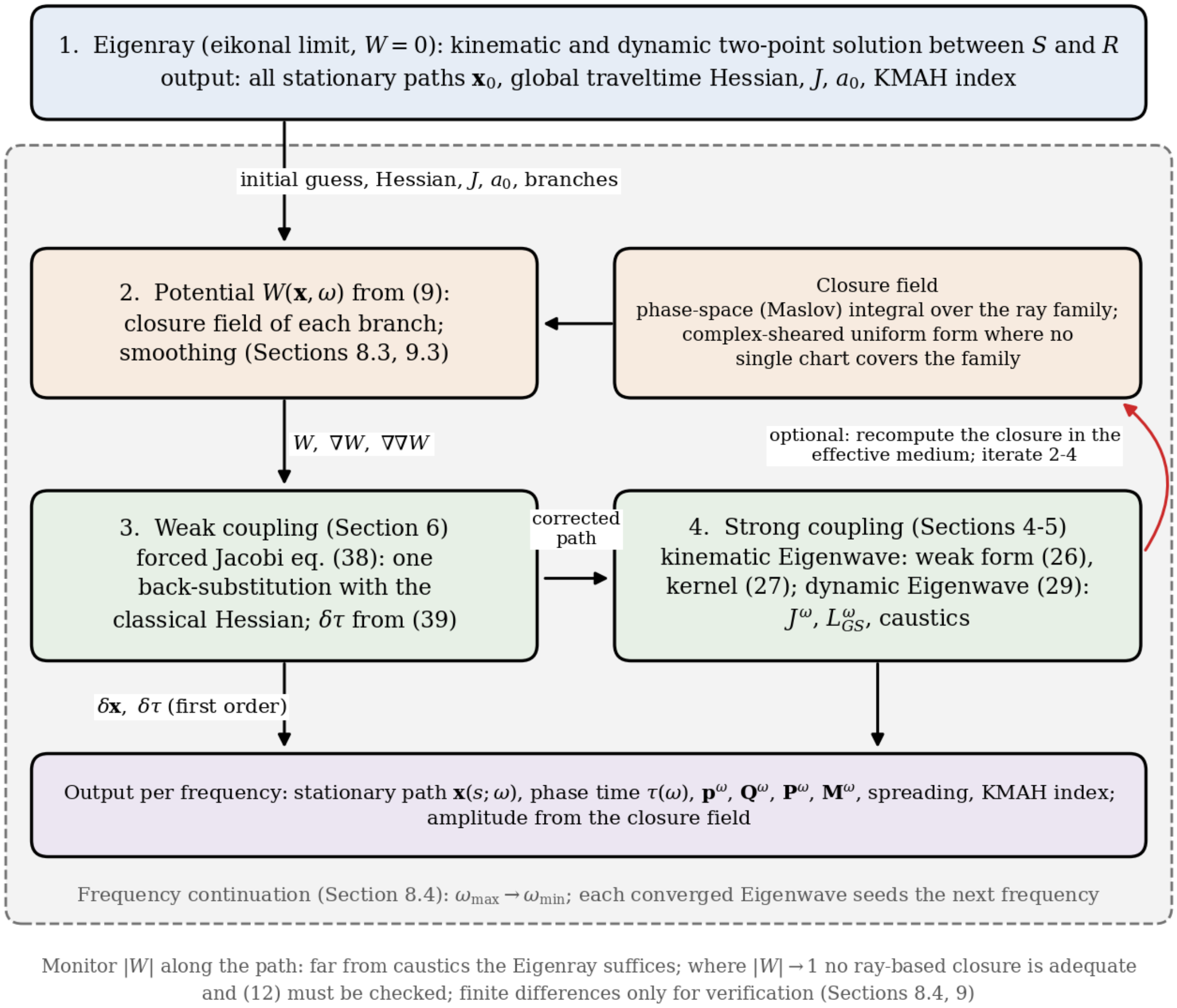

**Figure 2.** The workflow of the Eigenwave method (Section 7). The eikonal Eigenray (step 1) provides the initial path, the global traveltime Hessian and the ray-theoretical amplitude; the potential $W$ is assembled from the closure field of each branch (step 2, Sections 8.3 and 9.3); the weak-coupling correction requires one back-substitution with the classical Hessian (step 3); the strong-coupling kinematic and dynamic Eigenwave solvers (step 4) may be iterated with step 2 to self-consistency. The frequency band is traversed by continuation from high to low frequency (Section 8.4); the amplitude is taken from the closure field.

The frequency-dependent family of stationary paths, phase times and spreadings so obtained is the Eigenwave counterpart of the multi-frequency trajectories of Vasco and Nihei (2019a), with the two-point boundary conditions, the variational structure and the finite-element machinery of the Eigenray method.

## 8 Behaviour near caustics and regularization of the amplitude closure

The caustic difficulty identified in Sections 6 and 7 deserves a closer look, both because it is frequently misattributed to the trajectory-mechanics equations themselves and because its resolution determines how the closure of Section 7 must be implemented within the Eigenwave framework. Two questions arise naturally. First, does the singularity persist as the frequency is lowered - is the caustic “healed” by finite-frequency effects, as intuition about wavefront healing might suggest? Second, given that the dynamic ray tracing of Section 5 and Appendix B is available along each stationary path, what is the appropriate way to regularize the amplitude term feeding the frequency-dependent Hamiltonian (9)? The answers hinge on separating the exact trajectory system from its ray-closed approximation. For clarity we phrase the discussion in the scalar (isotropic acoustic) case with velocity $v$, where the potential of (9) reduces to $W = v^2 \nabla^2 a/(\omega^2 a)$; the conclusions carry over to the anisotropic elastic case with the qualifications noted at the end of the section.

### 8.1 The exact system

Together with the frequency-dependent eikonal equation $|\nabla_{\mathbf{x}}\tau|^2 = (1+W)/v^2$ (the scalar form of (13)), the imaginary part of the polar decomposition (7) yields the exact transport equation

$$\nabla_{\mathbf{x}} \cdot (a^2 \nabla_{\mathbf{x}}\tau) = 0, \tag{41}$$

in which $\nabla^2 a/a$ and $\nabla_{\mathbf{x}}\tau$ refer to the amplitude and phase time of the actual finite-frequency field, respectively, not to their ray-theoretical counterparts. The finite-frequency field is bounded at a geometrical caustic - it exhibits the familiar Airy pattern with a peak amplitude that grows only as $\omega^{1/6}$ - and its phase is single valued wherever the field does not vanish. A single-valued phase gradient means that the exact trajectories cannot cross, and this non-crossing is precisely what the potential $W$ enforces. The point is made explicit by the turning-point solution in a linear-gradient medium, which solves the Helmholtz equation exactly:

$$u = \mathrm{AI}(-\alpha\zeta)\exp(i\omega s/v_0), \qquad \frac{\nabla^2 a}{a} = -\alpha^3\zeta \Rightarrow \ W = -\frac{v_0^2\alpha^3}{\omega^2}\zeta, \tag{42}$$

where $s, \zeta$ are the coordinates along and normal to the caustic and $\alpha$ is a constant set by the gradient. The potential is linear in the normal distance $\zeta$ - finite everywhere, including on the caustic itself - and the exact trajectories on the illuminated side simply run parallel to the caustic, the potential term cancelling the transverse geometrical slowness. The genuine singularities of the exact formulation lie elsewhere: at the isolated points (in two dimensions) or lines (in three) where the field vanishes identically, $u = 0$, the phase gradient behaves as $1/r$ and $W$ diverges. These phase dislocations, generic in any multipathing interference field, constitute the “node problem” familiar from quantum trajectory methods (Wyatt, 2005). Lowering the frequency does not remove them; it merely spaces them by roughly one wavelength. Moreover, once the wavelength approaches the scale of the medium heterogeneity, $|W|$ becomes comparable to unity essentially everywhere: the equations remain exact, but the trajectories are no longer a perturbation of rays anywhere and the notion of a localized caustic dissolves into a diffractive field.

### 8.2 The ray-closed system

The situation is different for the ray-closed system that is actually solved, in which the amplitude term is evaluated from the zeroth-order ray amplitude $a_0$ delivered by the dynamic ray tracing (eq. (40) with $J$ from Section 5 or Appendix B). Near a fold caustic the ray Jacobian vanishes linearly in the distance $\zeta$ from the caustic, so that

$$a_0 \propto |J|^{-\frac{1}{2}}, \qquad J \propto \zeta \Rightarrow W_0 = \frac{v^2}{\omega^2}\frac{\nabla^2 a_0}{a_0} \simeq \frac{3v^2}{4\omega^2\zeta^2}\ . \tag{43}$$

Two consequences follow. The divergence at $\zeta = 0$ is set by geometry - the vanishing of the Jacobian - and not by frequency: the ratio $\nabla^2 a/a$ is scale free in the amplitude, so the fact that the peak enhancement of the true field weakens as the frequency drops (the $\omega^{1/6}$ law) offers no relief. Second, balancing the correction (43) against the transverse geometrical slowness of the two coalescing branches, $p_\perp^2$, with $\mathcal{R}$ the radius of curvature of the caustic,

$$p_\perp^2 \simeq \frac{2\zeta}{v^2\mathcal{R}} \sim \frac{W_0}{v^2} = \frac{3}{4\omega^2\zeta^2} \Rightarrow \zeta_A \sim (\lambda^2\mathcal{R})^{1/3}, \tag{44}$$

shows that the ray closure correctly predicts the onset of the Airy zone, of width $\zeta_A$, but is qualitatively wrong inside it - it blows up where the exact term (42) is linear and changes sign. Since $\zeta_A$ scales as $\omega^{-2/3}$, lowering the frequency enlarges, rather than shrinks, the region in which the closed system is unreliable, and below the frequency at which $(\lambda^2\mathcal{R})^{1/3}$ becomes comparable with $\mathcal{R}$ the "caustic zone" is the whole model. Reducing the frequency therefore relocates the difficulty; it does not remove it.

The practical conclusion is that the explosion of the amplitude term is a defect of the background fed to the closure, not of the stationarity principle, and that any remedy must be applied to the background amplitude before it is differentiated. This rules out clipping or tapering the correction after the fact: the finite-frequency Jacobi equation (29) along the Eigenwave requires the second spatial derivatives $\nabla_{\mathbf{x}}\nabla_{\mathbf{x}}W$ (through $\nabla_{\mathbf{x}}\nabla_{\mathbf{x}}n$ in (28)), and a term behaving as $\zeta^{-2}$ has a Hessian behaving as $\zeta^{-4}$, so a regularization that is not itself smooth to second order merely moves the singularity into the paraxial system. Within the Eigenray/Eigenwave framework three closures satisfy this requirement, in increasing order of fidelity.

## 8.3 Three admissible closures

*8.3.0.1 Complex-shifted dynamic ray tracing (Gaussian-beam-type closure).*

The dynamic ray tracing along each candidate stationary path is run with complex initial conditions, in the manner of the Gaussian-beam construction (Červený et al., 1982; Popov, 1982),

$$\mathbf{Q}(s_S) = \mathbf{I}, \qquad \mathbf{P}(s_S) = i\varepsilon\mathbf{I}, \qquad a \propto |\mathrm{det}\mathbf{Q}|^{-1/2} \Rightarrow |J| \to (\zeta^2 + \zeta_i^2)^{1/2}, \tag{45}$$

so that det $\mathbf{Q}$ never vanishes. Near a fold the vanishing Jacobian is replaced by a regularization of width $\zeta_i$ set by the beam parameter $\varepsilon$, the regularized amplitude is smooth to all orders, and the KMAH jump becomes the continuous arg det $\mathbf{Q}$. The amplitude Laplacian is available analytically from $\mathbf{Q}$ and $\mathbf{P}$, and on later iterations the same construction applies to the finite-frequency matrices of Appendix B. The width should be frequency aware; choosing it equal to the Airy scale at the caustic gives

$$\zeta_i \sim (\lambda^2 \mathcal{R})^{1/3} \Rightarrow |W|_{\max} \sim (\frac{\lambda}{\mathcal{R}})^{2/3}, \tag{46}$$

with $\mathcal{R}$ the radius of curvature of the caustic, obtainable from the arclength derivatives of the Jacobian; the peak of the correction then grows as the frequency is lowered, as it should. Because the beam is launched at the source, however, its width cannot fall below the Fresnel scale, and Section 9 shows that this limits the closure when the focusing anomaly is itself of Fresnel size.

Figure 3 compares, across a fold caustic, the exact potential (42) with the ray-closed term (43) and with its complex-shifted regularization (45), together with the amplitudes that feed them: the three agree outside the Airy zone and differ qualitatively inside it, and only the regularized amplitude is smooth enough to be differentiated twice.

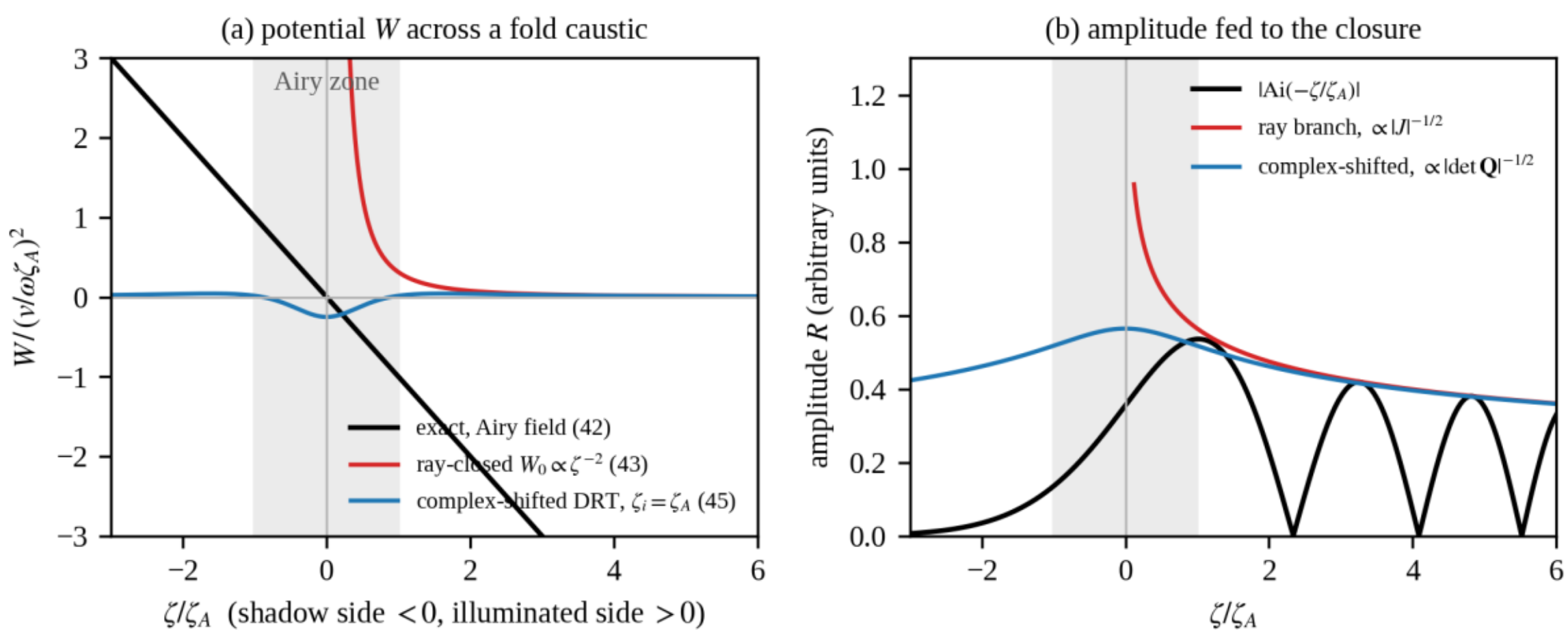

**Figure 3.** Behaviour of the amplitude closure across a fold caustic (Section 8), in the Airy-scaled normal coordinate $\zeta/\zeta_A$, $\zeta_A = 1/\alpha$; the illuminated side is $\zeta > 0$. (a) The potential $W$, in units of $(v/\omega\zeta_A)^2$: exact, from the Airy field (42), linear and sign-changing; ray-closed $W_0$ from the zeroth-order amplitude (43), diverging as $\zeta^{-2}$; and from the complex-shifted dynamic ray tracing (45) with the Airy-scaled width $\zeta_i = \zeta_A$ of (46), finite and smooth. The three curves agree outside the Airy zone (shaded) and differ qualitatively inside it. (b) The amplitudes that feed the closure: the exact Airy amplitude, the ray-theoretical branch amplitude $\propto |J|^{-1/2}$ (which envelopes the Airy maxima), and its complex-shifted regularization.

*8.3.0.2 Two-branch uniform (Airy) closure.*

The beam closure lacks the sign structure of the exact term (42): inside the Airy zone the true potential is linear in $\zeta$ and changes sign across the caustic, whereas the regularized amplitude produces a positive, symmetric bump. When the receiver lies near a fold, two stationary paths of the Eigenray stage are nearly coalescent, and their amplitudes and phase times $(a_1, \tau_1), (a_2, \tau_2)$ are the input of the Ludwig–Kravtsov uniform expansion (Kravtsov, 1964; Ludwig, 1966; Chapman, 2004). With

$$\bar{\tau} = 1/2\,(\tau_1 + \tau_2), \qquad \Delta\tau = \tau_2 - \tau_1 \geq 0, \qquad \xi = (3/4\,\omega\Delta\tau)^{2/3}, \tag{47}$$

the uniform representation reads, up to a constant phase factor,

$$u \simeq \sqrt{\pi} e^{i\omega\bar{\tau}} \left[ (a_1 + a_2)\xi^{1/4}\mathrm{Ai}(-\xi) - i(a_1 - a_2)\xi^{-1/4}\mathrm{Ai}'(-\xi) \right], \tag{48}$$

whose modulus provides the closure amplitude. It is regular at the caustic, reproduces the two branches, including the $-\pi/2$ KMAH shift, when $\omega\Delta\tau$ is large, and has the correct linear, sign-changing relative Laplacian inside the Airy zone; but it couples two stationary solutions whose Newton iteration is ill conditioned at coalescence, and it applies to folds only.

*8.3.0.3 Phase-space (mixed-representation) closure.*

Since the Eigenray formulation is already a phase-space one, the structurally most consistent option is that of Maslov (Maslov and Fedoriuk, 1981; Chapman, 2004): near a caustic in configuration space the Lagrangian manifold is regular in a mixed chart such as $(x_1, p_2)$ or $(p_1, x_2)$, the amplitude in that chart is proportional to the inverse square root of the corresponding mixed Jacobian and is finite, and so is its Laplacian. Transforming this amplitude back to a configuration-space potential $W$ requires a frequency-dependent stationary-phase step, which makes it the most laborious of the three; it is, however, the only closure that remains valid at cusps and higher catastrophes, where the fold-based Airy expression (48) no longer applies. Section 9 implements it level-wise, as a Maslov integral over the ray family in the mixed chart of horizontal slowness and depth, eq. (53); in the numerical example it is the most accurate of the closures tested.

### 8.4 Consequences for the frequency-dependent path

Three further points concern the Eigenwave itself. First, the frequency band should be traversed by continuation, from high to low frequency (Section 7): the stationary path is solved first at the upper end of the band, where $W$ is smallest and the Eigenray is an excellent initial guess (Section 6), and is then stepped downward in frequency, each converged path with its regularized dynamic quantities serving as the background for the next. Second, the dimensionless ratio $|W|$ should be monitored along the path; where it approaches unity the perturbative premise is lost and no regularization of $a_0$ will restore it - below that frequency the background must come from a wave-equation solution (in practice a phase-space representation of the ray family, or beam summation; a finite-difference solve, as in Vasco and Nihei, 2019a, is affordable only for verification along selected paths) rather than from any single-ray quantity, and the condition (12) must be checked explicitly. Third, in the anisotropic elastic case the same regularized $\mathbf{Q}$ should be applied to the polarization-gradient terms of the generalized Christoffel system (8), which share the $|J|^{-1/2}$ pathology; these terms are, in addition, undefined at conical and kiss singularities of the slowness surface irrespective of what is done to the Jacobian, so a slowness-surface singularity along a candidate path calls for a separate check.

For the finite-frequency extension of the Eigenray method the recommended closure is thus the phase-space one, implemented as a level-wise Maslov integral over the ray family (Section 9.3), in its complex-sheared uniform form where no single chart covers the family. The complex-shifted dynamic ray tracing (45) and the two-branch Airy closure (48) remain useful away from Fresnel-scale focusing; the numerical example of Section 9 shows, however, that beams launched from the source are at least a Fresnel zone wide and cannot resolve focusing by an anomaly of that size. The phase-space closure fits directly onto the dynamic machinery of Section 5 and Appendix B, and combines the kinematic fidelity of trajectory mechanics with a caustic-regular amplitude.

## 9 Numerical example: multipathing through a focusing low-velocity inclusion

We now apply Workflow A of Section 7 to a model that is deliberately hostile to ray theory, and compare the resulting multi-arrival Eigenwave seismogram with a finite-difference solution of the wave equation. Three difficulties are present at once: multipathing, a caustic located exactly at the receiver, and a velocity anomaly whose size is comparable with the Fresnel zone of the path. To keep the reference computation exact and inexpensive, the test is carried out in two dimensions and in the acoustic approximation of Appendix I with constant density. For an isotropic medium this is the scalar version of the qP problem: the kinematics are identical, the potential takes the form $W = \frac{v^2 \nabla^2 a}{\omega^2 a}$ and the amplitude obeys the transport equation (41). All fields are harmonic with time dependence $e^{-i\omega t}$, and the source is a line source, so that geometric spreading is cylindrical.

### 9.1 Model, geometry and finite-difference reference

The background is homogeneous with velocity $v_b = 2000\ \mathrm{m/s}$. It contains a circular inclusion of radius $R = 500\ \mathrm{m}$ and velocity $v_a = 1500\ \mathrm{m/s}$, centred at $\mathbf{x}_c = (0, 2000)\ \mathrm{m}$, i.e. at 2000 m depth. The source $\mathbf{x}_S = (0, 4000)$ m and the receiver $\mathbf{x}_R = (0, 0)$ on the surface lie on the vertical through the centre of the inclusion. The Eigenray and Eigenwave equations require second derivatives of the velocity, so the discontinuous boundary is replaced by a hyperbolic-tangent taper,

$$v(\mathbf{x}) = v_b + \frac{1}{2}(v_a - v_b)\left[1 - \tanh\left(\frac{|\mathbf{x} - \mathbf{x}_c| - R}{w}\right)\right], \qquad w = 20\ \mathrm{m}. \tag{49}$$

The half-width $w$ is small compared with the shortest wavelength of interest (about 60 m inside the inclusion at 25 Hz), so the taper is immaterial for the wavefield; this is verified below. It is not immaterial for ray quantities, as will also be shown.

The geometry is chosen so that the receiver sits on a caustic. In the paraxial approximation the inclusion is a converging cylindrical lens with relative index $n_r = v_b/v_a = 4/3$. Its focal length and the imaging condition are

$$f = \frac{n_r R}{2(n_r - 1)} = 1000\ \mathrm{m}, \qquad \frac{1}{d_S} + \frac{1}{d_R} = \frac{1}{f}, \tag{50}$$

where the distances $d_S$ and $d_R$ are measured from the centre of the lens (both principal planes of a ball lens pass through its centre). With $d_S = 2000$ m the paraxial image of the source lies at $d_R = 2000$ m, i.e. exactly at the receiver. Spherical aberration makes the marginal rays over-focus and cross the axis closer to the lens. The envelope of the lens family is therefore a cusp caustic whose tip coincides with the paraxial focus. The receiver thus sits at a cusp, the simplest catastrophe beyond the fold. There the two-branch Airy closure (48) is formally inapplicable (Section 8.3), and the canonical uniform field is the Pearcey integral (Pearcey, 1946; Berry and Upstill, 1980). In addition, at the peak frequency of 12 Hz the Fresnel radius at the lens is $r_F = \sqrt{\lambda d_S d_R/(d_S + d_R)} \approx 410$ m, comparable with $R$. Wavefront healing and diffraction around the inclusion are therefore expected to be strong.

The reference solution is computed with a finite-difference solver of the constant-density acoustic wave equation, $p_{tt} = v^2\big(\nabla^2 p + f(t)\delta(\mathbf{x} - \mathbf{x}_S)\big)$. It is eighth-order in space and second-order in time, with $\Delta x = 5$ m and $\Delta t = 1$ ms. The grid is surrounded by a 600 m wide absorbing sponge (Cerjan *et al.*, 1985) and padded so that boundary reflections reach the receiver only after 2.5 s. The source wavelet $f(t)$ is a Ricker wavelet with peak frequency 12 Hz, delayed by $t_0 = 0.12$ s. Against the analytic two-dimensional Green's function $G = (i/4)H_0^{(1)}(\omega r/v)$ at 1 km offset, the solver reproduces the peak amplitude to 1.6 per cent, with residual boundary reflections below 1 per cent. Two runs were made: one with the tapered model (49) and one with a nearly sharp boundary ($w = 2.5$ m, half a grid step). At the receiver, the main-arrival peaks of the two runs agree to 1 per cent and the peaks in the weak edge-precursor window agree to 9 per cent. This confirms that the taper is immaterial for the wavefield.

### 9.2 Eigenray stage: three stationary paths

All stationary paths of this model are monotonic in depth. We therefore use a two-dimensional reduction of the finite-element Eigenray solver of Section 4, with the path parametrized as a graph $x(z)$. The nodes are spaced $h = 2$ m apart in depth, the end points are fixed, and the discrete phase-time functional is

$$\tau_h = \sum_{i=0}^{N-1} \ell_i \frac{s_i + s_{i+1}}{2}, \quad \ell_i = \sqrt{(x_{i+1} - x_i)^2 + h^2}, \quad s = \frac{n}{v}, \tag{51}$$

with $n \equiv 1$ for the Eigenray and $n = \sqrt{1 + W}$ for the Eigenwave. Stationarity is solved by a damped Newton iteration. The Hessian of $\tau_h$ is tridiagonal: it is the reduced, two-dimensional counterpart of the global traveltime Hessian of the Eigenray method, and the number of its negative eigenvalues gives the Morse index of the path. Compared with the Hermite elements of Section 4.4, this discretization has no direction degrees of freedom. That is sufficient for a two-dimensional, depth-monotonic geometry. Initial guesses come from a Hamiltonian shooting fan of 321 rays over take-off angles $|\gamma| \leq 40°$.

Three stationary paths exist (Fig. 4 and Table 1). The *axial* Eigenray runs through the centre of the inclusion, with $\tau_0 = 2.1657$ s. It is a saddle point (Morse index 1): its conjugate point lies 5.4 m below the receiver, and at the receiver the point-source ray Jacobian is $J = -5.3$ m. Here $J$ is normalized so that $J = r$ in a homogeneous medium. The two symmetric *edge* Eigenrays leave the source at $\gamma = \pm 15.527°$ and graze the inclusion at 527 m from its centre, inside the taper, where $v \approx 1968$ m/s. They are minima (Morse index 0), with $\tau_0 = 2.0732$ s, and they form the first arrival, 93 ms ahead of the axial arrival. Their Jacobian is $J \approx 2.25 \times 10^6$ m. In the ray fan (Fig. 4b), rays with $|\gamma| \lesssim 14°$ are over-focused by the lens and cross the axis between about 1100 m and 5 m depth, forming the cusp. Rays between 14° and 15.5° are deflected sideways by near-grazing refraction and do not reach the receiver. Rays beyond 15.52° pass outside the inclusion. The paths obtained by the finite-element solver and by shooting agree to within $10^{-5}$ s in traveltime, and Newton converges in at most one iteration from the shooting guess.

**Table 1.** Eigenrays of the numerical example. $J(\mathbf{x}_R)$ is the point-source ray Jacobian at the receiver (normalized to $J = r$ in a homogeneous medium). The last column gives the ray-theoretical amplitude relative to that of the homogeneous background at the same distance, $(4000\ \mathrm{m}/|J|)^{1/2}$.

| Branch | Take-off angle | $\tau_0$ (s) | Morse index | $J(\mathbf{x}_R)$ (m) | Ray amplitude / homogeneous |
|---|---|---|---|---|---|
| Axial | 0° | 2.1657 | 1 | $-5.3$ | 27.5 |
| Edge (±) | ±15.527° | 2.0732 | 0 | $2.25 \times 10^6$ | 0.042 |

The ray-theoretical (Eigenray) seismogram follows from the transport equation (41) as

$$G_0(\mathbf{x}_R, \omega) = \sum_k \sqrt{\frac{v_R}{8\pi\omega|J_k|}} \exp\left[i\left(\omega\tau_{0k} + \frac{\pi}{4} - \kappa_k\frac{\pi}{2}\right)\right], \tag{52}$$

where $\kappa_k$ is the KMAH index, which here equals the Morse index. Relative to the homogeneous background, (52) predicts an axial amplitude 27.5 times larger and an edge amplitude 24 times smaller (Table 1). Both numbers are artefacts of the infinite-frequency limit. The first is caused by the caustic. The second is caused by an unresolved boundary: the edge Jacobian is controlled by the curvature of the taper at grazing, a sub-wavelength feature, and it grows without bound as $w \to 0$. The finite-difference gather (Fig. 5a) confirms the traveltime geometry. The healed flanks of the wavefront follow the edge families, and the strong axial arrival follows the lens family. The two edge branches intersect at the receiver at $\tau_0 + t_0 = 2.19$ s, where the wavefield shows only a weak precursor. The snapshot (Fig. 5b) shows the delayed, focusing lens wavefront nested within the undisturbed flanks, together with the diffractions from the inclusion edges.

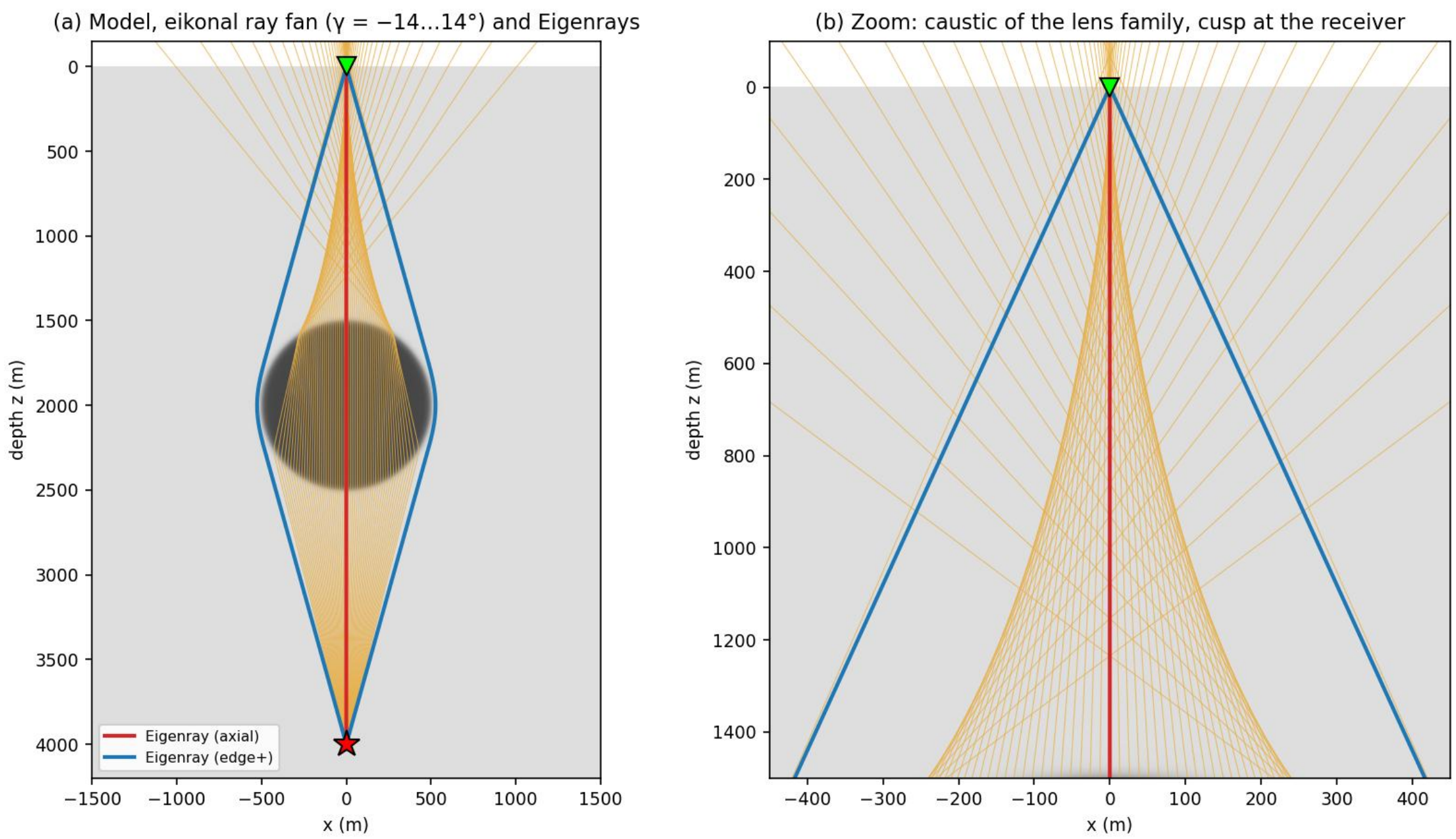

**Figure 4.** Numerical example of Section 9. (a) Model: homogeneous background $v_b = 2000$ m/s with a circular low-velocity inclusion ($v_a = 1500$ m/s, $R = 500$ m, tapered boundary with $w = 20$ m) centred at 2000 m depth. The source (star) is at 4000 m depth and the receiver (triangle) is at the surface. Thin lines: eikonal ray fan for $|\gamma| \leq 14°$ and the two grazing rays. Thick lines: the three Eigenrays (axial, red; edge, blue). (b) Upper part of the model: the over-focused rays of the lens family form a cusp caustic whose tip lies 5.4 m below the receiver.

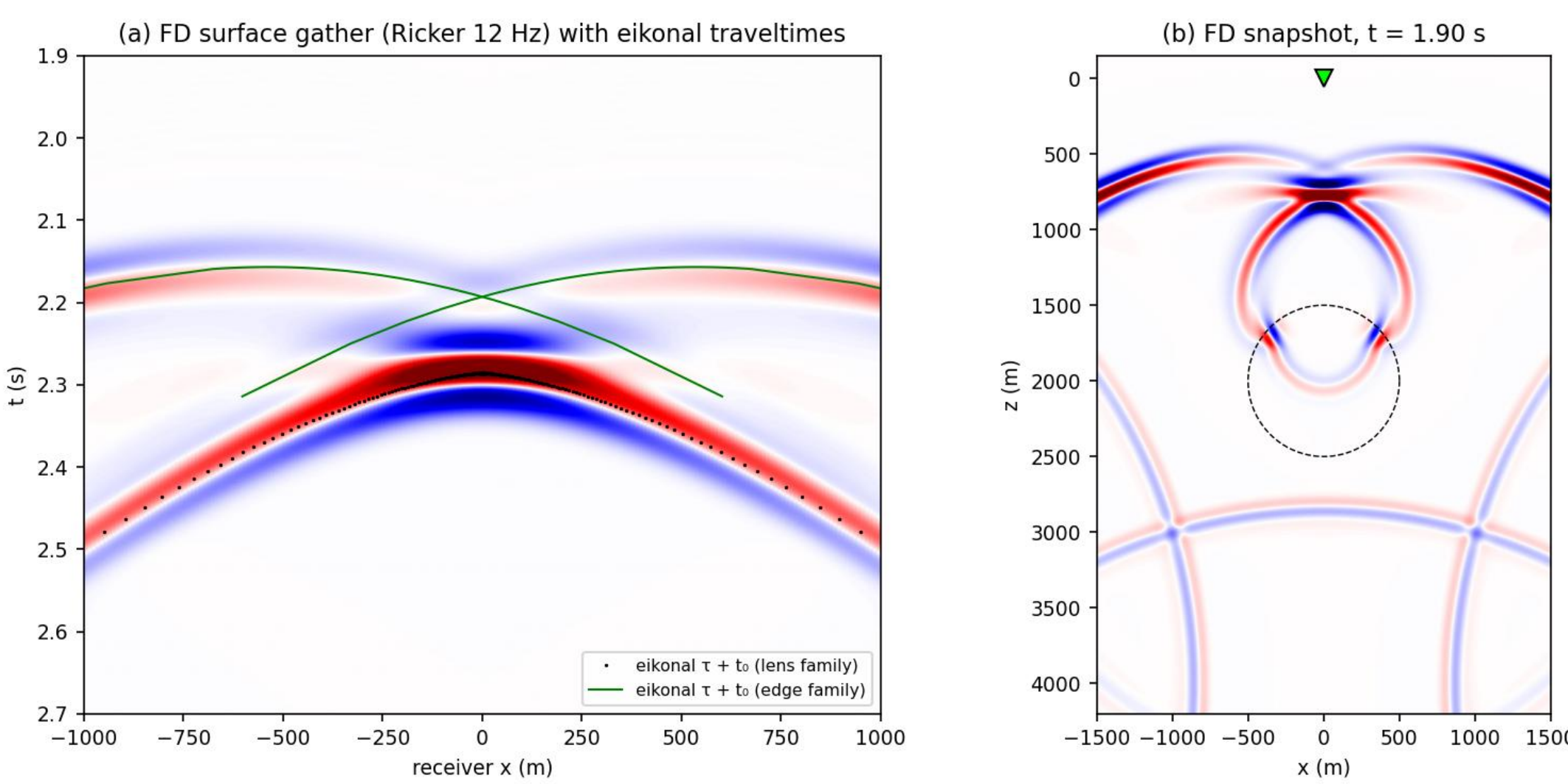

**Figure 5.** (a) Finite-difference surface gather (Ricker wavelet, 12 Hz, delay $t_0 = 0.12$ s) with the eikonal traveltimes $\tau_0 + t_0$ of the lens family (dots) and of the two edge families (solid lines). (b) Finite-difference snapshot at $t = 1.9$ s; the dashed circle is the nominal inclusion boundary. The faint arcs crossing the lower half are residual reflections from the absorbing boundaries and reach the receiver after 2.5 s.

## 9.3 Amplitude closure: the closure field and its phase-space representation

The Eigenwave equations of Sections 3–5 are exact for a given potential, but the potential is not a property of the medium. By (9) it is the relative curvature of the amplitude of the wavefield itself, $W = \frac{v^2 \nabla^2 a}{\omega^2 a}$. The amplitude is in turn governed by the transport equation (41), whose characteristics are the trajectories. Trajectories and amplitude therefore form a coupled system, and to compute an Eigenwave one must supply the amplitude field from outside, i.e. *close* the system. We call the rule that supplies the amplitude the *closure*. Three levels can be distinguished. In the *ray closure*, $a$ is the zeroth-order amplitude transported along neighbouring Eigenrays, eq. (40). It is inexpensive but singular at caustics (Section 8.2), which is precisely the situation of this example. In a *regularized closure*, $a$ is the modulus of an approximate finite-frequency wavefield that remains bounded at caustics. We call this auxiliary wavefield the *closure field*. In the *exact closure*, $a$ is the modulus of the exact wavefield. The Eigenwave equations are then an exact reformulation of the Helmholtz equation (Appendix I.1), and the Eigenwave phase is exact.

Two properties of the closure field are essential for what follows. First, the Eigenwave uses only its modulus. The phase of the closure field is discarded, and the Eigenwave recomputes the phase by integrating the effective slowness $n/v$ along its own stationary path. Comparing the two phases therefore provides an internal consistency check that needs no reference solution. Second, the closure field of a branch is not the exact field. The exact field contains all branches, their mutual interference and the waves reflected by the inclusion, and its potential is dominated by the resulting interference (Section 8.1). The closure field of a branch describes only that branch, or that family of rays. It is this per-branch potential that turns each Eigenray into its own Eigenwave, and multipathing is retained precisely because the branches are closed separately. The price is that the accuracy of each Eigenwave is bounded by that of its closure field.

Section 8.3 identifies the phase-space (Maslov) representation, summarized in Appendix K, as the only closure that remains valid at a cusp. A caustic in configuration space is generally regular in a mixed chart, and for the lens family the natural chart is $(p_x, z)$: at each depth level $z$, the rays are represented by their horizontal slowness $p_x$ rather than by their horizontal position $X$. With the rays of the family parametrized by the take-off angle $\gamma$, the closure field at level $z$ is the Maslov integral

$$
\begin{aligned}
u(x,z,\omega) &= \int \Phi(\gamma) B(\gamma,z) \exp\{i\omega(\tau + p_x(x - X))\} \mathrm{d}\gamma, \\
|B|^2 &= \frac{v|\partial_\gamma p_x|}{16\pi^2 \cos\theta}, \quad \arg B = \frac{\pi}{4} - \frac{\pi}{4}\operatorname{sgn}(-\partial_\gamma p_x),
\end{aligned}
\tag{53}
$$

where $X, \tau, p_x$ and the ray angle $\theta$ from the vertical are evaluated where the ray with take-off angle $\gamma$ crosses the level, and $\Phi(\gamma)$ is a smooth aperture taper (here from 11.5° to 14°). Evaluated by stationary phase, (53) reduces to the ray field (52): the stationary points are the rays through $(x, z)$, the Hessian of the phase is $-\partial_\gamma p_x \, \partial_\gamma X$, and a change of sign of $\partial_\gamma X$ at a caustic produces the KMAH shift automatically. The amplitude $B$ contains $\partial_\gamma p_x$ instead of $\partial_\gamma X$, so it stays finite on the caustics in $x$, including the cusp. The chart fails only where $\partial_\gamma p_x = 0$. Above the inclusion that never happens: $\partial_\gamma p_x$ is negative across the whole aperture on every level, so a single chart covers the entire caustic region. We therefore use (53) for $z \leq 1450$ m, just above the inclusion. Below, the ray closure of the near-axis rays is used; on the axis there is no caustic there. The integral uses only the kinematic and dynamic ray quantities that the Eigenray solver already provides, requires no beam parameter, and is evaluated only at the points where the potential is needed. An alternative closure field, a superposition of Gaussian beams launched from the source, was also tested. Such beams are at least a Fresnel zone wide, which in this model is comparable to the inclusion (about 410 m at 12 Hz), and the resulting amplitude was 20–25 per cent too low. It is not pursued further here.

Where no single chart covers a family, for instance when a family is split into branches by its fold rays (Section 9.7), the chart can be sheared with an imaginary coefficient, $p_x - \kappa x$ with $\kappa = i\kappa_i$:

$$u = \int \Phi B \exp\left\{i\omega\left(\tau + p_x(x - X) + \frac{i\kappa_i}{2}(x - X)^2\right)\right\} \mathrm{d}\gamma,$$
$$B = \sqrt{\frac{v}{8\pi|\partial_\gamma X|\cos\theta}}\frac{e^{i(\pi/4 - m\pi/2)}}{\sqrt{2\pi}}\sqrt{i\,\partial_\gamma X\,D},\quad D = \partial_\gamma p_x - i\kappa_i\,\partial_\gamma X, \tag{54}$$

where $m$ is the KMAH index of each ray and $\kappa_i = 2/(\omega w_b^2)$ sets a Gaussian window of width $w_b$ around each ray. Because $Im\,D = -\kappa_i\,\partial_\gamma X$, the factor $D$ never vanishes, and (54) is regular on caustics in $x$ and in $p_x$ alike. For the lens family of this example, (54) reproduces (53) to within 4 per cent in amplitude and 0.01 rad in phase at 12 Hz for any $w_b \geq 400$ m. The potential of a branch $k$ is then

$$W_k = \frac{v^2\nabla^2 a_k}{\omega^2 a_k} - \frac{v_S^2}{4\omega^2 r_S^2},\qquad a_k = |u_k|, \tag{55}$$

where $r_S$ is the distance from the source. The second term removes the cylindrical near-field potential of the line source, eq. (I8). Its integrated effect on the phase is already carried by the constant $\pi/4$ of the asymptotic Green's function (52), and counting it again would double it; the potential is also tapered to zero within 300–600 m of the source. On the axis of the cusp configuration the Laplacian is evaluated with a 5 m stencil. For the receiver line of Section 9.7, where the Eigenwaves bend, the potential is evaluated on a 10 m by 5 m grid, set to zero where the closure amplitude falls below 30 per cent of its maximum on the level, clipped to $|W| \leq 0.5$, and smoothed over a quarter of a wavelength, so that its first and second derivatives are stable. Sixteen frequencies are used, from 2 to 32 Hz in steps of 2 Hz. Fig. 6a,b shows the closure amplitude and the potential along the axis at 12 Hz. The potential is small and smooth, with $|W| \leq 0.06$ above the inclusion.

### 9.4 Eigenwave stage and comparison with finite differences

By symmetry the potential of the lens family is even in $x$, so the axial Eigenwave remains on the axis at all frequencies and the whole finite-frequency effect enters its phase time. The correction $\delta\tau = \tau_\omega - \tau_0$ is −27 ms at 2 Hz, −16 ms at 12 Hz and −5.3 ms at 32 Hz; its phase counterpart $\omega\delta\tau$ grows from −0.34 rad at 2 Hz to between −1.0 and −1.3 rad above 10 Hz. The Eigenwave carries no KMAH index here. Because the closure field contains the whole focal region, the potential produces continuously the focal (Gouy-type) phase anomaly that ray theory inserts as the discrete KMAH jump $-\pi/2$. This is the two-dimensional analogue of the Gouy phase of the Gaussian beam, eq. (I12), and the KMAH phase must therefore not be added on the axis. Sections 9.5 and 9.7 show that off the axis this bookkeeping is not reliable. The edge arrivals are diffractions around a boundary that is sharp on the scale of the wavelength; they are not treated by the closure and are represented by their Eigenrays. The Eigenwave seismogram is then

$$G_{\omega}(\mathbf{x}_R, \omega) = \sum_k a_k(\mathbf{x}_R, \omega) \exp\left[i\left(\omega\tau_k(\omega) + \frac{\pi}{4}\right)\right], \tag{56}$$

with the amplitude of each arrival taken from its closure field and the phase from its Eigenwave; the edge arrivals take the form (52). The amplitude is not taken from the transport equation (41) along the Eigenwave, because at the cusp tip the effective-medium Jacobian of the axial branch still nearly vanishes. The phase times and $a_k\sqrt{f}$ are interpolated between the 16 computed frequencies, and the result is convolved with the source wavelet. The closure field itself, with its own phase, provides a second synthetic seismogram.

*Phase reference.* Phase errors are measured by applying the same Gaussian time window, of width 20–30 ms and centred on the axial arrival, to the finite-difference trace and to the axial-only synthetic seismograms, and comparing their spectra between 6 and 26 Hz. The window then cancels, and the earlier edge arrival is suppressed. The reference is accurate to about 0.03 rad: on the homogeneous test of Section 9.1 the finite-difference phase error is 0.02 rad; the leakage of the edge arrival, even scaled by the factor of 3.5 by which the Eigenrays underestimate it, is below 0.02 rad; and varying the window width from 20 to 30 ms changes the results by at most 0.03 rad. A window applied to the finite-difference trace alone, and not to the synthetics, is not an acceptable substitute: it biases the phase by several tenths of a radian.

*Results.* Fig. 7 compares the seismograms. The finite-difference trace peaks at 2.281 s with amplitude 0.0347 (arbitrary units common to all traces). The Eigenray seismogram overestimates this amplitude by a factor of 12.4, peaks 4 ms early and has a distorted waveform, with a cross-correlation of 0.90 and a relative misfit of 12.3 in the window 2.10–2.45 s. The Eigenwave seismogram peaks at the correct time, 3.5 per cent high, with cross-correlation 0.981 and relative misfit 0.21; the closure field reaches 0.996 and 0.11. Against the matched-window reference the amplitude of the axial arrival is 1.02–1.05 times the finite-difference value for the Eigenwave and 1.03–1.07 for the closure field, and the rms phase error is 0.02–0.05 rad for the Eigenwave, 0.05–0.07 rad for the closure field and 0.37 rad for the Eigenray (Fig. 6c,d). These values refer to an aperture taper from 11.5° to 14°. With the full aperture of the family (Section 9.5) the closure field keeps its accuracy (0.06 rad), whereas the phase error of the Eigenwave rises to 0.17 rad: the phase of the closure field is robust, that of the Eigenwave is sensitive to the aperture of the closure. In the frequency domain the actual focusing gain is two to three times the homogeneous amplitude, not the 27.5 of ray theory. Over the whole band from 2 to 32 Hz the Eigenwave phase differs from that of its own closure field by 0.22 rad rms, mostly below 6 Hz, where the ray closure used below the inclusion is least adequate. The amplitude is insensitive to the aperture taper: moving it anywhere between 9–13° and 13–14° keeps the mean amplitude ratio (8–24 Hz) between 1.02 and 1.08.

The weak precursor at 2.174 s (amplitude 0.0024, 7 per cent of the main arrival) is the contribution of the two edge branches. It peaks about 20 ms before the eikonal edge time $\tau_0 + t_0 = 2.193$ s, and its Eigenray estimate (0.0013) is about half the finite-difference value. The finite-difference amplitude of this arrival hardly depends on $w$, whereas its ray amplitude scales with the taper (Fig. 7b).

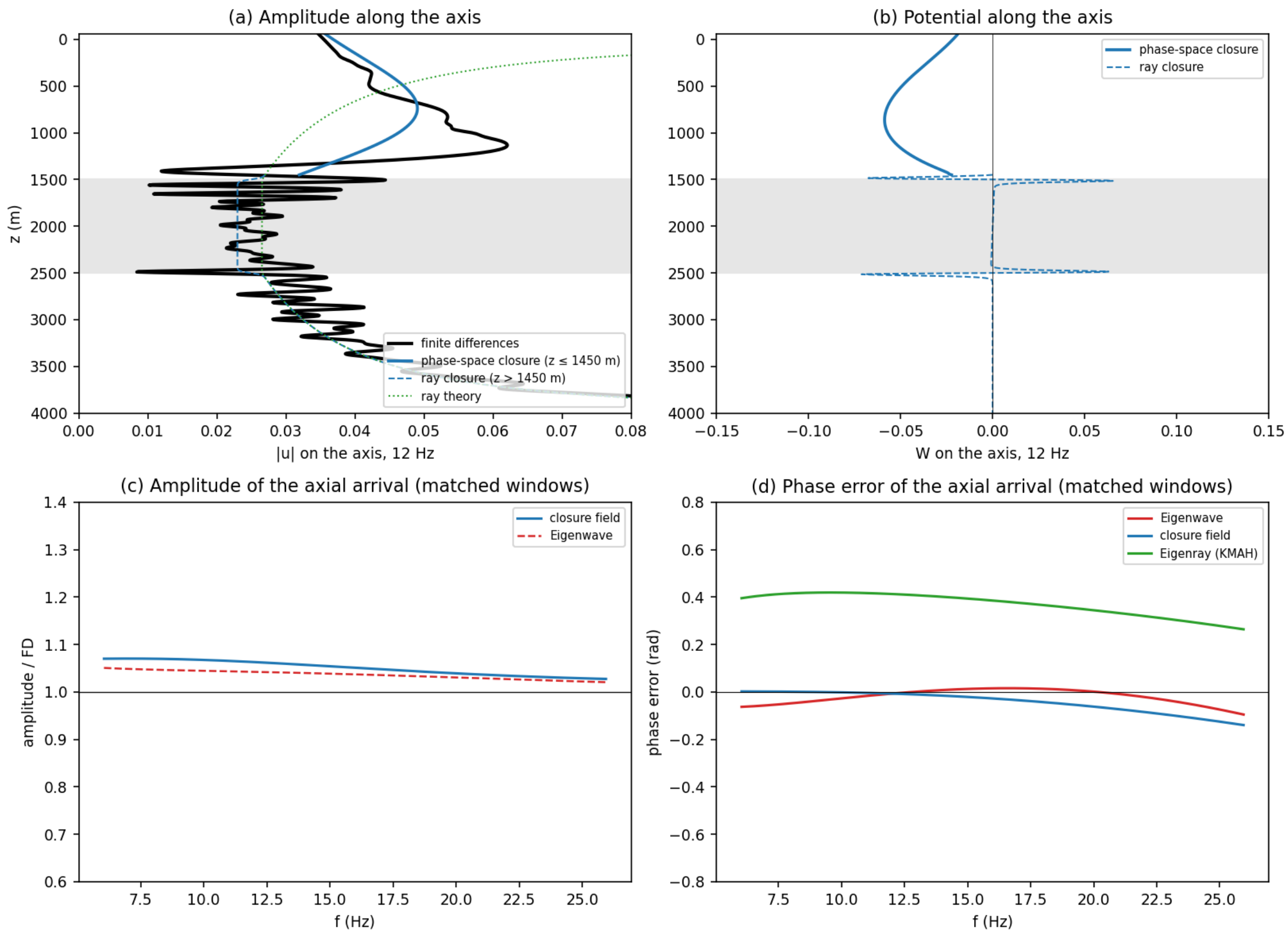

**Figure 6.** Phase-space closure at the cusp (Sections 9.3–9.4). (a) Amplitude along the axis at 12 Hz: finite differences, closure field (53) above 1450 m depth and ray closure below, and ray theory. (b) Potential (55) along the axis at 12 Hz. (c) Amplitude of the axial arrival relative to finite differences, with matched time windows. (d) Phase error of the axial arrival relative to finite differences, with matched time windows (width 25 ms), for the Eigenwave, the closure field and the Eigenray with its KMAH phase.

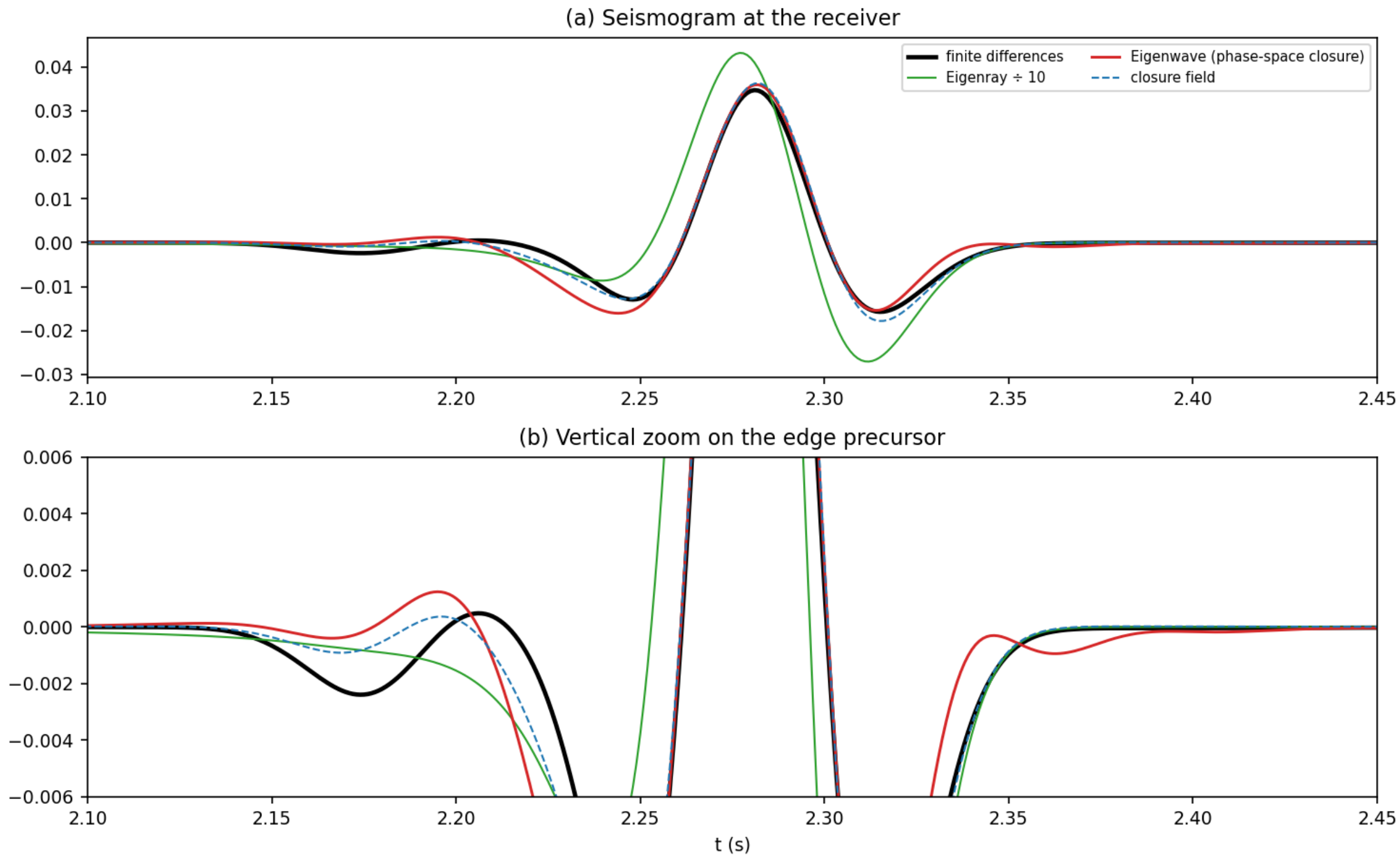

**Figure 7.** (a) Seismograms at the receiver: finite differences, Eigenray seismogram (52) divided by 10, Eigenwave seismogram (56) and closure field; the edge arrivals of the synthetic seismograms are those of their Eigenrays. (b) Vertical zoom on the edge precursor.

### 9.5 The full surface gather

To test the construction beyond a single receiver, the seismograms were computed for the 201 surface receivers of Fig. 5a, every 10 m between −1000 and 1000 m, and compared trace by trace with the finite-difference gather. Every receiver records one arrival of the lens family, whose rays all pass through the inclusion (take-off angles up to ±15.2°) and touch the cusp caustic (KMAH index 1), and one or two edge arrivals. The arrivals were located on a fan of 1477 rays. The closure field of the lens family, eq. (53), was evaluated once, on a 5 m grid and over the full aperture of the family, and for every receiver and frequency the Eigenwave was solved with the potential smoothed over a quarter of a wavelength (1616 path solves, about six minutes in all). Three gathers are compared in Fig. 8 and Table 2: the closure field, the Eigenwave and the Eigenray. In all three the edge arrivals are those of their Eigenrays.

The closure field reproduces the finite-difference gather along the whole line. Its median trace misfit is 0.20 (at most 0.24), its cross-correlation is at least 0.97, and the phase error of the lens arrival is 0.09 rad on the median and at most 0.13 rad. Within 200 m of the cusp its misfit is below 0.1; elsewhere the residual is dominated by the edge arrivals of the flanks (Fig. 8e). The Eigenray gather is accurate away from the cusp (median misfit 0.27) but fails at it, with a misfit of 12.4 on the axis. The Eigenwave gather is less accurate than both. With the KMAH index added off the axis its median misfit is 0.36 and its median phase error 0.29 rad; without it, the off-axis traces are wrong by 1–2 rad. In the transition zone 20–100 m from the axis neither choice works (0.6–0.9 rad): the paths cross the fine interference structure of the focal region, and their potential carries only part of the focal phase. On the axis itself, with the full aperture, the phase error of the Eigenwave is 0.17 rad, against 0.06 rad for the closure field. Keeping the closure amplitude but using the eikonal phase with its KMAH index gives a median misfit of 0.22 and a phase error of 0.15 rad, close to the Eigenray away from the cusp. For the seismogram, therefore, the closure field is the method that reproduces the finite-difference solution; the Eigenwave phase is reliable on the axis of the focusing but not yet near caustics off it.

**Table 2.** Comparison of the surface gather (201 traces, |x| ≤ 1000 m) with finite differences. Misfit and cross-correlation refer to a window around all arrivals of each trace; the phase error refers to the lens arrival between 6 and 26 Hz, measured with matched time windows of width 20 ms.

| Method | Relative misfit: median (max) | Cross-correlation: median (min) | Phase error (rad): median (max) |
|---|---|---|---|
| Closure field (53) | 0.20 (0.24) | 0.981 (0.971) | 0.09 (0.13) |
| Closure amplitude, eikonal phase with KMAH | 0.22 (0.43) | 0.976 (0.911) | 0.15 (0.37) |
| Eigenray, eq. (52) | 0.27 (12.4) | 0.975 (0.902) | 0.15 (0.37) |
| Eigenwave, KMAH added off the axis | 0.36 (1.2) | 0.935 (0.33) | 0.29 (1.3) |
| Eigenwave, no KMAH | 1.1–1.4 | 0.03–0.39 | 1.2–1.8 |

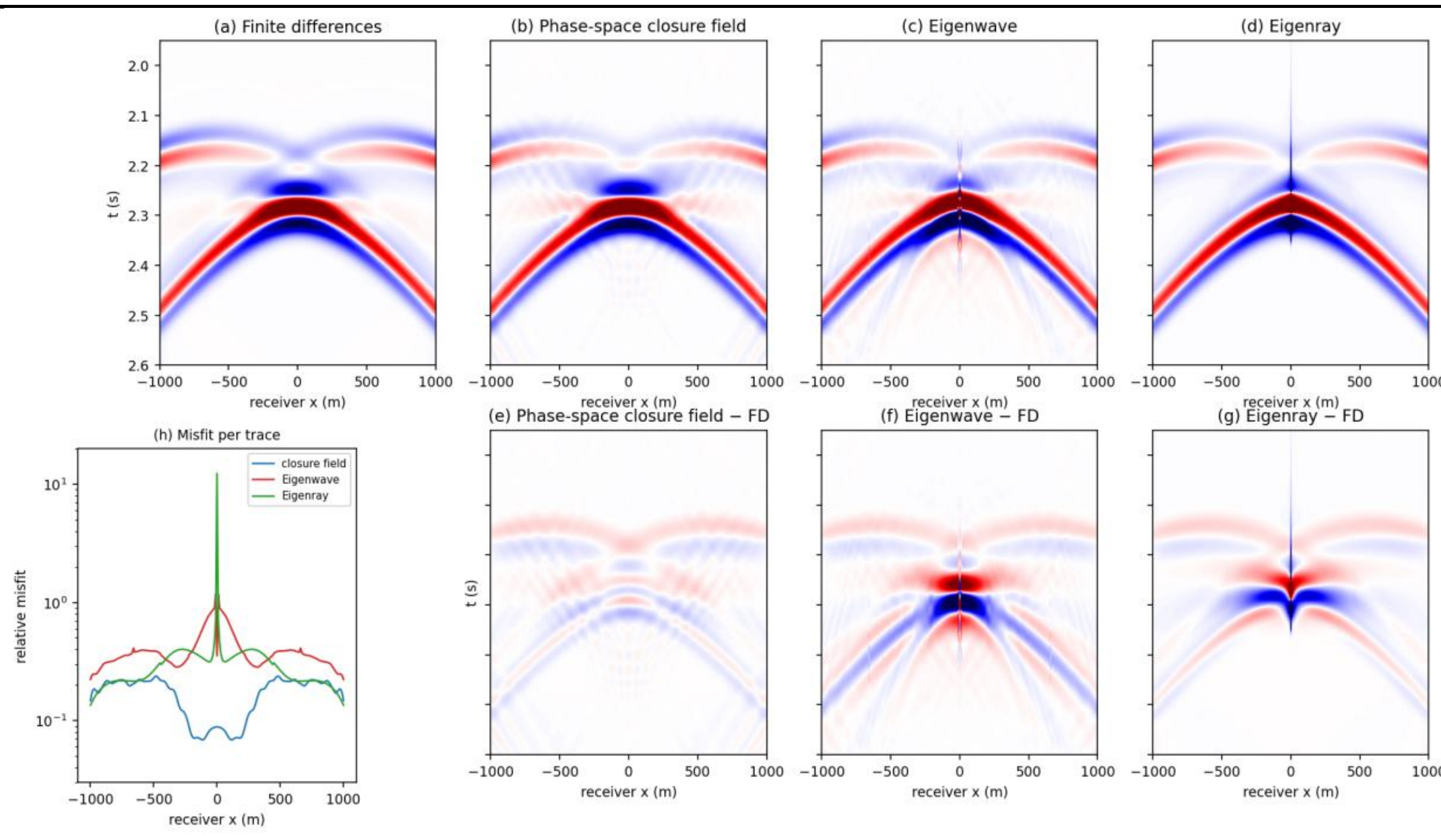

**Figure 8.** Surface gather of the cusp configuration (Section 9.5). (a) Finite differences. (b)–(d) Phase-space closure field (53), Eigenwave (with the KMAH index added off the axis) and Eigenray; the edge arrivals are those of their Eigenrays in all three. (e)–(g) Differences from finite differences, on the same colour scale as (a). (h) Relative misfit of each trace.

### 9.6 Verification with the exact closure

To separate the error of the closure from that of the stationary-path machinery, the finite-difference solver was rerun with an on-the-fly discrete Fourier transform of the wavefield in a 60 m wide strip around the axis, at the same 16 frequencies. On the axis the phase time of the axial Eigenwave reduces to $\int n/v\,\mathrm{d}z$.

The potential (55) is computed from the finite-difference field itself, $a = |u_{\mathrm{FD}}|$, with a 5 m Laplacian stencil and 5 m smoothing. The exact potential is very different from the closure potential. In and below the inclusion it oscillates with a period of about half a wavelength and amplitudes of order one. These oscillations are the interference of the transmitted wave with the waves reflected inside the inclusion and scattered by its boundary. They appear as the ripple of the finite-difference amplitude in Fig. 6a and are absent from every per-branch closure field. Nevertheless, the Eigenwave in the exact potential reproduces the finite-difference phase at the receiver with an rms error of 0.19 rad between 2 and 22 Hz, because the oscillations of $W$ largely cancel in the phase integral. The residual is numerical, and it reveals an intrinsic sensitivity. To first order the phase error is $\delta\varphi \approx \pi N_\lambda \delta\overline{W}$, where $N_\lambda$ is the path length in wavelengths and $\delta\overline{W}$ the path-averaged error of $W$. For $N_\lambda \approx 24$ (12 Hz), a phase accuracy of 0.1 rad requires $W$ to be correct on average to about $1.3 \times 10^{-3}$. Accordingly, changing the differentiation stencil (5–15 m) or the smoothing length (0–10 m) raises the rms error over the band to between 0.5 and 1.4 rad. At 24, 30 and 32 Hz the test fails outright. The finite-difference field then has a phase singularity on or next to the axis between about 950 and 1250 m depth, where $\partial_z\varphi$ changes sign, the exact potential diverges and the exact trajectory to the receiver is no longer the axis. This is the node problem of Section 8.1. With an accurate potential, then, the stationary-path machinery reproduces the wave-equation phase, and the remaining errors of Section 9.4 originate in the closure.

### 9.7 Receivers away from the cusp

In the configuration of Sections 9.1–9.6 the receiver sits at a cusp, which is close to a worst case. To measure how the accuracy depends on the distance from a caustic, the inclusion was moved to 3000 m depth, and the comparison was repeated for receivers along the surface. Everything else was kept unchanged. The source is now 1000 m below the centre of the inclusion, at its paraxial focal distance, so the near-axis rays emerge almost parallel, and the over-focused marginal rays form a cusp whose folds reach the surface at $x = \pm 173$ m. Receivers with $|x| < 173$ m record three arrivals of the lens family: a central branch and two outer branches, all of which have touched a caustic on their way (KMAH index 1), except the central one. Beyond the folds, a single arrival remains, from the outer branch that meets the central one at the opposite fold. No edge-family ray reaches the receivers used ($|x| \leq 800$ m). The distance from the caustic is measured in Airy units. Near the fold at $x_c = 173$ m the traveltime difference of the two coalescing arrivals is $\Delta\tau \approx c(x_c - x)^{3/2}$, with $c \approx 3.15 \times 10^{-6}\,\mathrm{s\,m^{-3/2}}$, so that the argument of the uniform Airy asymptotics is $\xi = (3\omega\Delta\tau/4)^{2/3} = (x_c - x)/\ell_A$, with Airy length $\ell_A = (3\omega c/4)^{-2/3} \approx 315$ m at 12 Hz. Positive $\xi$ lies inside the cusp and negative $\xi$ on the shadow side of the fold. The receivers used span $-2 \leq \xi \leq 0.55$.

One finite-difference run records the whole line. For each receiver the Eigenrays were computed. Each branch was given its own uniform closure field (54), with its take-off window bounded by the fold rays ($|\gamma| = 14.3°$), and with $w_b = 800$ m. The potential was obtained with masking at 30 per cent of the level maximum, clipping at $|W| \leq 0.5$ and smoothing over a quarter of a wavelength, and the Eigenwave of each arrival was solved by the finite-element solver, single pass. Three synthetic seismograms are compared: the Eigenray seismogram (52); the Eigenwave seismogram, with the amplitude of each arrival from its closure field and the phase from its Eigenwave; and the sum of the closure fields. Table 3 and Fig. 9 summarize the results. For the single-arrival receivers the misfit is measured in a window around the lens arrival, for reasons given below.

**Table 3.** Relative misfit to finite differences along the receiver line (inclusion at 3000 m depth). $\xi$: Airy distance from the fold at 173 m at 12 Hz. For the Eigenray at the three-arrival receivers the peak-amplitude ratio is given in parentheses.

| x (m) | ξ (12 Hz) | Arrivals | Eigenray | Eigenwave | Closure field |
|---|---|---|---|---|---|
| 0 | +0.55 | 3 | 0.34 (×1.29) | 0.35 | 0.12 |
| 100 | +0.23 | 3 | 0.44 (×1.42) | 0.33 | 0.12 |
| 170 | ≈0 | 3 | 1.81 (×2.84) | 0.21 | 0.13 |
| 250 | −0.24 | 1 | 0.77 | 0.54 | 0.15 |
| 400 | −0.72 | 1 | 0.81 | 0.52 | 0.38 |
| 600 | −1.36 | 1 | 0.27 | 0.43 | 0.26 |
| 800 | −2.0 | 1 | 0.17 | 0.37 | 0.17 |

Three regimes appear. Within about one Airy length of the caustic ($x \leq 250$ m), the Eigenray fails in amplitude, overestimating the peak by factors of 1.3–2.8. The Eigenwave reduces its misfit substantially, most strongly at the fold (0.21 against 1.81), and the closure field is best everywhere (0.12–0.15). Far from the caustic ($\xi \lesssim -1.4$), ray theory is already accurate, with misfit 0.17–0.27, and the single-pass Eigenwave is slightly worse (0.37–0.43). There the true finite-frequency correction is smaller than the numerical noise in $W$. The Eigenwave still shifts the phase time by 3–6 ms at 12 Hz, which the finite-difference solution does not support. This is the ill-conditioning identified in Section 9.6: the phase error scales as $\pi N_\lambda$ times the mean error of $W$, and here $N_\lambda \approx 24$. The noise also generates spurious conjugate points: at the far receivers and the highest frequencies the Eigenwave Morse index rises to 2 or 3. Between these regimes, at 0.3–1.5 Airy lengths on the shadow side, the error is dominated by an arrival that belongs to no ray branch. The wave diffracted around the inclusion, together with the evanescent tail of the fold, precedes the lens arrival (Fig. 9a,e). It is missed by the Eigenray and the Eigenwave and only partly captured by the closure field. In a window that includes it, the misfits at 400 m rise to 0.92, 0.73 and 0.36, respectively.

*Phase bookkeeping.* Every arrival of the outer branches has touched a caustic before reaching the surface. If the Eigenwave phase is taken as $\omega\tau_\omega + \pi/4$ alone, as for the axial branch of Section 9.4, the Eigenwave seismograms are wrong outright, with cross-correlations with finite differences between −0.15 and 0.74. With the finite-frequency KMAH index added, taken as the Morse index of each Eigenwave in accordance with (B20), they give the values of Table 3. In the cusp configuration, by contrast, adding the Morse index of the axial Eigenwave would have produced a focal phase of about −2.9 rad, against −1.1 to −1.5 rad in the finite-difference solution. The clean phase comparison of Section 9.4, applied along the line, refines this picture. Inside the cusp (x = 0 and 100 m) the rms phase error is 0.07–0.14 rad for the Eigenray and 0.22–0.26 rad for the Eigenwave. The phase-time corrections of the outer-branch Eigenwaves, −0.1 to −0.7 rad at 12 Hz, are not supported by the finite-difference solution: keeping the closure amplitude but replacing the Eigenwave phase by the eikonal phase with its KMAH index gives 0.06 rad, the best of all methods there. Far from the caustic (600–800 m) the phase errors are 0.13–0.15 rad for the Eigenray, 0.11–0.13 rad for the closure field and 0.22–0.37 rad for the Eigenwave, and recomputing the Eigenwave phase from the unsmoothed potential does not change this (0.21–0.37 rad). On the axis of the cusp configuration, where the closure field contains the whole focal region, the potential carries the focal phase and the KMAH index must not be added (Section 9.4); for the split caustics of the receiver line it must be added. The full gather of Section 9.6 shows, however, that even a caustic contained in one closure field is not handled reliably off the axis, so that no simple rule emerges. Near caustics the Eigenwave phase correction is not yet reliable, and the eikonal phase with its KMAH index, combined with the closure amplitude, is the safer estimate. A theoretical treatment of the phase bookkeeping is required.

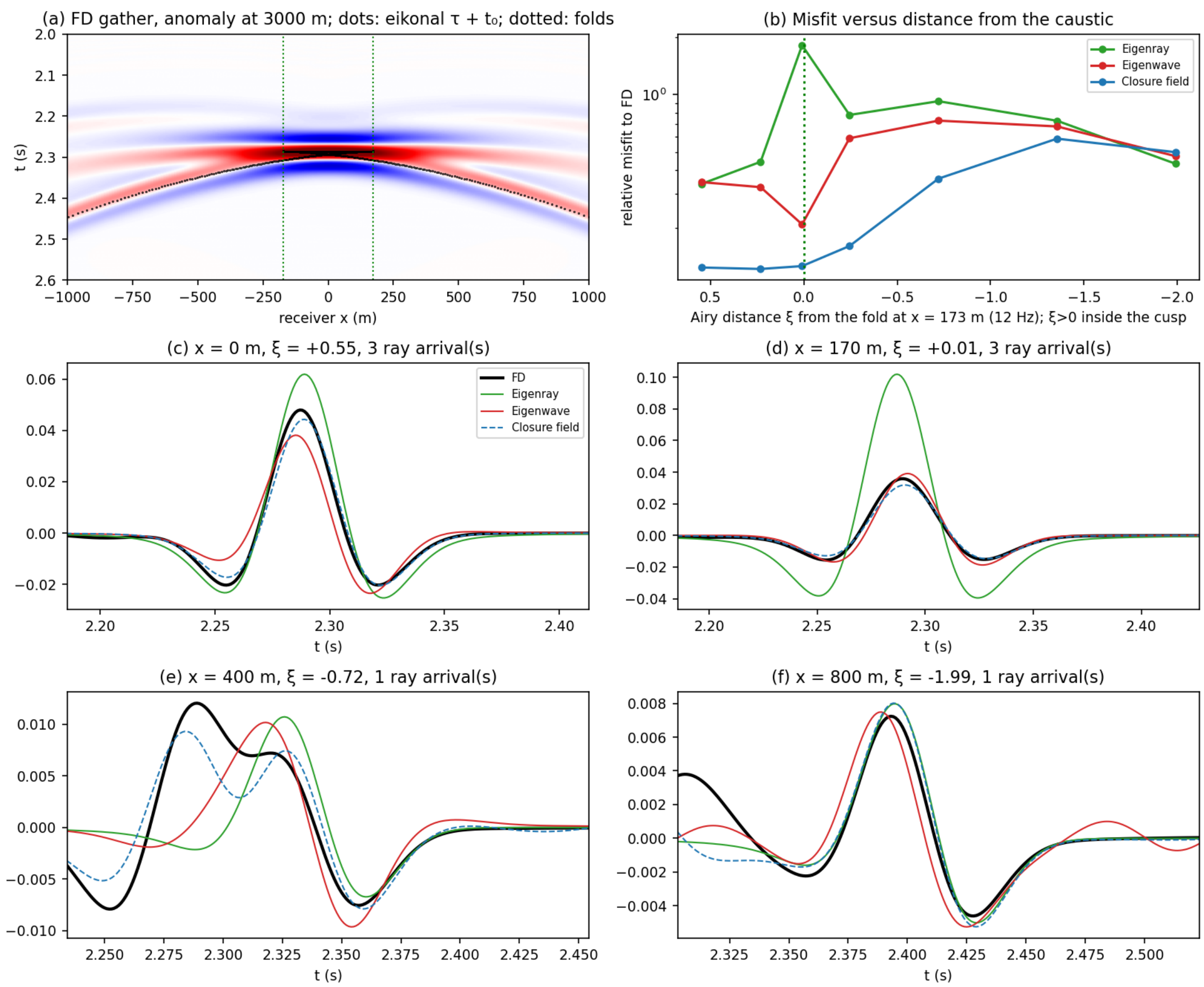

**Figure 9.** Receiver line with the inclusion at 3000 m depth (Section 9.7). (a) Finite-difference surface gather with the eikonal traveltimes of the lens family (dots); the dotted lines mark the folds at $x = \pm 173$ m. (b) Relative misfit to finite differences against the Airy distance $\xi$ from the fold at 12 Hz, in the full window around all arrivals. (c)–(f) Seismograms at $x = 0$, 170, 400 and 800 m: finite differences, Eigenray, Eigenwave and closure field.

## 9.8 Advantages demonstrated by the example

The example demonstrates the following strengths of the Eigenwave approach.

- **Complete, classified multipathing.** The two-point formulation finds all stationary paths from coarse shooting guesses and converges in one or two Newton iterations. The same Hessian that resolves the path supplies its Morse index, and hence the KMAH index, without tracking caustics along the ray.

- **Caustic-regular seismograms.** At the cusp the Eigenray overestimates the amplitude by a factor of 12. The phase-space closure field reproduces the finite-difference seismograms along the whole receiver line, with a median trace misfit of 0.20 and phase errors below 0.13 rad. On the axis of the focusing the potential converts the discrete KMAH jump into a continuous, frequency-dependent focal phase.

- **A closure native to the framework.** The phase-space closure (53), and its uniform form (54), use only the kinematic and dynamic ray quantities of the family that the Eigenray solver already computes. They need no beam parameter, are evaluated only where the potential is needed, and remain valid at the cusp.

- **Low incremental cost per frequency.** By the effective-medium equivalence (Proposition 1), the Eigenwave is computed by the unchanged Eigenray solver with a rescaled slowness. Each frequency is a warm-started Newton solve of typically one to four iterations. No wave-equation solve is needed along the path.

- **Built-in diagnostics.** The magnitude of $W$ along the path indicates the validity of the construction and shows where the finite-frequency phase is acquired. At the cusp, that is the caustic region above the inclusion, not the receiver itself.

- **Frequency-dependent paths.** Each arrival is associated with a frequency-dependent path along which its phase accumulates. Such paths are the natural carriers of finite-frequency sensitivity kernels (Appendix I.3) and of path-based tomography.

## 9.9 Limitations revealed by the example

The example also exposes the following limitations.

- **The Eigenwave is only as good as its closure.** With the exact closure the Eigenwave reproduces the wave-equation phase (0.19 rad rms, Section 9.6), and with the phase-space closure its phase on the axis of the cusp is accurate to 0.02–0.17 rad, depending on the aperture of the closure. Its amplitude, however, is taken from the closure field, and its phase differs from that of the closure field by 0.22 rad rms over the full band, mostly below 6 Hz. No approximate closure is an exact solution of the Helmholtz equation, and the Eigenwave inherits its inconsistencies. Closure fields built from beams launched at the source are limited by the Fresnel width of the beams.

- **Caustics beyond the fold.** At a cusp tip the transport-equation amplitude along a one-iteration Eigenwave remains singular and the Morse index is not robust, so the amplitude must be taken from the closure field. A single mixed chart covers the caustic family here, but where caustics in $p_x$ also occur, charts must be combined or the uniform form (54) used, and at the joint with the ray closure the fields are not continuous. Iterating the closure to self-consistency, which was also tested, converges only with regularization of the potential. It moves the focus of the near-axis rays from the receiver towards the lens, close to where the finite-difference field focuses, but it does not improve the accuracy, because the regularization removes part of the focal phase.

- **Diffraction-dominated arrivals.** The edge precursor, a diffraction around a boundary that is sharp on the scale of the wavelength, is underestimated by a factor of about two, and on the shadow side of a fold the wave diffracted around the inclusion belongs to no ray branch (Section 9.7). Neither is represented by a ray-family closure.

- **Sensitivity of the phase to the potential.** The phase is an integral of second derivatives of the amplitude over many wavelengths. Its error scales as $\pi N_\lambda$ times the average error of $W$, so an accuracy of 0.1 rad requires $W$ to about $10^{-3}$ in this example. The removal of the source term, the source taper, and the clipping, differentiation and smoothing of $W$ all affect the phase times at the level of milliseconds.

- **No gain far from caustics.** More than about 1.5 Airy lengths from any caustic, the finite-frequency phase correction is smaller than the numerical noise of the potential, and the single-pass Eigenwave is slightly less accurate than the Eigenray (Section 9.7).

- **Eigenwave phase near caustics.** Whether the KMAH index must be added to the Eigenwave phase near a caustic could not be decided by a simple rule: omitting it is correct on the axis of a symmetric focus, adding it is correct for split caustics, and neither is correct close to the axis (Sections 9.5 and 9.7). Near caustics the Eigenwave phase correction is therefore not yet fully reliable, and the eikonal phase with its KMAH index is the safer estimate.

- **Locality in three dimensions.** The phase-space closure is path-local and inexpensive, but it requires a fan of rays around each Eigenray, and in three dimensions the mixed chart becomes a choice among several partial Legendre transforms whose automatic selection along a path remains to be developed.

- **Scope.** The test is two-dimensional, acoustic and isotropic. Anisotropy, shear modes and mode coupling (Section 10) are not exercised.

### 9.10 Summary and recommended configuration

On a model combining multipathing, a cusp caustic at the receiver and a Fresnel-scale anomaly, the phase-space closure field reproduces the finite-difference seismograms along the whole receiver line (Table 2): median trace misfit 0.20, cross-correlation at least 0.97 and phase errors below 0.13 rad, whereas the Eigenray seismogram is wrong by an order of magnitude at the cusp. The remaining error of the Eigenwave does not come from the stationary-path machinery: with the exact closure its phase agrees with the wave equation to 0.19 rad rms, except where the total field has nodes. It comes from the closure and from the phase bookkeeping near caustics. The Eigenwave phase is accurate on the axis of the focusing but sensitive to the aperture of the closure, and it is not yet reliable near caustics off the axis or for split caustics. Far from caustics (Section 9.7) the Eigenray is already accurate, and the finite-frequency correction is smaller than the numerical noise of the potential. Diffraction around features that are sharp on the scale of the wavelength belongs to no ray family and is represented by none of the methods.

The recommended configuration is therefore as follows. Use the Eigenray solver to find and classify all stationary paths, and use the Eigenray alone where the arrivals are several Airy lengths from any caustic. Near caustics, compute the seismogram from the closure field, built from the phase-space representation (53) of the ray family, or from its uniform form (54) where no single chart covers the family. Use the Eigenwave for what the closure field cannot provide: the frequency-dependent stationary paths, the phase times away from caustics, the sensitivity kernels, and the magnitude of W along the path as a validity indicator. Appendix L compares the two constructions in detail.

## 10 Discussion and conclusions

The Eigenray method solves the two-point boundary-value ray tracing problem in 3D smooth heterogeneous general anisotropic elastic media. Its kinematics rest on Fermat's principle with an arclength-related Lagrangian depending on both the location and the direction of the trajectory, admitting an efficient finite-element implementation with the Galerkin method, weak formulation and Hermite interpolation. Starting from an initial non-stationary discretized trajectory, a constrained target function is optimized by computing corrections to the spatial locations and directions that yield a stationary traveltime solution, with the node-distribution and direction-normalization constraints closing the system. The dynamics follow from the same variational structure at second order, with the Jacobi equation solved by identical finite elements, delivering paraxial rays, the ray Jacobian and relative geometric spreading. Numerical examples in Koren and Ravve (2021); Ravve and Koren (2021) demonstrate reliable ray paths and controlled, stable amplitudes in configurations challenging for conventional shooting: multipathing through velocity anomalies, head waves beyond the crossover offset, thin high-velocity layers and gas-cloud models.

That formulation was deliberately arranged as the kinematic backbone of a finite-frequency generalization. In the trajectory-mechanics approach (Vasco and Nihei, 2019a), an exponential (polar) representation of the wavefield leads, without invoking the infinite-frequency limit, to frequency-dependent eikonal and trajectory equations in which the paths themselves depend on frequency through amplitude-gradient coupling terms; broad-band, finite-frequency trajectories tunnel through and are deflected by velocity anomalies in a manner that smoothly interpolates between the ray-theoretical and scattering regimes. The variational, direction-explicit finite-element machinery - in particular the Hermite discretization carrying both location and direction as degrees of freedom, the constrained target-function optimization and the reuse of the assembled second-derivative operators - extends naturally to such frequency-dependent Lagrangians, in which the ray-velocity kernel is augmented by finite-frequency correction terms, with the amplitude closure regularized through the phase-space (Maslov) closure of Section 8.3, implemented in Section 9.3. This paper has carried out that extension.

Specifically, we have extended the Eigenray two-point ray tracing method to the finite-frequency trajectories of trajectory mechanics. The extension rests on a single structural result: for a fixed wave mode and a given amplitude field, the finite-frequency Hamiltonian $\lambda(\mathbf{x},\mathbf{p}) - W(\mathbf{x},\omega) = 1$ is dynamically equivalent to the classical Christoffel Hamiltonian of an effective medium with the conformally rescaled stiffness $\tilde{\mathbf{C}}^{\omega} = \tilde{\mathbf{C}}/(1+W)$. The Eigenwave Lagrangian is therefore $L^{\omega} = nL$ with $n = \sqrt{1+W}$, and it inherits the first-degree homogeneity in the ray direction on which the entire Eigenray theory is built. From this we derived, in explicit form:

- the finite-frequency Euler-Lagrange equation (22), with the additional force $\nabla_{\mathbf{x}} W/\left(2nv_{\text{ray}}\right)$, the finite-frequency slowness $\mathbf{p}^{\omega} = n\mathbf{p}$, the matching arclength Hamiltonian and Legendre transformation, and the unchanged weak formulation with the modified ray-velocity kernel (27);
- the finite-frequency Hessians (28) and the Jacobi dynamic ray tracing equation (29), whose finite-frequency terms are driven by $\nabla_{\mathbf{x}} n$ and $\nabla_{\mathbf{x}}\nabla_{\mathbf{x}} n$, together with the unchanged point-source/plane-wave initial conditions, the ray Jacobian, the caustic criteria, the geometric spreading (34) and the equivalence with the Hamiltonian dynamic ray tracing (36)-(37);
- the weak-coupling theory: the first-order Eigenwave path is the solution of the classical Jacobi equation forced by $\nabla_{\mathbf{x}} W$ (38), and the phase-time correction is $1/2 \int W/v_{\text{ray}} \, \mathrm{d}s$ along the Eigenray (39), both scaling as $\omega^{-2}$.

Several by-products deserve mention. The source-point normalized geometric spreading is frequency independent (35). The frequency dependence of the phase and ray velocities is a pure scalar factor, so the anisotropic "machinery" - the ray-to-phase direction map, the polarizations, the spatial and directional derivatives of the ray velocity (Ravve and Koren, 2019) - is reused without modification. The finite-frequency phase time reduces to the eikonal traveltime as $\omega \to \infty$ and coincides with the phase-time expression of Vasco and Nihei (2019a). Finally, the finite-element resolving matrix of the dynamic Eigenwave problem is again the (finite-frequency) traveltime Hessian of the kinematic stage.

Two aspects call for further work. The first is the closure near caustics. The Eigenwave equations are exact for an exact potential, and the numerical example of Section 9 confirms that, given such a potential, the Eigenwave reproduces the wave-equation phase at a cusp. In their ray-closed form, however, they do not by themselves provide an accurate amplitude at caustics and cusps. There the Eigenwave must be combined with a uniform closure, of which the phase-space (Maslov) closure is the natural choice. With it, the closure field reproduces the finite-difference seismograms of Section 9 along the whole receiver line (median trace misfit 0.20), where the Eigenray fails at the cusp. The resulting method therefore keeps the Eigenwave formulation and closes it with a Maslov phase-space closure field: the Eigenray solver finds and classifies the arrivals, the closure field supplies the potential and the amplitude, the Eigenwave supplies the frequency-dependent paths, phase times and sensitivity kernels, and near caustics the seismogram is taken from the closure field rather than from the Eigenwave phase. The Eigenwave phase, however, is reliable on the axis of the focusing but not yet off it: near caustics it depends on the aperture of the closure and on the bookkeeping of the KMAH index, and far from caustics the finite-frequency correction is smaller than the numerical noise of the potential. Iterating the closure to self-consistency, that is, the self-consistent determination of $W$ within the two-point framework, moves the focus towards its finite-frequency position but does not remove the caustics of the family: the non-crossing property holds for the exact trajectories of the total field (Section 8.1), not for the ray-closed, per-branch Eigenwaves. A formulation that keeps the family intact, for example a superposition of finite-frequency members of the Maslov integral, may make the Eigenwave phase reliable near caustics and remains the main open problem. The second is the treatment of quasi-shear modes and mode coupling: since $W$ is defined through the polarization of the mode, the effective-medium equivalence holds mode by mode, and rapidly varying regions where the modes couple require a matrix-valued generalization of $W$ (the full vector $\mathbf{w}$ of (8)). Systematic validation against the finite-difference-derived trajectories of Vasco and Nihei (2019a) for the layer and well-log models will be presented in a companion paper.

**Acknowledgements**

The author is grateful to Emerson-AspenTech for the support of this study and for the permission to publish its results. The Eigenray method, on which this work is built, was developed jointly with Igor Ravve, and the author gratefully acknowledges his considerable contribution to it. Although Igor was unable to take part in the present research, much of what is reported here rests on our long and fruitful collaboration.

## Appendix A. Rules and conventions of the tensor-form notation

This appendix collects, in one place, the rules that govern the coordinate-free (tensor-form) equations used throughout the paper. All equations of the main text can be expanded into Cartesian components by the conventions listed here.

### A.1 Typefaces and basic objects

Scalars are set in italics ($L$, $H$, $W$, $n$, $v_{\mathrm{ray}}$, $\lambda$, $J$, $\tau$, $s$, $\omega$).

Boldface lowercase letters denote three-component spatial vectors: the location $\mathbf{x}$, the ray direction $\mathbf{r}$, the slowness $\mathbf{p}$, the paraxial shift $\mathbf{y}$, gradients such as $\nabla_{\mathbf{x}} W$.

Boldface uppercase letters denote second-order tensors ($3 \times 3$ matrices), e.g. the identity $\mathbf{I}$, the Christoffel tensor $\mathbf{\Gamma}$, the ray-velocity tensor $\widetilde{\mathbf{A}}$, the Lagrangian Hessians $L_{\mathbf{xx}}$, $L_{\mathbf{xr}}$, $L_{\mathbf{rx}}$, $L_{\mathbf{rr}}$, or reduced $2 \times 2$ matrices in the ray-centred frame ($\mathbf{Q}$, $\mathbf{P}$, $\mathbf{M}$, $\mathbf{X}$, $\mathbf{Y}$, $\mathbf{Z}$).

The density-normalized stiffness $\tilde{\mathbf{C}}$ is a fourth-order tensor with the units of velocity squared, $\tilde{c}_{ijkl} = c_{ijkl}/\rho$, obeying the elastic symmetries $\tilde{c}_{ijkl} = \tilde{c}_{jikl} = \tilde{c}_{ijlk} = \tilde{c}_{klij}$.

A hat denotes a unit vector ($\hat{\mathbf{g}}$, $\hat{\mathbf{a}}$, $\mathbf{n} = \mathbf{p}/|\mathbf{p}|$); a tilde marks density normalization ($\tilde{\mathbf{C}}$) or, in Appendix B, transverse (frame-projected) quantities ($\tilde{\mathbf{p}}$, $\widetilde{\mathbf{X}}$).

An overdot is the total derivative with respect to the arclength along the trajectory, $\dot{(\ )} = d(\ )/ds$.

The superscript $\omega$ marks finite-frequency (effective-medium) quantities ($L^{\omega}$, $\mathbf{p}^{\omega}$, $v_{\mathrm{ray}}^{\omega}$, $\tilde{\mathbf{C}}^{\omega}$), the subscript $S$ evaluation at the source point, and the superscript $\mathsf{T}$ the transpose.

### A.2 Derivatives

The operators $\nabla_{\mathbf{x}}$ and $\nabla_{\mathbf{r}}$ are the gradients with respect to the location and the ray direction, taken at fixed value of the other argument; $\nabla_{\mathbf{p}}$ is the gradient in slowness space.

Subscripts attached to the Lagrangian and the Hamiltonian are shorthand for these gradients,

$$L_{\mathbf{x}} = \nabla_{\mathbf{x}} L, \;\; L_{\mathbf{r}} = \nabla_{\mathbf{r}} L, \;\; H_{\mathbf{x}} = \nabla_{\mathbf{x}} H, \;\; H_{\mathbf{p}} = \nabla_{\mathbf{p}} H. \tag{A1}$$

Double subscripts denote the second-derivative (Hessian) tensors, with the *first* subscript labelling the row index and the *second* the column index,

$$(L_{\mathbf{xr}})_{ij} = \frac{\partial^2 L}{\partial x_i \, \partial r_j}, \qquad L_{\mathbf{rx}} = (L_{\mathbf{xr}})^{\mathsf{T}}, \;\; (L_{\mathbf{xx}})_{ij} = \frac{\partial^2 L}{\partial x_i \, \partial x_j}, \;\; (L_{\mathbf{rr}})_{ij} = \frac{\partial^2 L}{\partial r_i \, \partial r_j}, \tag{A2}$$

so that $L_{\mathbf{xx}}$ and $L_{\mathbf{rr}}$ are symmetric while $L_{\mathbf{xr}}$ and $L_{\mathbf{rx}}$ are not, individually.

The same convention applies to $\nabla_{\mathbf{x}}\nabla_{\mathbf{x}} v_{\text{ray}}$, $\nabla_{\mathbf{x}}\nabla_{\mathbf{r}} v_{\text{ray}}$, $\nabla_{\mathbf{x}}\nabla_{\mathbf{x}} n$, etc.

The divergence contracts the gradient with the adjacent index of its (tensor-valued) argument: for a second-order field $\boldsymbol{\sigma}$,

$$(\nabla_{\mathbf{x}} \cdot \boldsymbol{\sigma})_i = \frac{\partial \sigma_{ij}}{\partial x_j}, \tag{A3}$$

which is unambiguous for the (symmetric) stress used in the text.

### A.3 Products and contractions

A single dot contracts one pair of adjacent indices: $\mathbf{b} \cdot \mathbf{c} = b_i c_i$, $(\mathbf{A} \cdot \mathbf{c})_i = A_{ij} c_j$, $(\mathbf{b} \cdot \mathbf{A})_j = b_i A_{ij}$, $\mathbf{b} \cdot \mathbf{A} \cdot \mathbf{c} = b_i A_{ij} c_j$, $(\mathbf{A} \cdot \mathbf{B})_{ik} = A_{ij} B_{jk}$. The dot is frequently omitted between a Hessian and a vector ($L_{\mathbf{rr}} \dot{\mathbf{y}} \equiv L_{\mathbf{rr}} \cdot \dot{\mathbf{y}}$) and between conformable matrices. For the fourth-order stiffness, the sandwich contraction defining the Christoffel tensor is

$$\boldsymbol{\Gamma} = \mathbf{p} \cdot \tilde{\mathbf{C}} \cdot \mathbf{p}, \qquad \Gamma_{ik} = p_j \, \tilde{c}_{ijkl} \, p_l, \tag{A4}$$

i.e. the two dots contract the slowness with the second and the fourth index of $\tilde{\mathbf{C}}$; owing to the minor symmetries of $\tilde{\mathbf{C}}$ the result does not depend on which index of each symmetric pair is contracted, and $\boldsymbol{\Gamma}$ is symmetric. A double dot (colon) contracts two pairs of indices,

$$\left(\tilde{\mathbf{C}} \colon \nabla_{\mathbf{x}} \mathbf{u}\right)_{ij} = \tilde{c}_{ijkl} \, \frac{\partial u_k}{\partial x_l}, \qquad (\hat{\mathbf{g}} \otimes \mathbf{b}) \colon \tilde{\mathbf{C}} \colon (\hat{\mathbf{g}} \otimes \mathbf{c}) = \hat{g}_i \, b_j \, \tilde{c}_{ijkl} \, \hat{g}_k \, c_l. \tag{A5}$$

The dyadic (outer, tensor) product of two vectors is the rank-one second-order tensor

$$(\mathbf{b} \otimes \mathbf{c})_{ij} = b_i c_j, \qquad (\mathbf{b} \otimes \mathbf{c}) \cdot \mathbf{d} = (\mathbf{c} \cdot \mathbf{d}) \mathbf{b}, \tag{A6}$$

$$\mathbf{d}\cdot(\mathbf{b}\otimes\mathbf{c})=(\mathbf{d}\cdot\mathbf{b})\mathbf{c},\qquad(\mathbf{b}\otimes\mathbf{c})^{\mathsf{T}}=\mathbf{c}\otimes\mathbf{b}.$$

These rules are used repeatedly to expand rank-one Hessian corrections such as $(\nabla_{\mathbf{x}}n\otimes\mathbf{p})\dot{\mathbf{y}}=(\mathbf{p}\cdot\dot{\mathbf{y}})\nabla_{\mathbf{x}}n$.

The cross product and the scalar triple product have their usual meaning, with $J=\mathbf{y}_1\times\mathbf{y}_2\cdot\mathbf{r}$ the (signed) triple product.

The vector triple-product ("BAC–CAB") identity

$$\mathbf{b}\times(\mathbf{c}\times\mathbf{d})=\mathbf{c}(\mathbf{b}\cdot\mathbf{d})-\mathbf{d}(\mathbf{b}\cdot\mathbf{c})\tag{A7}$$

is used in equations (24) and (38).

The double cross product with the unit ray direction is the shorthand for the projection onto the plane normal to the ray,

$$\mathbf{r}\times\mathbf{p}\times\mathbf{r}\equiv\mathbf{r}\times(\mathbf{p}\times\mathbf{r})=(\mathbf{I}-\mathbf{r}\otimes\mathbf{r})\mathbf{p}=\mathbf{p}-(\mathbf{p}\cdot\mathbf{r})\mathbf{r}\quad(\mathbf{r}\cdot\mathbf{r}=1),\tag{A8}$$

and $\mathbf{I}-\mathbf{r}\otimes\mathbf{r}$ is the corresponding (symmetric, idempotent) normal projector.

Finally, det, $(\ )^{-1}$ and tr denote the determinant, the inverse and the trace of square matrices, respectively.

### A.4 Homogeneity and the Euler (Bliss) identities

A function $F(\mathbf{x},\mathbf{r})$ is homogeneous of degree $m$ in $\mathbf{r}$ if $F(\mathbf{x},k\mathbf{r})=k^mF(\mathbf{x},\mathbf{r})$ for all $k>0$. Euler's theorem states $\nabla_{\mathbf{r}}F\cdot\mathbf{r}=mF$; differentiating it once more with respect to $\mathbf{r}$ and $\mathbf{x}$ generates the identities used in the text. The arclength Lagrangians $L$ and $L^{\omega}=nL$ are homogeneous of degree one in $\mathbf{r}$ (the factor $n$ depends on $\mathbf{x}$ only), whence the Bliss identities (2) and (19),

$$L_{\mathbf{r}}\cdot\mathbf{r}=L,\quad L_{\mathbf{xr}}\cdot\mathbf{r}=\mathbf{r}\cdot L_{\mathbf{rx}}=L_{\mathbf{x}},\qquad L_{\mathbf{rr}}\cdot\mathbf{r}=\mathbf{0},\tag{A9}$$

the last of which makes $L_{\mathbf{rr}}$ singular, with $\mathbf{r}$ the eigenvector of its zero eigenvalue; this degeneracy is the source of the tangential (reparametrization) freedom of the Jacobi equation.

The ray velocity $v_{\text{ray}}(\mathbf{x},\mathbf{r})$ and the polarization $\hat{\mathbf{g}}$ are homogeneous of degree zero in the direction argument, so that

$$\nabla_{\mathbf{r}}v_{\text{ray}}\cdot\mathbf{r}=0,\qquad\nabla_{\mathbf{r}}\nabla_{\mathbf{r}}v_{\text{ray}}\cdot\mathbf{r}=-\nabla_{\mathbf{r}}v_{\text{ray}},\tag{A10}$$

while the Christoffel eigenvalue $\lambda(\mathbf{x},\mathbf{p})$ is homogeneous of degree two in the slowness, $\nabla_{\mathbf{p}}\lambda \cdot \mathbf{p} = 2\lambda$. These homogeneity degrees are preserved by the conformal rescaling $\tilde{\mathbf{C}}^{\omega} = \tilde{\mathbf{C}}/n^2$, since $n$ is independent of both $\mathbf{r}$ and $\mathbf{p}$.

### A.5 Ray-centred (reduced) quantities

Along a resolved trajectory, $\{\mathbf{e}_1,\mathbf{e}_2,\mathbf{r}\}$ is the right-handed orthonormal ray-centred triad with minimal-rotation transport (B1). Greek indices $\alpha,\beta \in \{1,2\}$ label the two normal legs. Projections onto the frame produce reduced two-component vectors and $2\times 2$ matrices, e.g.

$$q_\alpha = \mathbf{e}_\alpha \cdot \mathbf{y}, \qquad \nu_\alpha = \mathbf{e}_\alpha \cdot \nabla_{\mathbf{x}} n, \qquad \tilde{p}_\alpha = \mathbf{e}_\alpha \cdot \mathbf{p}, \qquad X_{\alpha\beta} = \mathbf{e}_\alpha \cdot L_{\mathbf{xx}} \cdot \mathbf{e}_\beta, \tag{A11}$$

and similarly $\mathbf{Y}$, $\mathbf{Z}$, $\mathbf{g}$, $\mathbf{z}$ and their finite-frequency counterparts (B4)–(B5).

Reduced objects obey the same dot, dyadic and transpose rules as their three-dimensional parents, with $\boldsymbol{\nu}\,\tilde{\mathbf{p}}^{\mathrm{T}}$ etc. the $2\times 2$ rank-one dyads. Matrices carrying a caret ($\widehat{\mathbf{Q}}$, $\widehat{\mathbf{P}}$, $\widehat{\mathbf{M}}$) are the tangent-augmented Cartesian $3\times 3$ companions of the ray-centred $2\times 2$ blocks, assembled column-wise from vectors as in (B25), where $[\mathbf{b}\ \mathbf{c}\ \mathbf{d}]$ denotes the matrix with columns $\mathbf{b}$, $\mathbf{c}$, $\mathbf{d}$.

### A.6 Operators, variations and order symbols

Calligraphic symbols denote operators acting on trajectories: $\mathcal{E}[\mathbf{x}] = dL_{\mathbf{r}}/ds - L_{\mathbf{x}}$ and $\mathcal{E}^{\omega}[\mathbf{x}] = dL^{\omega}_{\mathbf{r}}/ds - L^{\omega}_{\mathbf{x}}$ are the classical and finite-frequency Euler–Lagrange operators, and $\mathcal{J}_0[\mathbf{y}]$ is the classical Jacobi operator (the Fréchet derivative of $\mathcal{E}$ at the stationary path).

The symbol $\delta$ denotes a first variation ($\delta\mathbf{x}$, $\delta\tau$, $\delta L$, $\delta\mathbf{p}^{\omega}$) and $\delta^2$ a second variation; square brackets enclose functional arguments. Landau symbols $O(\cdot)$ have their usual asymptotic meaning, here mostly with respect to the small parameters $W$, $|\delta\mathbf{x}|$, $|\mathbf{q}|$ or $\omega^{-2}$.

## Appendix B. Finite-frequency ray-centred matrices $\mathbf{Q}^{\omega}$, $\mathbf{P}^{\omega}$ and the paraxial matrix $\mathbf{M}^{\omega} = \mathbf{P}^{\omega}(\mathbf{Q}^{\omega})^{-1}$

This appendix extends to the finite-frequency (Eigenwave) trajectories the dictionary between the Lagrangian shift-vector formulation and the classical ray-centred paraxial matrices $\mathbf{Q}$, $\mathbf{P}$ and $\mathbf{M} = \mathbf{P}\mathbf{Q}^{-1}$ (Červený, 2000), established for Eigenrays in Koren (2026). Throughout, all quantities are evaluated along the resolved Eigenwave $\mathbf{x}(s)$ of Section 4, and the classical objects ($L$, $\mathbf{p}$, $L_{\mathbf{xx}}$, …) denote the eikonal expressions (3)-(4) evaluated at the points and directions of that trajectory. We write $\boldsymbol{\nu} \equiv \nabla_{\mathbf{x}} n$ for the gradient of the index and $\mathbf{N} \equiv \nabla_{\mathbf{x}}\nabla_{\mathbf{x}} n$ for its Hessian, eq. (16).

### B.1 Ray-centred frame along the Eigenwave

Along the Eigenwave introduce the right-handed orthonormal triad $\{\mathbf{e}_1,\mathbf{e}_2,\mathbf{r}\}$ with minimal-rotation transport of the normal legs,

$$\dot{\mathbf{e}}_\alpha = -(\mathbf{e}_\alpha \cdot \dot{\mathbf{r}})\mathbf{r}, \alpha \in \{1,2\}, \tag{B1}$$

which preserves orthonormality because $\dot{\mathbf{r}} \perp \mathbf{r}$. The normal shift of Section 5 is expanded as $\mathbf{y} = q_1\mathbf{e}_1 + q_2\mathbf{e}_2$, $\mathbf{q} = (q_1, q_2)^\mathsf{T}$, and

$$\dot{\mathbf{y}} = \sum_{\alpha=1}^{2} \dot{q}_\alpha \, \mathbf{e}_\alpha - (\mathbf{q} \cdot \mathbf{z})\mathbf{r}, z_\alpha = \mathbf{e}_\alpha \cdot \dot{\mathbf{r}}. \tag{B2}$$

Note that $\dot{\mathbf{r}}$, and hence $\mathbf{z}$, is the curvature of the *Eigenwave*, which by (24) contains the finite-frequency bending $\mathbf{r} \times (\boldsymbol{\nu} \times \mathbf{p})/n$ in addition to the eikonal curvature.

**B.2 Conjugate momentum: the finite-frequency slowness perturbation**

Since $L^\omega_{\mathbf{r}} = \mathbf{p}^\omega = n\mathbf{p}$, the first-order perturbation of the finite-frequency slowness along a paraxial Eigenwave is the argument of the total derivative in (29),

$$\delta\mathbf{p}^\omega = L^\omega_{\mathbf{rx}} \cdot \mathbf{y} + L^\omega_{\mathbf{rr}} \cdot \dot{\mathbf{y}} = n(L_{\mathbf{rx}} \cdot \mathbf{y} + L_{\mathbf{rr}} \cdot \dot{\mathbf{y}}) + (\nu \cdot \mathbf{y})\mathbf{p}, \tag{B3}$$

by (28). The finite-frequency Euler identities (19), $L^\omega_{\mathbf{rr}} \cdot \mathbf{r} = \mathbf{0}$ and $L^\omega_{\mathbf{xr}} \cdot \mathbf{r} = L^\omega_{\mathbf{x}}$, hold as in the eikonal case. Define the reduced $2 \times 2$ blocks and 2-vectors ($\alpha, \beta \in \{1,2\}$)

$$X^\omega_{\alpha\beta} = \mathbf{e}_\alpha \cdot L^\omega_{\mathbf{xx}} \cdot \mathbf{e}_\beta, Y^\omega_{\alpha\beta} = \mathbf{e}_\alpha \cdot L^\omega_{\mathbf{xr}} \cdot \mathbf{e}_\beta, Z^\omega_{\alpha\beta} = \mathbf{e}_\alpha \cdot L^\omega_{\mathbf{rr}} \cdot \mathbf{e}_\beta, g^\omega_\alpha = \mathbf{e}_\alpha \cdot L^\omega_{\mathbf{x}}, \tag{B4}$$

and their eikonal counterparts $\mathbf{X}, \mathbf{Y}, \mathbf{Z}, \mathbf{g}$ (same expressions with $L$), together with the transverse index gradient and transverse slowness

$$\nu_\alpha = \mathbf{e}_\alpha \cdot \nabla_{\mathbf{x}} n, \tilde{p}_\alpha = \mathbf{e}_\alpha \cdot \mathbf{p}, N_{\alpha\beta} = \mathbf{e}_\alpha \cdot \nabla_{\mathbf{x}}\nabla_{\mathbf{x}} n \cdot \mathbf{e}_\beta. \tag{B5}$$

Projecting (28) and (20) on the frame gives the explicit finite-frequency blocks

$$\mathbf{Z}^\omega = n\mathbf{Z}, \mathbf{Y}^\omega = n\mathbf{Y} + \boldsymbol{\nu}\tilde{\mathbf{p}}^\mathsf{T}, \mathbf{X}^\omega = n\mathbf{X} + \boldsymbol{\nu}\mathbf{g}^\mathsf{T} + \mathbf{g}\boldsymbol{\nu}^\mathsf{T} + L\mathbf{N}, \mathbf{g}^\omega = n\mathbf{g} + L\boldsymbol{\nu}. \tag{B6}$$

Projecting (B3) on $\mathbf{e}_\alpha$, using (B2) and $L^\omega_{\mathbf{rr}}\mathbf{r} = \mathbf{0}$, yields the reduced finite-frequency momentum

$$\pi^\omega_\alpha = \mathbf{e}_\alpha \cdot \delta\mathbf{p}^\omega, \boldsymbol{\pi}^\omega = (\mathbf{Y}^\omega)^\mathsf{T}\mathbf{q} + \mathbf{Z}^\omega\dot{\mathbf{q}} = n(\mathbf{Y}^\mathsf{T}\mathbf{q} + \mathbf{Z}\dot{\mathbf{q}}) + (\boldsymbol{\nu} \cdot \mathbf{q})\tilde{\mathbf{p}}. \tag{B7}$$

The finite-frequency reduced momentum is thus $n$ times the eikonal form, plus a term proportional to the transverse slowness $\tilde{\mathbf{p}}$ and to the component of the shift along the transverse index gradient; in isotropic media $\tilde{\mathbf{p}} = \mathbf{0}$ and the last term disappears.

### B.3 Reduced accessory Lagrangian and the linear Hamiltonian system

Substituting (B2) into the second variation of the phase-time functional (Appendix C) and using (19) to eliminate the tangential terms, the finite-frequency accessory Lagrangian is

$$\Omega^{\omega}(\mathbf{q},\dot{\mathbf{q}}) = 1/2\,\mathbf{q}\cdot\widetilde{\mathbf{X}}^{\omega}\mathbf{q} + \mathbf{q}\cdot\mathbf{Y}^{\omega}\dot{\mathbf{q}} + 1/2\,\dot{\mathbf{q}}\cdot\mathbf{Z}^{\omega}\dot{\mathbf{q}},\widetilde{\mathbf{X}}^{\omega} = \mathbf{X}^{\omega} - \mathbf{z}(\mathbf{g}^{\omega})^{\mathsf{T}} - \mathbf{g}^{\omega}\mathbf{z}^{\mathsf{T}}, \tag{B8}$$

with $\boldsymbol{\pi}^{\omega} = \partial\Omega^{\omega}/\,\partial\dot{\mathbf{q}}$ recovering (B7). Solving (B7) for $\dot{\mathbf{q}}$ and substituting into the Euler-Lagrange equation of (B8) gives the linear Hamiltonian system

$$\dot{\mathbf{q}} = \mathbf{A}^{\omega}\mathbf{q} + \mathbf{B}^{\omega}\boldsymbol{\pi}^{\omega},\quad \dot{\boldsymbol{\pi}}^{\omega} = \mathbf{C}^{\omega}\mathbf{q} - (\mathbf{A}^{\omega})^{\mathsf{T}}\boldsymbol{\pi}^{\omega}, \tag{B9}$$

$$\mathbf{A}^{\omega} = -(\mathbf{Z}^{\omega})^{-1}(\mathbf{Y}^{\omega})^{\mathsf{T}},\quad \mathbf{B}^{\omega} = (\mathbf{Z}^{\omega})^{-1},\quad \mathbf{C}^{\omega} = \widetilde{\mathbf{X}}^{\omega} - \mathbf{Y}^{\omega}(\mathbf{Z}^{\omega})^{-1}(\mathbf{Y}^{\omega})^{\mathsf{T}}. \tag{B10}$$

Inserting (B6), and denoting by $\mathbf{A},\mathbf{B},\mathbf{C}$ the eikonal coefficient matrices (same formulae with $\mathbf{X},\mathbf{Y},\mathbf{Z},\mathbf{g}$), the finite-frequency coefficients read explicitly

$$\begin{aligned} \mathbf{A}^{\omega} &= \mathbf{A} - \frac{1}{n}(\mathbf{Z}^{-1}\widetilde{\mathbf{p}})\boldsymbol{\nu}^{\mathsf{T}},\qquad \mathbf{B}^{\omega} = \frac{1}{n}\mathbf{B},\\ \mathbf{C}^{\omega} &= n\mathbf{C} + \boldsymbol{\nu}\mathbf{d}^{\mathsf{T}} + \mathbf{d}\boldsymbol{\nu}^{\mathsf{T}} + L\mathbf{N} - \frac{\widetilde{\mathbf{p}}\cdot\mathbf{Z}^{-1}\widetilde{\mathbf{p}}}{n}\boldsymbol{\nu}\boldsymbol{\nu}^{\mathsf{T}}, \mathbf{d} \equiv \mathbf{g} - \mathbf{Y}\mathbf{Z}^{-1}\widetilde{\mathbf{p}} - L\mathbf{z}, \end{aligned} \tag{B11}$$

where $\mathbf{B}^{\omega}$ and $\mathbf{C}^{\omega}$ are symmetric. The finite-frequency corrections to the dynamic ray tracing coefficients are therefore of three types: a uniform rescaling ($n\mathbf{C}$, $\mathbf{B}/n$), rank-one and rank-two terms driven by the transverse index gradient $\boldsymbol{\nu}$ (which vanish where $W$ is transversely constant, and are further reduced in isotropic media where $\widetilde{\mathbf{p}} = \mathbf{0}$), and the transverse curvature $L\,\mathbf{N}$ of the index, which is the finite-frequency analogue of the velocity-Hessian term in $\mathbf{C}$: focusing/defocusing of the finite-frequency ray tube by the transverse curvature of $W$. All are $O(\omega^{-2})$ when $W = \kappa/\omega^{2}$.

### B.4 The matrices $\mathbf{Q}^{\omega}$, $\mathbf{P}^{\omega}$, initial conditions and the symplectic invariant

Stack the ray-centred coordinates and momenta of the two point-source basic solutions of (29) column-wise,

$$\mathbf{Q}^{\omega}(s) = \left[\mathbf{q}^{(1)}\mathbf{q}^{(2)}\right],\qquad \mathbf{P}^{\omega}(s) = \left[\boldsymbol{\pi}^{\omega(1)}\boldsymbol{\pi}^{\omega(2)}\right], \tag{B12}$$

and analogously the plane-wave pair. Both pairs satisfy the matrix form of (B9),

$$\dot{\mathbf{Q}}^{\omega} = \mathbf{A}^{\omega}\mathbf{Q}^{\omega} + \mathbf{B}^{\omega}\mathbf{P}^{\omega},\qquad \dot{\mathbf{P}}^{\omega} = \mathbf{C}^{\omega}\mathbf{Q}^{\omega} - (\mathbf{A}^{\omega})^{\mathsf{T}}\mathbf{P}^{\omega}. \tag{B13}$$

The initial conditions (32) translate directly. For the point-source pair $\mathbf{Q}^\omega(s_S) = \mathbf{O}$, and since $\mathbf{q} = \mathbf{0}$ at $S$ the extra term of (B7) vanishes there, so that $\mathbf{P}^\omega(s_S) = \mathbf{Z}_S^\omega \dot{\mathbf{Q}}(s_S) = n_S \mathbf{E}_S \mathbf{\Lambda}_S$, with $\mathbf{E}_S$ the unit eigenvectors of $\mathbf{Z}_S$ (unit Wronskian) and $\mathbf{\Lambda}_S = \mathrm{diag}(\lambda_{1,S}, \lambda_{2,S})$; hence

$$\det\mathbf{P}^\omega(s_S) = n_S^2 \lambda_{1,S}\lambda_{2,S} = \lambda_{1,S}^\omega \lambda_{2,S}^\omega, \tag{B14}$$

consistent with $L_{\mathbf{rr}}^\omega = nL_{\mathbf{rr}}$. For the plane-wave pair, $\mathbf{Q}^\omega(s_S) = \mathbf{I}$ up to the eigenvector rotation and, the initial shift derivatives being purely tangential, $\mathbf{P}^\omega(s_S) = (\mathbf{Y}_S^\omega)^\mathsf{T}\mathbf{Q}^\omega(s_S) = \left(n_S \mathbf{Y}_S^\mathsf{T} + \widetilde{\mathbf{p}}_S \mathbf{v}_S^\mathsf{T}\right)\mathbf{Q}^\omega(s_S)$: unlike the point-source pair, the plane-wave initial momentum *does* feel the transverse index gradient at the source. Any two solution pairs of (B13) conserve the symplectic invariant; for the point-source pair

$$(\mathbf{Q}^\omega)^\mathsf{T}\mathbf{P}^\omega - (\mathbf{P}^\omega)^\mathsf{T}\mathbf{Q}^\omega = \mathbf{O}. \tag{B15}$$

### B.5 The matrix $\mathbf{M}^\omega$: Riccati equation, symmetry and the paraxial phase time

Away from finite-frequency caustics define

$$\mathbf{M}^\omega(s) = \mathbf{P}^\omega(s)(\mathbf{Q}^\omega(s))^{-1}, \tag{B16}$$

which satisfies the Riccati equation with the finite-frequency coefficients (B11),

$$\dot{\mathbf{M}}^\omega = \mathbf{C}^\omega - (\mathbf{A}^\omega)^\mathsf{T}\mathbf{M}^\omega - \mathbf{M}^\omega\mathbf{A}^\omega - \mathbf{M}^\omega\mathbf{B}^\omega\mathbf{M}^\omega, \tag{B17}$$

and is symmetric by (B15). Since $\Omega^\omega$ is homogeneous of degree two, $2\Omega^\omega = \mathrm{d}(\mathbf{q}\cdot\boldsymbol{\pi}^\omega)/\mathrm{d}s$ along solutions, and the second variation of the phase time between the source and a receiver displaced by $\mathbf{q}$ in the normal plane is $1/2\ \mathbf{q}\cdot\mathbf{M}^\omega\mathbf{q}$. Adding the first-variation boundary term $\mathbf{p}^\omega\cdot\mathbf{y}(R)$ gives the paraxial expansion of the finite-frequency phase time in ray-centred transverse coordinates,

$$\tau(\mathbf{q};\omega) = \tau_0(\omega) + n\widetilde{\mathbf{p}}\cdot\mathbf{q} + 1/2\ \mathbf{q}\cdot\mathbf{M}^\omega\mathbf{q} + O(|\mathbf{q}|^3), \qquad \delta\mathbf{p}_\perp^\omega = \mathbf{M}^\omega\mathbf{q}. \tag{B18}$$

The linear term is the transverse component of the finite-frequency slowness $\mathbf{p}^\omega = n\mathbf{p}$ (it vanishes in isotropic media), and $\mathbf{M}^\omega$ is the transverse Hessian of the finite-frequency phase time: the finite-frequency wavefront-curvature matrix. In isotropic media the wavefront curvature is $\mathbf{K}^\omega = v_{\mathrm{ray}}^\omega\mathbf{M}^\omega = (v/n)\mathbf{M}^\omega$. A useful pointwise reading of (B7) is obtained by writing $\mathbf{P}^\omega = n\mathbf{P}_* + \widetilde{\mathbf{p}}\mathbf{v}^\mathsf{T}\mathbf{Q}^\omega$, where $\mathbf{P}_*$ collects the eikonal-form momenta $\mathbf{Y}^\mathsf{T}\mathbf{q} + \mathbf{Z}\dot{\mathbf{q}}$ of the same (Eigenwave) shifts; then

$$\mathbf{M}^\omega = n\mathbf{M}_* + \widetilde{\mathbf{p}}\mathbf{v}^\mathsf{T}, \qquad \mathbf{M}_* = \mathbf{P}_*(\mathbf{Q}^\omega)^{-1}, \tag{B19}$$

so that the finite-frequency curvature matrix is the eikonal-form matrix of the Eigenwave tube, scaled by the index, plus a rank-one anisotropic term coupling the transverse slowness to the transverse index gradient (since $\mathbf{M}^{\omega}$ is symmetric, $\mathbf{M}_*$ is not, its skew part being $(\boldsymbol{\nu}\tilde{\mathbf{p}}^{\mathsf{T}} - \tilde{\mathbf{p}}\boldsymbol{\nu}^{\mathsf{T}})/n$).

**B.6 Ray Jacobian, geometric spreading and canonical blocks**

Expanding the point-source shifts in the frame, the triple product (33) is the determinant of the ray-centred matrix,

$$J^{\omega}(s) = \mathbf{y}_1 \times \mathbf{y}_2 \cdot \mathbf{r} = \det\mathbf{Q}^{\omega}(s), \tag{B20}$$

so finite-frequency caustics are the zeros of $\det\mathbf{Q}^{\omega}$, where $\mathbf{M}^{\omega}$ becomes singular and the KMAH index is incremented. Combining (B14) and (B20) with (34), the relative geometric spreading takes the propagator form

$$L_{\mathrm{GS}}^{\omega}(s) = \sqrt{\frac{v_{\mathrm{ray},S}}{v_{\mathrm{phs},S}} \frac{v_{\mathrm{ray}}(s)}{v_{\mathrm{phs}}(s)} \frac{|\det\mathbf{Q}^{\omega}(s)|}{\det\mathbf{P}^{\omega}(s_S)}}, \tag{B21}$$

formally identical to the eikonal expression, the index having been absorbed in $\det\mathbf{P}^{\omega}(s_S)$ (recall that $v_{\mathrm{ray}}/v_{\mathrm{phs}}$ is frequency independent). Defining the canonical plane-wave and point-source blocks by identity/zero initial conditions,

$$\begin{aligned} \mathbf{Q}_1^{\omega}(s_S) &= \mathbf{I}, \quad \mathbf{P}_1^{\omega}(s_S) = \mathbf{O}(\text{plane wave}); \\ \mathbf{Q}_2^{\omega}(s_S) &= \mathbf{O}, \quad \mathbf{P}_2^{\omega}(s_S) = \mathbf{I}(\text{point source}), \end{aligned} \tag{B22}$$

they assemble the symplectic $4 \times 4$ finite-frequency propagator $\boldsymbol{\Pi}^{\omega}(s, s_S) = \begin{bmatrix} \mathbf{Q}_1^{\omega} & \mathbf{Q}_2^{\omega} \\ \mathbf{P}_1^{\omega} & \mathbf{P}_2^{\omega} \end{bmatrix}$ with the block identities $(\mathbf{Q}_1^{\omega})^{\mathsf{T}}\mathbf{P}_2^{\omega} - (\mathbf{P}_1^{\omega})^{\mathsf{T}}\mathbf{Q}_2^{\omega} = \mathbf{I}$, etc. The eigen-normalized point-source pair is $\mathbf{Q}^{\omega} = \mathbf{Q}_2^{\omega}\, n_S \mathbf{E}_S \boldsymbol{\Lambda}_S$, so $J^{\omega} = n_S^2 \lambda_{1,S}\lambda_{2,S}\det\mathbf{Q}_2^{\omega}$ and

$$\begin{aligned} L_{\mathrm{GS}}^{\omega}(s) &= \sqrt{\frac{v_{\mathrm{ray},S}}{v_{\mathrm{phs},S}} \frac{v_{\mathrm{ray}}(s)}{v_{\mathrm{phs}}(s)} |\det\mathbf{Q}_2^{\omega}(s)|}, \\ \mathbf{M}^{\omega}(s) &= [\mathbf{P}_1^{\omega} + \mathbf{P}_2^{\omega}\mathbf{M}_S^{\omega}][\mathbf{Q}_1^{\omega} + \mathbf{Q}_2^{\omega}\mathbf{M}_S^{\omega}]^{-1}, \end{aligned} \tag{B23}$$

the second relation being the finite-frequency $\mathbf{M}$-continuation for a paraxial field launched with an arbitrary symmetric $\mathbf{M}_S^{\omega}$; it is the natural ingredient of finite-frequency (trajectory-mechanics-based) Gaussian beams, in which the beam is attached to an Eigenwave rather than to an Eigenray and its complex curvature $\mathbf{M}_S^{\omega}$ is continued with (B23).

### B.7 The Cartesian 3 × 3 matrices

Linearizing the arclength Hamilton system of $H^\omega$ (Appendix D) gives the Cartesian paraxial system (36). The linearized flow conserves $\delta H^\omega = H^\omega_{\mathbf{x}} \cdot \delta\mathbf{x} + H^\omega_{\mathbf{p}} \cdot \delta\mathbf{p}$; physical, eikonal-preserving paraxial Eigenwaves satisfy $\delta H^\omega = 0$ and, since $H^\omega_{\mathbf{p}} = \mathbf{r}$, their tangential slowness perturbation is fixed by

$$\mathbf{r} \cdot \delta\mathbf{p}^\omega = -H^\omega_{\mathbf{x}} \cdot \delta\mathbf{x} = L^\omega_{\mathbf{x}} \cdot \delta\mathbf{x} = \left(-n\frac{\nabla_{\mathbf{x}} v_{\mathrm{ray}}}{v_{\mathrm{ray}}^2} + \frac{\nabla_{\mathbf{x}} W}{2n v_{\mathrm{ray}}}\right) \cdot \delta\mathbf{x}, \tag{B24}$$

where (20) was used; the finite-frequency force appears explicitly. The ray-tangent pair $(\delta\mathbf{x}, \delta\mathbf{p}^\omega) = (\mathbf{r}, -H^\omega_{\mathbf{x}}) = (\mathbf{r}, \mathrm{d}\mathbf{p}^\omega/\mathrm{d}s)$ is an exact solution, and the tangent-augmented point-source matrices are

$$\widehat{\mathbf{Q}}^\omega = [\mathbf{y}_1 \mathbf{y}_2 \mathbf{r}], \widehat{\mathbf{P}}^\omega = \left[\delta\mathbf{p}^\omega_1 \delta\mathbf{p}^\omega_2 \dot{\mathbf{p}^\omega}\right], \qquad \det\widehat{\mathbf{Q}}^\omega = J^\omega, \tag{B25}$$

with the dictionary to the ray-centred blocks, $\widehat{\mathbf{E}} = [\mathbf{e}_1\ \mathbf{e}_2]$, $\mathbf{h}^\omega = \widehat{\mathbf{E}}^{\mathsf{T}} H^\omega_{\mathbf{x}} = -\mathbf{g}^\omega$,

$$\widehat{\mathbf{Q}}^\omega = \left[\widehat{\mathbf{E}}\mathbf{Q}^\omega \mathbf{r}\right], \widehat{\mathbf{P}}^\omega = \left[\widehat{\mathbf{E}}\mathbf{P}^\omega - \mathbf{r}(\mathbf{h}^\omega)^{\mathsf{T}}\mathbf{Q}^\omega \dot{\mathbf{p}^\omega}\right]. \tag{B26}$$

The Cartesian matrix $\widehat{\mathbf{M}}^\omega = \widehat{\mathbf{P}}^\omega\left(\widehat{\mathbf{Q}}^\omega\right)^{-1}$ is the full second-derivative matrix of the point-source finite-frequency phase-time field $\tau(\mathbf{x};\omega)$, whose gradient is the finite-frequency slowness $\nabla_{\mathbf{x}}\tau = \mathbf{p}^\omega = n\mathbf{p}$:

$$\begin{gathered}\tau(\mathbf{x};\omega) = \tau_0 + \mathbf{p}^\omega \cdot \delta\mathbf{x} + 1/2\, \delta\mathbf{x} \cdot \widehat{\mathbf{M}}^\omega \cdot \delta\mathbf{x} + O(|\delta\mathbf{x}|^3), \\ \widehat{\mathbf{M}}^\omega \mathbf{y}_i = \delta\mathbf{p}^\omega_i, \widehat{\mathbf{M}}^\omega \mathbf{r} = \dot{\mathbf{p}^\omega}.\end{gathered} \tag{B27}$$

It is symmetric, satisfies the Cartesian Riccati equation built from the blocks $H^\omega_{\mathbf{xx}}, H^\omega_{\mathbf{xp}}, H^\omega_{\mathbf{pp}}$ (obtainable from (28) through the inverse Hessian relations of Ravve and Koren, 2021, eq. 38, with $L \to L^\omega$), and inherits its singularities from the zeros of $J^\omega$. In the weak-coupling regime the finite-frequency correction to $\widehat{\mathbf{M}}$ follows by linearizing the Riccati equation around the eikonal solution with the $O(W)$ parts of (B11) as sources; alternatively $\widehat{\mathbf{M}}^\omega$ is read directly from the finite-element Hessian of the phase-time functional condensed to the endpoints, exactly as in Ravve and Koren (2020a).

### B.8 Role in the amplitude closure

The finite-frequency propagator provides the leading amplitude of the Eigenwave family, $a \propto \left(\rho\, n\, v_{\mathrm{ray}}\, |\det\mathbf{Q}^\omega|\right)^{-1/2}$ (eq. (40)), which is the input to $W$ through (9). The transverse second derivatives of $a$ required by $W$ are second-order paraxial quantities - derivatives of $\mathbf{Q}^\omega$ with respect to the ray coordinates - and are not contained in the first-order propagator; they must be evaluated either by finite differences across a fan of neighbouring Eigenwaves or by carrying the second-order paraxial system, and, near caustics of the current iterate, from a regularized (Gaussian-beam or phase-space) superposition built with (B23). This closes the loop between the finite-frequency $\mathbf{Q}^{\omega}, \mathbf{P}^{\omega}$ and the potential $W$ that shapes them.

## Appendix C. Hessians of the Eigenwave Lagrangian and the Jacobi equation

With $L^{\omega} = n(\mathbf{x})L(\mathbf{x}, \mathbf{r})$ (we suppress $\omega$), the chain rule gives $L^{\omega}_{\mathbf{x}} = nL_{\mathbf{x}} + L\nabla_{\mathbf{x}}n$ and $L^{\omega}_{\mathbf{r}} = nL_{\mathbf{r}}$. Differentiating again,

$$\begin{aligned} L^{\omega}_{\mathbf{xx}} &= nL_{\mathbf{xx}} + \nabla_{\mathbf{x}}n \otimes L_{\mathbf{x}} + L_{\mathbf{x}} \otimes \nabla_{\mathbf{x}}n + L\,\nabla_{\mathbf{x}}\nabla_{\mathbf{x}}n, \\ L^{\omega}_{\mathbf{xr}} &= (L^{\omega}_{\mathbf{rx}})^{\mathsf{T}} = nL_{\mathbf{xr}} + \nabla_{\mathbf{x}}n \otimes L_{\mathbf{r}}, \\ L^{\omega}_{\mathbf{rr}} &= nL_{\mathbf{rr}}, \end{aligned} \tag{C1}$$

which, with $L_{\mathbf{r}} = \mathbf{p}$, is (28). The second variation of the phase-time functional (18) for a shift $\mathbf{y}(s)$ with $\mathbf{y}(S) = \mathbf{y}(R) = \mathbf{0}$ is

$$\delta^2\tau = \int_S^R (\mathbf{y} \cdot L^{\omega}_{\mathbf{xx}}\mathbf{y} + 2\,\mathbf{y} \cdot L^{\omega}_{\mathbf{xr}}\dot{\mathbf{y}} + \dot{\mathbf{y}} \cdot L^{\omega}_{\mathbf{rr}}\dot{\mathbf{y}})\,\mathrm{d}s, \tag{C2}$$

and its Euler-Lagrange equation with respect to $\mathbf{y}$ is (29) . The identities (19) follow from Euler's theorem for the first-degree homogeneous function $L^{\omega}(\mathbf{x}, k\mathbf{r}) = kL^{\omega}(\mathbf{x}, \mathbf{r})$; e.g. differentiating $L^{\omega}_{\mathbf{r}} \cdot \mathbf{r} = L^{\omega}$ with respect to $\mathbf{r}$ gives $L^{\omega}_{\mathbf{rr}}\mathbf{r} + L^{\omega}_{\mathbf{r}} = L^{\omega}_{\mathbf{r}}$, i.e. $L^{\omega}_{\mathbf{rr}}\mathbf{r} = \mathbf{0}$. Because of these identities, the contraction of (29) with $\mathbf{r}$ is an identity, and $\rho_t(s)\mathbf{r}(s)$ solves (29) for any $\rho_t$ (Bliss, 1916).

The expanded form (30) follows from (28) using $(\mathbf{p} \otimes \nabla_{\mathbf{x}}n)\mathbf{y} = (\nabla_{\mathbf{x}}n \cdot \mathbf{y})\mathbf{p}$, $(\nabla_{\mathbf{x}}n \otimes \mathbf{p})\dot{\mathbf{y}} = (\mathbf{p} \cdot \dot{\mathbf{y}})\nabla_{\mathbf{x}}n$, $(\nabla_{\mathbf{x}}n \otimes L_{\mathbf{x}})\mathbf{y} = (L_{\mathbf{x}} \cdot \mathbf{y})\nabla_{\mathbf{x}}n$ and $(L_{\mathbf{x}} \otimes \nabla_{\mathbf{x}}n)\mathbf{y} = (\nabla_{\mathbf{x}}n \cdot \mathbf{y})L_{\mathbf{x}}$.

## Appendix D. Hamiltonian form and consistency with the trajectory equations

*9.0.0.1 Consistency of (22) with (11).*

Along a trajectory of (11), $\mathrm{d}\mathbf{x}/\mathrm{d}T = 1/2\,\nabla_{\mathbf{p}}\lambda$ and, on $\lambda = n^2$, $\mathbf{p} \cdot \mathrm{d}\mathbf{x}/\mathrm{d}T = n^2$; hence $\mathbf{r} = \mathrm{d}\mathbf{x}/\mathrm{d}s$ is the direction of $\nabla_{\mathbf{p}}\lambda$, $\mathrm{d}s/\mathrm{d}T = |1/2\,\nabla_{\mathbf{p}}\lambda| = nv_{\mathrm{ray}}$ (since $v^{\omega}_{\mathrm{ray}} = |\mathrm{d}\mathbf{x}/\mathrm{d}\tau|$ with $\mathrm{d}\tau = n^2\mathrm{d}T$, so $|\mathrm{d}\mathbf{x}/\mathrm{d}T| = n^2 v^{\omega}_{\mathrm{ray}} = nv_{\mathrm{ray}}$), and the slowness $\mathbf{p}$ of (11) is the finite-frequency slowness $\mathbf{p}^{\omega}$ of this paper. The second equation of (11) then reads

$$\frac{\mathrm{d}\mathbf{p}^{\omega}}{\mathrm{d}s} = \frac{1}{nv_{\mathrm{ray}}}[-\,1/2\,\nabla_{\mathbf{x}}\lambda(\mathbf{x}, \mathbf{p}^{\omega}) + 1/2\,\nabla_{\mathbf{x}}W]. \tag{D1}$$

On the other hand, in the effective medium, the classical relation $\mathrm{d}\mathbf{p}^{\omega}/\mathrm{d}s = -\nabla_{\mathbf{x}}v^{\omega}_{\mathrm{ray}}/\left(v^{\omega}_{\mathrm{ray}}\right)^2$ holds (eq. 8 Koren and Ravve, 2021), and for a Christoffel-type Hamiltonian $\lambda_{\mathrm{eff}} = \lambda/n^2$ the classical identity $\mathrm{d}\mathbf{p}/\mathrm{d}\tau = -\,1/2\,\nabla_{\mathbf{x}}\lambda_{\mathrm{eff}}$ (flow parameter $\tau$, $\mathrm{d}s = v^{\omega}_{\mathrm{ray}}\mathrm{d}\tau$) gives $\mathrm{d}\mathbf{p}^{\omega}/\mathrm{d}s =$ $-\nabla_{\mathbf{x}}\lambda_{\text{eff}}/(2v^{\omega}_{\text{ray}}) = -[\nabla_{\mathbf{x}}\lambda - \lambda\nabla_{\mathbf{x}}n^2/n^2]/(2n^2 v^{\omega}_{\text{ray}}) = [-\nabla_{\mathbf{x}}\lambda + \nabla_{\mathbf{x}}W]/(2nv_{\text{ray}})$, using $\lambda = n^2$ and $n^2 v^{\omega}_{\text{ray}} = nv_{\text{ray}}$. The two expressions coincide, which proves the equivalence of (22) and (11) in the parametrization $\mathrm{d}s = nv_{\text{ray}}\,\mathrm{d}T$.

*9.0.0.2 Hamiltonian dynamic ray tracing.*

Differentiating the arclength Hamilton equations $\dot{\mathbf{x}} = H^{\omega}_{\mathbf{p}}$, $\dot{\mathbf{p}} = -H^{\omega}_{\mathbf{x}}$ with respect to a ray coordinate $\gamma$ yields (36); the constraint is the derivative of $H^{\omega} = 0$. The Legendre transformation $L^{\omega}(\mathbf{x},\mathbf{r}) = \mathbf{p}\cdot\mathbf{r} - H^{\omega}(\mathbf{x},\mathbf{p})$ with $\mathbf{r} = H^{\omega}_{\mathbf{p}}$ implies $L^{\omega}_{\mathbf{r}} = \mathbf{p}$, $L^{\omega}_{\mathbf{x}} = -H^{\omega}_{\mathbf{x}}$; differentiating $\mathbf{r} = H^{\omega}_{\mathbf{p}}(\mathbf{x}, L^{\omega}_{\mathbf{r}}(\mathbf{x},\mathbf{r}))$ with respect to $\mathbf{r}$ and $\mathbf{x}$ gives $\mathbf{I} = H^{\omega}_{\mathbf{pp}}L^{\omega}_{\mathbf{rr}} + \lambda_{\mathbf{r}}\mathbf{r}\otimes\mathbf{r}$ (the projector on the normal plane arises because $L^{\omega}_{\mathbf{rr}}\mathbf{r} = 0$) and $\mathbf{0} = H^{\omega}_{\mathbf{px}} + H^{\omega}_{\mathbf{pp}}L^{\omega}_{\mathbf{rx}}$, from which (37) follows exactly as in Ravve and Koren (2020b, Appendices H-I); the derivation uses only the homogeneity of $L^{\omega}$ and is therefore identical for the Eigenwave.

## Appendix E. Linearization of the finite-frequency Euler-Lagrange operator

Write $L^{\omega} = nL$ with $n = 1 + 1/2\,W + O(W^2)$. Then $\mathcal{E}^{\omega}[\mathbf{x}] = \mathcal{E}[\mathbf{x}] + \mathcal{E}_1[\mathbf{x}] + O(W^2)$ with

$$\begin{aligned}\mathcal{E}_1[\mathbf{x}] &= \frac{\mathrm{d}}{\mathrm{d}s}(1/2\,WL_{\mathbf{r}}) - 1/2\,WL_{\mathbf{x}} - L\ 1/2\,\nabla_{\mathbf{x}}W \\ &= 1/2\,(\nabla_{\mathbf{x}}W\cdot\mathbf{r})\mathbf{p} + 1/2\,W(\dot{\mathbf{p}} - L_{\mathbf{x}}) - \frac{\nabla_{\mathbf{x}}W}{2v_{\text{ray}}},\end{aligned} \tag{E1}$$

where $L_{\mathbf{r}} = \mathbf{p}$, $\dot{W} = \nabla_{\mathbf{x}}W\cdot\mathbf{r}$ and $L = 1/v_{\text{ray}}$ were used. Evaluate at $\mathbf{x} = \mathbf{x}_0 + \delta\mathbf{x}$: $\mathcal{E}[\mathbf{x}_0 + \delta\mathbf{x}] = \mathcal{E}[\mathbf{x}_0] + \mathcal{J}_0[\delta\mathbf{x}] + O(\delta\mathbf{x}^2)$ with $\mathcal{E}[\mathbf{x}_0] = \mathbf{0}$ and $\mathcal{J}_0[\mathbf{y}] = \mathrm{d}(L_{\mathbf{rx}}\mathbf{y} + L_{\mathbf{rr}}\dot{\mathbf{y}})/\mathrm{d}s - L_{\mathbf{xx}}\mathbf{y} - L_{\mathbf{xr}}\dot{\mathbf{y}}$ the classical Jacobi operator (the Fréchet derivative of the Euler-Lagrange operator), while $\mathcal{E}_1[\mathbf{x}_0 + \delta\mathbf{x}] = \mathcal{E}_1[\mathbf{x}_0] + O(W\delta\mathbf{x})$ and, along $\mathbf{x}_0$, $\dot{\mathbf{p}}_0 = L_{\mathbf{x}}$ so that the middle term of $\mathcal{E}_1$ vanishes. Collecting the first-order terms of $\mathcal{E}^{\omega}[\mathbf{x}] = \mathbf{0}$:

$$\mathcal{J}_0[\delta\mathbf{x}] = \frac{\nabla_{\mathbf{x}}W}{2v_{\text{ray}}} - 1/2\,(\nabla_{\mathbf{x}}W\cdot\mathbf{r}_0)\mathbf{p}_0, \tag{E2}$$

which is (38). Its tangential component vanishes identically ($\mathbf{p}_0\cdot\mathbf{r}_0 = 1/v_{\text{ray}}$), so the two independent normal components determine $\delta\mathbf{x}\perp\mathbf{r}_0$ under the two-point conditions $\delta\mathbf{x}(S) = \delta\mathbf{x}(R) = \mathbf{0}$; the tangential part of $\delta\mathbf{x}$ is a reparametrization and is set to zero. The phase-time correction (39) follows from $\tau = \int nL\,\mathrm{d}s = \int L\,\mathrm{d}s + 1/2\int WL\,\mathrm{d}s + O(W^2)$ and the stationarity of $\int L\,\mathrm{d}s$ at $\mathbf{x}_0$, which removes the $O(\delta\mathbf{x})$ contribution of the first integral; the $O(W\delta\mathbf{x})$ contribution of the second is of second order.

## Appendix F. Illustration: horizontally layered potential in a homogeneous background

Consider a homogeneous isotropic background of velocity $v$ (or, more generally, a homogeneous VTI medium) and a potential depending on depth only, $W = W(z, \omega)$, as in the gradational-layer example of Vasco and Nihei (2019a). The Eigenwave Lagrangian $L^{\omega} = n(z)\sqrt{\mathbf{r}\cdot\mathbf{r}}/v_{\mathrm{ray}}(\mathbf{r})$ does not depend on the horizontal coordinates, hence the horizontal components of $L^{\omega}_{\mathbf{r}} = \mathbf{p}^{\omega}$ are conserved along the Eigenwave: $p^{\omega}_h = n(z)\, p_h(\mathbf{r}) = \text{const}$. In the isotropic case, $p_h = \sin\theta/v$ and

$$\frac{n(z,\omega)\sin\theta(z)}{v} = \text{const}, \tag{F1}$$

a finite-frequency Snell law: the trajectory refracts through the layer as if the velocity were $v/n(z,\omega)$, i.e. it bends *towards* the vertical where $W > 0$ and away from it where $W < 0$, and the effect scales as $\omega^{-2}$. The finite-frequency Jacobi equation (29) in this case has $\nabla_{\mathbf{x}} n = (0,0,n')$, $\nabla_{\mathbf{x}}\nabla_{\mathbf{x}} n = n''\mathbf{e}_3 \otimes \mathbf{e}_3$, and reduces, in the vertical plane of the ray, to a scalar equation for the in-plane normal shift with coefficients depending on $z$ only, which can be integrated in quadratures for a piecewise-analytic $W(z)$; the phase-time correction (39) along the straight background ray is $\delta\tau = 1/2v \int W(z,\omega)\, \mathrm{d}s = 1/2v\cos\theta \int W(z,\omega)\, \mathrm{d}z$. This provides a simple analytic benchmark for the finite-element Eigenwave solver.

Figure 10 shows this benchmark for a Gaussian layer in a homogeneous background: the Eigenwaves refract towards the vertical inside the layer, the lateral shift and the phase-time correction scale as $\omega^{-2}$, and the first-order correction (39) is within a few per cent of the exact quadrature of (F1) down to the lowest frequency, where the peak of the potential reaches $W = 0.35$.

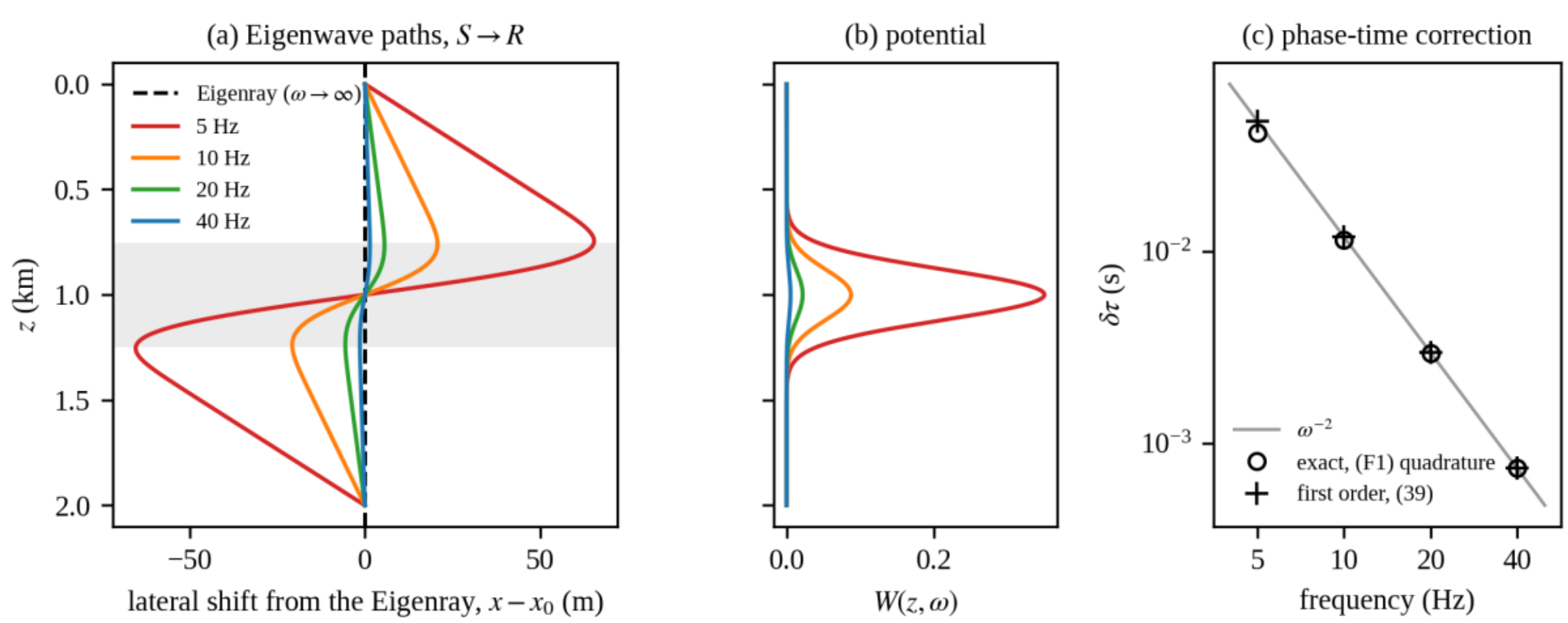

**Figure 10.** Analytic benchmark of Appendix F: homogeneous background $v = 2$ km/s, source at the origin, receiver at $x = 3$ km, $z = 2$ km, and a Gaussian layered potential $W(z,\omega) = \kappa(z)/\omega^2$ centred at $z = 1$ km (shaded: $\pm 2$ standard deviations, 0.12 km) with $W_{\max} = 0.35$ at 5 Hz. (a) Lateral shift of the two-point Eigenwaves from the Eigenray at 5, 10, 20 and 40 Hz, obtained from the finite-frequency Snell law (F1): the trajectories refract towards the vertical inside the layer,

where $W > 0$. (b) The potential at the four frequencies. (c) Phase-time correction $\delta\tau$: exact, from the quadrature of (F1), and first order, from (39); the line is the $\omega^{-2}$ law.

## Appendix G. Summary of the main equations for isotropic elastic media

This appendix specializes the main equations of the paper to isotropic elastic media, keeping the tensor-form notation. Isotropy brings three simplifications: the ray and phase velocities coincide and are direction independent, the slowness is aligned with the ray, and all directional derivatives of the velocity vanish.

### G.1 Medium, Christoffel tensor and mode velocities

The density-normalized isotropic stiffness tensor acts on an arbitrary second-order tensor $\mathbf{A}$ as

$$\tilde{\mathbf{C}}\colon \mathbf{A} = (v_P^2 - 2v_S^2)(\mathrm{tr}\,\mathbf{A})\mathbf{I} + v_S^2(\mathbf{A} + \mathbf{A}^{\mathrm{T}}), \tag{G1}$$

where $v_P(\mathbf{x})$ and $v_S(\mathbf{x})$ are the compressional and shear velocities. The Christoffel tensor of equation (5) becomes

$$\mathbf{\Gamma}(\mathbf{x},\mathbf{p}) = \mathbf{p}\cdot\tilde{\mathbf{C}}\cdot\mathbf{p} = v_S^2(\mathbf{p}\cdot\mathbf{p})\mathbf{I} + (v_P^2 - v_S^2)\mathbf{p}\otimes\mathbf{p}, \tag{G2}$$

with the P eigenvalue $v_P^2(\mathbf{p}\cdot\mathbf{p})$ and longitudinal polarization $\hat{\mathbf{g}} = \mathbf{p}/|\mathbf{p}|$, and the doubly degenerate S eigenvalue $v_S^2(\mathbf{p}\cdot\mathbf{p})$ with transverse polarizations. For a fixed mode with velocity $v(\mathbf{x})$ (either $v_P$ or $v_S$),

$$\lambda(\mathbf{x},\mathbf{p}) = v^2(\mathbf{x})(\mathbf{p}\cdot\mathbf{p}), \quad v_{\mathrm{phs}} = v_{\mathrm{ray}} = v(\mathbf{x}), \quad \nabla_{\mathbf{r}} v_{\mathrm{ray}} = \mathbf{0}, \quad \mathbf{p} = \frac{\mathbf{r}}{v}, \tag{G3}$$

so the slowness surface is a sphere, the phase and ray directions coincide, $\mathbf{n} = \mathbf{r}$, and the ray-velocity tensor reduces to $\tilde{\mathbf{A}}(\hat{\mathbf{g}}) = v^2\mathbf{I}$ for the selected mode.

### G.2 Eikonal (Eigenray) kinematics and dynamics

With $\mathbf{r}\cdot\mathbf{r} = 1$, the Lagrangian (1) and its gradients (3) reduce to

$$L = \frac{\sqrt{\mathbf{r}\cdot\mathbf{r}}}{v(\mathbf{x})}, \quad L_{\mathbf{r}} = \mathbf{p} = \frac{\mathbf{r}}{v}, \quad L_{\mathbf{x}} = -\frac{\nabla_{\mathbf{x}} v}{v^2}, \tag{G4}$$

and the Euler–Lagrange equation (3) becomes the classical isotropic ray equation, in the two equivalent tensor forms

$$\frac{d}{ds}\left(\frac{\mathbf{r}}{v}\right) = -\frac{\nabla_{\mathbf{x}} v}{v^2} \quad \Leftrightarrow \quad \dot{\mathbf{r}} = -(\mathbf{I} - \mathbf{r} \otimes \mathbf{r})\frac{\nabla_{\mathbf{x}} v}{v}: \tag{G5}$$

only the transverse component of the velocity gradient bends the ray. The Hessians (4) simplify to

$$L_{\mathbf{xx}} = 2\frac{\nabla_{\mathbf{x}} v \otimes \nabla_{\mathbf{x}} v}{v^3} - \frac{\nabla_{\mathbf{x}} \nabla_{\mathbf{x}} v}{v^2}, \quad L_{\mathbf{xr}} = (L_{\mathbf{rx}})^{\mathsf{T}} = -\frac{\nabla_{\mathbf{x}} v \otimes \mathbf{r}}{v^2}, \quad L_{\mathbf{rr}} = \frac{\mathbf{I} - \mathbf{r} \otimes \mathbf{r}}{v}, \tag{G6}$$

and enter the Jacobi equation (6) unchanged in form. The two nonzero eigenvalues of $L_{\mathbf{rr}}$ are degenerate, $\lambda_1 = \lambda_2 = 1/v$, with any orthonormal pair in the normal plane as eigenvectors (the isotropic degenerate case of the initial conditions (32)). The point-source ray Jacobian is $J = \mathbf{y}_1 \times \mathbf{y}_2 \cdot \mathbf{r}$, and, since $v_{\text{ray}} = v_{\text{phs}}$ and $\lambda_{1,S}\lambda_{2,S} = 1/v_S^2$, the relative geometric spreading (34) reduces to

$$L_{\text{GS}}(s) = v_S\sqrt{|J(s)|}, \quad \sigma = \int_S^R v \, ds, \quad \left(\frac{L_{\text{GS}}}{\sigma}\right)_S = 1. \tag{G7}$$

**G.3 Finite-frequency potential, index of refraction and effective medium**

The potential (9), evaluated with the amplitude field $\mathbf{a} = a\hat{\mathbf{a}}$ of the selected mode, is

$$W(\mathbf{x}, \omega) = \frac{\hat{\mathbf{a}} \cdot \nabla_{\mathbf{x}} \cdot \left(\rho \, \tilde{\mathbf{C}} \colon \nabla_{\mathbf{x}} \mathbf{a}\right)}{\rho \, \omega^2 \, a}, \tag{G8}$$

which, in the scalar (acoustic) analogue with smooth density, reduces to $W = v^2 \nabla_{\mathbf{x}} \cdot \nabla_{\mathbf{x}} a/(\omega^2 a)$, the form used in Section 8. The index of refraction and the effective medium (12), (14) are

$$n(\mathbf{x}, \omega) = \sqrt{1 + W(\mathbf{x}, \omega)}, \quad v^{\omega}(\mathbf{x}, \omega) = \frac{v(\mathbf{x})}{n(\mathbf{x}, \omega)}, \quad \tilde{\mathbf{C}}^{\omega} = \frac{\tilde{\mathbf{C}}}{n^2}, \tag{G9}$$

i.e. the effective isotropic medium simply carries the frequency-dependent velocity $v/n$, exactly as in the exact ray theory of Foreman (1989) for the Helmholtz equation (Remark 2). The gradients of the index follow from (16).

**G.4 Finite-frequency (Eigenwave) kinematics**

The Eigenwave Lagrangian (17), momentum (20)–(21) and Euler–Lagrange equation (22)–(23) become, with $\mathbf{r} \cdot \mathbf{r} = 1$,

$$L^{\omega} = \frac{n(\mathbf{x},\omega)}{v(\mathbf{x})} = \frac{1}{v^{\omega}}, \qquad L^{\omega}_{\mathbf{r}} = \mathbf{p}^{\omega} = \frac{n\,\mathbf{r}}{v} = \frac{\mathbf{r}}{v^{\omega}},$$
$$L^{\omega}_{\mathbf{x}} = -n\frac{\nabla_{\mathbf{x}} v}{v^2} + \frac{\nabla_{\mathbf{x}} W}{2nv} = -\frac{\nabla_{\mathbf{x}} v^{\omega}}{(v^{\omega})^2}, \tag{G10}$$

$$\frac{d}{ds}\left(\frac{n\,\mathbf{r}}{v}\right) = -n\frac{\nabla_{\mathbf{x}} v}{v^2} + \frac{\nabla_{\mathbf{x}} W}{2nv} \quad \Leftrightarrow \quad \dot{\mathbf{r}} = -(\mathbf{I} - \mathbf{r}\otimes\mathbf{r})\frac{\nabla_{\mathbf{x}} v^{\omega}}{v^{\omega}}, \tag{G11}$$

the classical ray equation of the medium $v^{\omega} = v/n$, as required by Proposition 1. Since $\mathbf{p} = \mathbf{r}/v$ is aligned with the ray, the compact form (24) reduces to

$$n[\dot{\mathbf{p}} - L_{\mathbf{x}}] = \mathbf{r}\times(\nabla_{\mathbf{x}} n \times \mathbf{p}) = (\mathbf{I} - \mathbf{r}\otimes\mathbf{r})\frac{\nabla_{\mathbf{x}} n}{v}: \tag{G12}$$

the finite-frequency force is the transverse gradient of the index divided by the velocity. For a laterally homogeneous potential and medium (Appendix F), the horizontal slowness conservation yields the finite-frequency Snell law $n(z,\omega)\sin\theta(z)/v(z) = \text{const}$.

### G.5 Finite-frequency (Eigenwave) dynamics

The Hessians (28) of the Eigenwave Lagrangian specialize to

$$L^{\omega}_{\mathbf{xx}} = nL_{\mathbf{xx}} + \nabla_{\mathbf{x}} n \otimes L_{\mathbf{x}} + L_{\mathbf{x}} \otimes \nabla_{\mathbf{x}} n + \frac{\nabla_{\mathbf{x}}\nabla_{\mathbf{x}} n}{v}, \tag{G13}$$

$$L^{\omega}_{\mathbf{xr}} = (L^{\omega}_{\mathbf{rx}})^{\mathsf{T}} = -n\frac{\nabla_{\mathbf{x}} v\otimes\mathbf{r}}{v^2} + \frac{\nabla_{\mathbf{x}} n\otimes\mathbf{r}}{v}, \qquad L^{\omega}_{\mathbf{rr}} = n\frac{\mathbf{I} - \mathbf{r}\otimes\mathbf{r}}{v}, \tag{G14}$$

with $L_{\mathbf{xx}}$, $L_{\mathbf{x}}$ from Appendix G.2, and enter the finite-frequency Jacobi equation (29) unchanged in form. The point-source eigenvalues are $\lambda^{\omega}_{1,S} = \lambda^{\omega}_{2,S} = n_S/v_S$, so the ray Jacobian, geometric spreading (34) and the leading amplitude (40) read

$$J^{\omega} = \mathbf{y}_1 \times \mathbf{y}_2 \cdot \mathbf{r},$$
$$L^{\omega}_{\mathrm{GS}}(s) = \frac{v_S}{n_S}\sqrt{|J^{\omega}(s)|} = v^{\omega}_S\sqrt{|J^{\omega}(s)|}, \tag{G15}$$
$$a \propto (\rho\, n\, v\, |J^{\omega}|)^{-1/2}.$$

In the ray-centred frame the transverse slowness vanishes, $\tilde{\mathbf{p}} = \mathbf{0}$, so all the rank-one terms of Appendix B that are driven by $\tilde{\mathbf{p}}$ drop out: with $\mathbf{Z} = \mathbf{I}_2/v$, the dynamic ray tracing coefficients (B11) reduce to

$$\mathbf{A}^\omega = \mathbf{A}, \quad \mathbf{B}^\omega = \frac{\mathbf{B}}{n}, \qquad \mathbf{C}^\omega = n\,\mathbf{C} + \boldsymbol{\nu}\,\mathbf{d}^\mathsf{T} + \mathbf{d}\,\boldsymbol{\nu}^\mathsf{T} + \frac{\mathbf{N}}{v}, \qquad \mathbf{d} = \mathbf{g} - \frac{\mathbf{z}}{v}, \tag{G16}$$

the wavefront-curvature matrix is $\mathbf{M}^\omega = n\,\mathbf{M}_*$ (equation (B19) without the anisotropic rank-one term, $\mathbf{M}_*$ then being symmetric), and the physical wavefront curvature is $\mathbf{K}^\omega = v^\omega \mathbf{M}^\omega$. The paraxial phase-time expansion (B18) loses its linear term.

### G.6 Weak coupling

For $|W| \ll 1$, the first-order Eigenwave path solves the classical isotropic Jacobi equation with the forcing (38), which, using $\mathbf{p}_0 = \mathbf{r}_0/v$, becomes purely transverse in an explicitly projected form,

$$\mathbf{f}(s) = \frac{\nabla_\mathbf{x} W}{2v} - \frac{(\nabla_\mathbf{x} W \cdot \mathbf{r}_0)\mathbf{r}_0}{2v} = (\mathbf{I} - \mathbf{r}_0 \otimes \mathbf{r}_0)\frac{\nabla_\mathbf{x} W}{2v}, \tag{G17}$$

and the finite-frequency phase-time correction (39) is

$$\delta\tau(\omega) = \frac{1}{2}\int_{\mathbf{x}_0} \frac{W(\mathbf{x},\omega)}{v(\mathbf{x})}\,ds, \tag{G18}$$

evaluated along the classical ray $\mathbf{x}_0(s)$; both the path and the phase-time corrections scale as $\omega^{-2}$ when $W = \kappa/\omega^2$.

## Appendix H. Summary of the main equations for transversely isotropic media with a tilted axis of symmetry (TTI)

This appendix specializes the main equations of the paper to transversely isotropic media with a tilted axis of symmetry (TTI), keeping the tensor-form notation. The medium is described by the five density-normalized stiffnesses $\tilde{c}_{11}, \tilde{c}_{13}, \tilde{c}_{33}, \tilde{c}_{44}, \tilde{c}_{66}$ of the axis-aligned (crystal) frame, with the units of velocity squared, and by the unit symmetry-axis field $\hat{\mathbf{k}}(\mathbf{x})$; all six fields may vary spatially. In contrast with the isotropic case of Appendix G, the phase and ray velocities differ and are direction dependent, and the slowness is not aligned with the ray, so the general equations of the main text apply in full force; the purpose of this appendix is to provide their TTI-specific ingredients in closed, coordinate-free form.

### H.1 Medium, Christoffel tensor and mode velocities

Let $\mathbf{K} = \hat{\mathbf{k}} \otimes \hat{\mathbf{k}}$ and $\mathbf{T} = \mathbf{I} - \mathbf{K}$ be the axial and transverse projectors, with $\mathbf{K}\!:\mathbf{A} = \hat{\mathbf{k}} \cdot \mathbf{A}\,\hat{\mathbf{k}}$ and $\mathbf{T}\!:\mathbf{A} = \operatorname{tr}\mathbf{A} - \hat{\mathbf{k}} \cdot \mathbf{A}\,\hat{\mathbf{k}}$ for an arbitrary second-order tensor $\mathbf{A}$. The density-normalized TTI stiffness tensor acts on $\mathbf{A}$ as

$$\tilde{\mathbf{C}}:\mathbf{A} = (\tilde{c}_{11} - 2\tilde{c}_{66})(\mathbf{T}:\mathbf{A})\mathbf{T} + \tilde{c}_{66}(\mathbf{TAT} + \mathbf{TA}^{\mathsf{T}}\mathbf{T}) + \tilde{c}_{13}[(\mathbf{K}:\mathbf{A})\mathbf{T} + (\mathbf{T}:\mathbf{A})\mathbf{K}] + \tilde{c}_{33}(\mathbf{K}:\mathbf{A})\mathbf{K} + \tilde{c}_{44}(\mathbf{TAK} + \mathbf{TA}^{\mathsf{T}}\mathbf{K} + \mathbf{KAT} + \mathbf{KA}^{\mathsf{T}}\mathbf{T}), \tag{H1}$$

which reduces to the isotropic form (G1) for $\tilde{c}_{11} = \tilde{c}_{33} = v_P^2$, $\tilde{c}_{44} = \tilde{c}_{66} = v_S^2$, $\tilde{c}_{13} = v_P^2 - 2v_S^2$. Decomposing the slowness into its axial and transverse parts, $p_\parallel = \mathbf{p}\cdot\hat{\mathbf{k}}$, $\mathbf{p}_t = \mathbf{T}\mathbf{p}$, $p_\perp^2 = \mathbf{p}_t\cdot\mathbf{p}_t$, the Christoffel tensor of equation (5) becomes

$$\mathbf{\Gamma}(\mathbf{x},\mathbf{p}) = (\tilde{c}_{11} - \tilde{c}_{66})\mathbf{p}_t \otimes \mathbf{p}_t + \left(\tilde{c}_{66}\, p_\perp^2 + \tilde{c}_{44}\, p_\parallel^2\right)\mathbf{T} + \left(\tilde{c}_{33}\, p_\parallel^2 + \tilde{c}_{44}\, p_\perp^2\right)\mathbf{K} + (\tilde{c}_{13} + \tilde{c}_{44})p_\parallel\left(\mathbf{p}_t \otimes \hat{\mathbf{k}} + \hat{\mathbf{k}} \otimes \mathbf{p}_t\right). \tag{H2}$$

The polarization normal to the axial plane spanned by $\hat{\mathbf{k}}$ and $\mathbf{p}$ is an exact eigenvector of (H2): the SH mode decouples, with the quadratic (elliptical) eigenvalue

$$\mathbf{\Gamma}\,\hat{\mathbf{g}}_{SH} = \lambda_{SH}\hat{\mathbf{g}}_{SH}, \quad \hat{\mathbf{g}}_{SH} = \frac{\hat{\mathbf{k}}\times\mathbf{p}}{|\hat{\mathbf{k}}\times\mathbf{p}|}, \quad \lambda_{SH} = \mathbf{p}\cdot\widetilde{\mathbf{A}}_{SH}\,\mathbf{p}, \quad \widetilde{\mathbf{A}}_{SH} = \tilde{c}_{66}\mathbf{T} + \tilde{c}_{44}\mathbf{K}, \tag{H3}$$

while the coupled qP and qSV eigenvalues and polarizations live in the axial plane. Introducing the scalar components $\Gamma_{tt} = \hat{\mathbf{p}}_t\cdot\mathbf{\Gamma}\,\hat{\mathbf{p}}_t$, $\Gamma_{kk} = \hat{\mathbf{k}}\cdot\mathbf{\Gamma}\,\hat{\mathbf{k}}$ and $\Gamma_{tk} = \hat{\mathbf{p}}_t\cdot\mathbf{\Gamma}\,\hat{\mathbf{k}}$, with $\hat{\mathbf{p}}_t = \mathbf{p}_t/p_\perp$,

$$\Gamma_{tt} = \tilde{c}_{11}p_\perp^2 + \tilde{c}_{44}p_\parallel^2, \quad \Gamma_{kk} = \tilde{c}_{33}p_\parallel^2 + \tilde{c}_{44}p_\perp^2, \quad \Gamma_{tk} = (\tilde{c}_{13} + \tilde{c}_{44})p_\parallel\, p_\perp,$$
$$\lambda_{qP,qSV} = \frac{\Gamma_{tt} + \Gamma_{kk}}{2} \pm \sqrt{\frac{(\Gamma_{tt} - \Gamma_{kk})^2}{4} + \Gamma_{tk}^2}, \tag{H4}$$

with the plus sign for qP and the minus sign for qSV. All three eigenvalues are homogeneous of degree two in the slowness, in accordance with (10) and the identities of A.4. Owing to the minor symmetries of $\tilde{\mathbf{C}}$, the ray-velocity tensor $\widetilde{\mathbf{A}}(\hat{\mathbf{g}})$ of Section 3 has exactly the same closed form as (H2), with the slowness replaced by the polarization ($p_\parallel \to \hat{\mathbf{g}}\cdot\hat{\mathbf{k}}$, $\mathbf{p}_t \to \mathbf{T}\hat{\mathbf{g}}$), so that $\lambda = \mathbf{p}\cdot\widetilde{\mathbf{A}}(\hat{\mathbf{g}})\mathbf{p}$ and the group velocity is $\widetilde{\mathbf{A}}(\hat{\mathbf{g}})\mathbf{p}$, as in (10)–(11). In the Thomsen (1986) parametrization,

$$\tilde{c}_{33} = v_{P0}^2, \quad \tilde{c}_{44} = v_{S0}^2, \quad \tilde{c}_{11} = v_{P0}^2(1 + 2\varepsilon),$$
$$\tilde{c}_{66} = v_{S0}^2(1 + 2\gamma), \quad (\tilde{c}_{13} + \tilde{c}_{44})^2 = (v_{P0}^2 - v_{S0}^2)[v_{P0}^2(1 + 2\delta) - v_{S0}^2], \tag{H5}$$

where $v_{P0}(\mathbf{x})$ and $v_{S0}(\mathbf{x})$ are the axial compressional and shear velocities, and $\varepsilon, \delta, \gamma$ are the dimensionless anisotropy parameters.

### H.2 Eikonal (Eigenray) kinematics and dynamics

By the rotational symmetry about the axis, for a fixed mode the ray velocity depends on the ray direction only through the axial angle: with $\mathbf{r}\cdot\mathbf{r} = 1$,

$$v_{\rm ray}(\mathbf{x},\mathbf{r}) = v_{\rm ray}[m(\mathbf{x}),\chi], \quad \chi = \mathbf{r}\cdot\hat{\mathbf{k}}, \quad \nabla_{\mathbf{r}} v_{\rm ray} = v_\chi\big(\hat{\mathbf{k}} - \chi\,\mathbf{r}\big),$$
$$v_\chi \equiv \frac{\partial v_{\rm ray}}{\partial \chi}, \tag{H6}$$

where $m(\mathbf{x}) = \{v_{P0}, v_{S0}, \varepsilon, \delta, \gamma\}$ collects the local medium parameters; the identity $\nabla_{\mathbf{r}} v_{\rm ray}\cdot\mathbf{r} = 0$ (homogeneity of degree zero, A.4) is manifest. The slowness (3) acquires the closed form

$$\mathbf{p} = L_{\mathbf{r}} = \frac{\mathbf{r}}{v_{\rm ray}} - \frac{\nabla_{\mathbf{r}} v_{\rm ray}}{v_{\rm ray}^2} = \frac{1}{v_{\rm ray}}\left[\left(1 + \chi\frac{v_\chi}{v_{\rm ray}}\right)\mathbf{r} - \frac{v_\chi}{v_{\rm ray}}\hat{\mathbf{k}}\right], \qquad \mathbf{p}\cdot\mathbf{r} = \frac{1}{v_{\rm ray}}: \tag{H7}$$

the slowness lies in the axial plane $\mathrm{span}\{\mathbf{r},\hat{\mathbf{k}}\}$, and its deviation from the ray direction vanishes for propagation along and normal to the axis and, identically, in the isotropy limit. The spatial gradient follows by the chain rule through the parameter and tilt fields,

$$\nabla_{\mathbf{x}} v_{\rm ray} = \sum_m \frac{\partial v_{\rm ray}}{\partial m}\nabla_{\mathbf{x}} m + v_\chi\big(\nabla_{\mathbf{x}}\hat{\mathbf{k}}\big)\mathbf{r}, \tag{H8}$$

where $\nabla_{\mathbf{x}}\hat{\mathbf{k}}$ is the second-order tensor of the tilt gradients (with the row/column convention of A.2): the tilt variation enters the ray bending on the same footing as the parameter gradients. One further differentiation of (H7)–(H8) yields the Hessians (4) that drive the Jacobi equation (6); the directional derivatives of the TTI ray velocity, for all three modes, are given in closed form by Ravve and Koren (2019). For the decoupled SH mode everything is explicit, since (H3) is a quadratic form with the direction-independent tensor $\widetilde{\mathbf{A}}_{SH}$:

$$v_{{\rm phs},SH}(\mathbf{x},\mathbf{n}) = \sqrt{\mathbf{n}\cdot\widetilde{\mathbf{A}}_{SH}\,\mathbf{n}},$$
$$v_{{\rm ray},SH}(\mathbf{x},\mathbf{r}) = \big(\mathbf{r}\cdot\widetilde{\mathbf{A}}_{SH}^{-1}\,\mathbf{r}\big)^{-1/2}, \qquad \widetilde{\mathbf{A}}_{SH}^{-1} = \frac{\mathbf{T}}{\tilde{c}_{66}} + \frac{\mathbf{K}}{\tilde{c}_{44}}, \tag{H9}$$

the ellipsoidal (elliptical-anisotropy) case.

### H.3 Finite-frequency potential, index of refraction and effective medium

The potential $W(\mathbf{x},\omega)$ retains its general form (9), evaluated with the amplitude field of the selected mode, and the index of refraction is $n = \sqrt{1+W}$ (12). The conformal rescaling (14) divides all five stiffnesses by $n^2$ and leaves the axis untouched,

$$\tilde{\mathbf{C}}^\omega = \frac{\tilde{\mathbf{C}}}{n^2} \quad \Leftrightarrow \quad \tilde{c}_{ij}^{\,\omega} = \frac{\tilde{c}_{ij}}{n^2}, \quad ij \in \{11,13,33,44,66\}, \quad \hat{\mathbf{k}}^\omega = \hat{\mathbf{k}}. \tag{H10}$$

Since $\varepsilon, \delta, \gamma$ are ratios of stiffnesses, they are invariant under this common rescaling: the effective medium is again TTI, with the same tilt and the same anisotropy parameters, and only the axial velocities carry the index of refraction,

$$v_{P0}^{\omega} = \frac{v_{P0}}{n}, \qquad v_{S0}^{\omega} = \frac{v_{S0}}{n}, \qquad \varepsilon^{\omega} = \varepsilon, \qquad \delta^{\omega} = \delta, \qquad \gamma^{\omega} = \gamma. \tag{H11}$$

Consequently $v_{\text{phs}}^{\omega} = v_{\text{phs}}/n$, $v_{\text{ray}}^{\omega} = v_{\text{ray}}/n$, $\mathbf{p}^{\omega} = n\,\mathbf{p}$ and $\hat{\mathbf{g}}^{\omega} = \hat{\mathbf{g}}$, as in (15): the finite-frequency coupling does not rotate the polarizations, does not tilt the axis, and neither creates nor destroys anisotropy. The gradients of $n$ follow from (16).

**H.4 Finite-frequency (Eigenwave) kinematics**

The Eigenwave Lagrangian (17) and momentum (20)–(21) are evaluated with the TTI ray velocity,

$$\begin{gathered} L^{\omega} = n\,L = \frac{n(\mathbf{x},\omega)\sqrt{\mathbf{r}\cdot\mathbf{r}}}{v_{\text{ray}}(\mathbf{x},\mathbf{r})}, \\ \mathbf{p}^{\omega} = n\,\mathbf{p} = \frac{n}{v_{\text{ray}}}\left[\left(1 + \chi\frac{v_{\chi}}{v_{\text{ray}}}\right)\mathbf{r} - \frac{v_{\chi}}{v_{\text{ray}}}\hat{\mathbf{k}}\right], \\ \mathbf{p}^{\omega}\cdot\mathbf{r} = \frac{1}{v_{\text{ray}}^{\omega}}, \end{gathered} \tag{H12}$$

and the Euler-Lagrange equation retains the compact form (24),

$$n[\dot{\mathbf{p}} - L_{\mathbf{x}}] = \mathbf{r}\times(\nabla_{\mathbf{x}} n \times \mathbf{p}), \tag{H13}$$

where the left-hand side measures the deviation from the classical TTI ray equation $\dot{\mathbf{p}} = L_{\mathbf{x}}$, and the finite-frequency force on the right couples the transverse index gradient to the full slowness (H7) rather than to $\mathbf{r}/v$ as in Appendix G. Equivalently, by Proposition 1 and (H10)–(H11), the Eigenwave kinematics is standard TTI ray tracing in the effective medium: an existing TTI two-point (Eigenray) solver is extended to finite frequency by the mere substitution of the rescaled axial velocities $v_{P0}/n$ and $v_{S0}/n$, with unchanged $\varepsilon, \delta, \gamma$ and $\hat{\mathbf{k}}$.

**H.5 Finite-frequency (Eigenwave) dynamics**

The Hessians (28) and the finite-frequency Jacobi equation (29) apply verbatim with the classical TTI Hessians (4); none of the isotropic simplifications of G.5 occurs. In particular, in the ray-centred frame the transverse slowness does not vanish: projecting (H7) onto the normal legs,

$$\tilde{p}_{\alpha} = \mathbf{e}_{\alpha}\cdot\mathbf{p} = -\frac{v_{\chi}}{v_{\text{ray}}^{2}}\left(\mathbf{e}_{\alpha}\cdot\hat{\mathbf{k}}\right), \qquad \alpha \in \{1,2\}, \tag{H14}$$

it is driven by the transverse projection of the symmetry axis. All the rank-one $\tilde{\mathbf{p}}$-driven terms of Appendix B therefore remain active: the coefficient $\mathbf{Y}^{\omega}$ of (B6) retains its $\boldsymbol{\nu}\,\tilde{\mathbf{p}}^{\mathsf{T}}$ correction, the dynamic ray tracing coefficients (B11) keep their full anisotropic structure, and the paraxial matrix of (B19) acquires the rank-one anisotropic term,

$$\mathbf{M}^{\omega} = n\,\mathbf{M}_{*} + \tilde{\mathbf{p}}\,\boldsymbol{\nu}^{\mathsf{T}}, \qquad \nu_{\alpha} = \mathbf{e}_{\alpha} \cdot \nabla_{\mathbf{x}} n. \tag{H15}$$

At a point source the two eigenvalues $\lambda_{1,S}, \lambda_{2,S}$ of $L_{\mathbf{rr},S}$ are distinct in general (the qSV-SH splitting of the slowness-surface curvatures), so the initial conditions (32) apply in their general, nondegenerate form, and the geometric spreading (34) retains the ray-to-phase velocity ratios at the source and along the path.

### H.6 Weak coupling

For $|W| \ll 1$, the forcing of the first-order Eigenwave path (38) can no longer be reduced to the orthogonal transverse projection of Appendix G: using $\mathbf{p}_0 \cdot \mathbf{r}_0 = 1/v_{\mathrm{ray}}$,

$$\mathbf{f}(s) = \frac{1}{2}\,\mathbf{r}_0 \times (\nabla_{\mathbf{x}} W \times \mathbf{p}_0) = \frac{1}{2}\left(\frac{\mathbf{I}}{v_{\mathrm{ray}}} - \mathbf{p}_0 \otimes \mathbf{r}_0\right)\nabla_{\mathbf{x}} W, \qquad \mathbf{f} \cdot \mathbf{r}_0 = 0, \tag{H16}$$

a transverse but oblique projection of the potential gradient, reflecting the misalignment of the slowness and the ray. The phase-time correction (39) is unchanged in form,

$$\delta\tau(\omega) = \frac{1}{2}\int_{\mathbf{x}_0} \frac{W(\mathbf{x}, \omega)}{v_{\mathrm{ray}}(\mathbf{x}, \mathbf{r}_0)}\, ds, \tag{H17}$$

evaluated along the classical TTI central ray $\mathbf{x}_0(s)$. For $\varepsilon = \delta = \gamma = 0$ the axis becomes immaterial, and all the formulas of this appendix reduce to those of Appendix G.

## Appendix I. Acoustic approximation and homogeneous medium: frequency-dependent trajectories and the banana-doughnut pattern

This appendix specializes the theory to the scalar acoustic wave equation and then to a homogeneous medium, where the Eigenwaves are available in closed form. It serves two purposes. First, it provides the simplest analytic benchmarks for the finite-element Eigenwave solver, complementing the layered example of Appendix F. Second, it makes explicit the relation between the frequency-dependent trajectories of trajectory mechanics and the finite-frequency traveltime sensitivity kernels of Marquering et al. (1999) and Dahlen et al. (2000), whose “banana-doughnut” shape - a region of sensitivity that surrounds the geometrical ray but vanishes on it - appears paradoxical from the ray-theoretical viewpoint. In the Eigenwave framework the banana-doughnut kernel turns out to be the ray integral of the interference part of the potential $W$, so that the paradox is resolved rather than merely circumvented.

### I.1 The acoustic approximation

Let $u(\mathbf{x},\omega)$ be the scalar (pressure) wavefield of the acoustic wave equation in the frequency domain,

$$\rho\,\nabla_{\mathbf{x}}\cdot\left(\frac{\nabla_{\mathbf{x}}u}{\rho}\right)+\frac{\omega^2}{v^2}u=0, \tag{I1}$$

with velocity $v(\mathbf{x})$ and density $\rho(\mathbf{x})$. Formally, (I1) is the isotropic elastodynamic equation of Section 3.1 with the shear velocity set to zero, $\tilde{\mathbf{C}}\colon\mathbf{A}=v^2(\mathrm{tr}\,\mathbf{A})\mathbf{I}$ in (G1), so that the tensor objects of the main text collapse to

$$\begin{gathered}\boldsymbol{\Gamma}(\mathbf{x},\mathbf{p})=v^2\,\mathbf{p}\otimes\mathbf{p},\qquad \lambda(\mathbf{x},\mathbf{p})=v^2(\mathbf{p}\cdot\mathbf{p}),\\ \hat{\mathbf{g}}=\frac{\mathbf{p}}{|\mathbf{p}|},\qquad \widetilde{\mathbf{A}}(\hat{\mathbf{g}})\mathbf{p}=v^2\mathbf{p},\qquad v_{\mathrm{phs}}=v_{\mathrm{ray}}=v,\qquad \mathbf{p}=\frac{\mathbf{r}}{v},\end{gathered} \tag{I2}$$

the Christoffel tensor being of rank one, with the single nonzero eigenvalue $v^2(\mathbf{p}\cdot\mathbf{p})$ and longitudinal polarization. The polar decomposition (7), $u=a\,e^{i\varphi}$ with real $a$ and $\varphi$ and $\mathbf{p}=\nabla_{\mathbf{x}}\varphi/\omega$, splits (I1) into the frequency-dependent eikonal equation and the transport equation,

$$\begin{gathered}v^2(\mathbf{p}\cdot\mathbf{p})=1+W(\mathbf{x},\omega),\\ W(\mathbf{x},\omega)=\frac{v^2}{\omega^2}\frac{\rho\,\nabla_{\mathbf{x}}\cdot(\rho^{-1}\nabla_{\mathbf{x}}a)}{a}=\frac{v^2}{\omega^2}\left(\frac{\nabla^2 a}{a}-\nabla_{\mathbf{x}}\ln\rho\cdot\frac{\nabla_{\mathbf{x}}a}{a}\right),\end{gathered} \tag{I3}$$

$$\nabla_{\mathbf{x}}\cdot\left(\frac{a^2\,\mathbf{p}}{\rho}\right)=0, \tag{I4}$$

where $\nabla^2\equiv\nabla_{\mathbf{x}}\cdot\nabla_{\mathbf{x}}$. For constant density, (I3)-(I4) reduce to the forms used in Section 8, $W=v^2\nabla^2 a/(\omega^2 a)$ and (41). Two remarks are in order. First, no polarization approximation is involved: in the elastic case the scalar potential (10) is obtained from the vector equation (8) by replacing the amplitude direction $\hat{\mathbf{a}}$ with the Christoffel polarization $\hat{\mathbf{g}}$ (Remark 1), whereas in the acoustic case the wavefield is scalar and (I3) is exact. The acoustic Eigenwave equations are therefore an exact reformulation of the Helmholtz equation for any given amplitude field, in the sense of Foreman (1989) and Protasov and Gadylshin (2017), and (I3) is formally identical to the quantum Hamilton-Jacobi equation of Bohm (1952), $W$ playing the role of the quantum potential (Wyatt, 2005). Second, since the polarization is longitudinal and the slowness is aligned with the ray, the kinematic, dynamic and weak-coupling Eigenwave equations (G9)-(G18) of Appendix G apply verbatim with $v$ the acoustic velocity; in particular, the Lagrangian, the momentum and the Euler-Lagrange equation read

$$L^{\omega} = \frac{n(\mathbf{x},\omega)}{v(\mathbf{x})}, \qquad \mathbf{p}^{\omega} = \frac{n\,\mathbf{r}}{v},$$
$$\dot{\mathbf{r}} = -(\mathbf{I} - \mathbf{r} \otimes \mathbf{r})\left[\frac{\nabla_{\mathbf{x}} v}{v} - \frac{\nabla_{\mathbf{x}} W}{2(1+W)}\right], \quad n = \sqrt{1+W}, \tag{I5}$$

the density entering only through the potential (I3) and the transport equation (I4).

**I.2 Homogeneous medium: exact and first-order Eigenwaves**

Let $v$ and $\rho$ be constant. The eikonal rays are straight lines, and (I5) reduces to

$$\dot{\mathbf{r}} = \frac{(\mathbf{I} - \mathbf{r} \otimes \mathbf{r})\nabla_{\mathbf{x}} W}{2(1+W)}: \tag{I6}$$

the Eigenwave is bent only by the transverse gradient of the potential, its curvature vector pointing towards larger $W$ (larger effective slowness). Whether the trajectories depart from straight lines is thus a property of the wavefield rather than of the medium, and it is instructive to examine the exact potential of three elementary solutions of (I1).

*Plane wave and point source.* For a plane wave $u = \exp(i\omega\,\hat{\mathbf{r}} \cdot \mathbf{x}/v)$ the amplitude is constant and $W \equiv 0$. For a point source at $\mathbf{x}_S$,

$$u = \frac{e^{i\omega r_S/v}}{4\pi r_S}, \qquad r_S = |\mathbf{x} - \mathbf{x}_S|, \qquad \nabla^2 \frac{1}{r_S} = 0 \ \ (r_S > 0) \ \ \Rightarrow$$
$$W \equiv 0, \quad \tau = \frac{r_S}{v} \ \text{ for all } \omega: \tag{I7}$$

the amplitude of the exact spherical wave is a harmonic function, the potential vanishes identically, and the Eigenwaves coincide with the straight radial rays at every frequency, with the eikonal traveltime as phase time. For a point source in a three-dimensional homogeneous acoustic medium, ray theory is not an approximation but the exact solution, and no frequency dependence of the trajectories can arise. Conversely, any frequency dependence of the trajectories in a homogeneous medium is due to the geometry of the wavefield - the finite aperture of the source, or the interference with scattered waves - as in the examples that follow.

*Line source.* In two dimensions the exact field of a line source is $u = (i/4)H_0^{(1)}(\omega r/v)$, whose amplitude is not harmonic. From the large-argument expansion of the Hankel function,

$$W = \frac{v^2}{4\omega^2 r^2} + O\left(\frac{v^4}{\omega^4 r^4}\right), \qquad \varphi = \frac{\omega r}{v} - \frac{\pi}{4} - \frac{v}{8\omega r} + O\left(\frac{v^3}{\omega^3 r^3}\right), \tag{I8}$$

and indeed $d\varphi/dr = (\omega/v)\sqrt{1+W}$, as required by (I3). The potential is positive and radially symmetric, so by (I6) the Eigenwaves remain the straight radial lines; $W$ accounts exactly for the near-field phase of the cylindrical wave, i.e. for the departure of the phase from its eikonal value $\omega r/v$, which is concentrated within a few wavelengths of the source. The corresponding phase-time correction, $\delta\tau = -v/(8\omega^2 r)$ up to the constant source phase $-\pi/4$, is recovered from (39) as $\frac{1}{2}\int W\, ds/v$ and carries the $\omega^{-2}$ signature of Section 6.

*Gaussian beam.* A beam of waist $b_0$ launched along the $s$-axis is described, to paraxial accuracy, by

$$u = \frac{b_0}{b(s)}\exp\left(-\frac{\zeta^2}{b^2(s)}\right)\exp\left\{i\left[\frac{\omega s}{v} + \frac{\omega\,\zeta^2 s}{2v(s^2+\ell_0^2)} - \arctan\frac{s}{\ell_0}\right]\right\},$$
$$b^2(s) = b_0^2\left(1+\frac{s^2}{\ell_0^2}\right),\quad \ell_0 = \frac{\omega b_0^2}{2v}, \tag{I9}$$

where $\zeta$ is the distance from the axis, $b(s)$ the beam half-width and $\ell_0$ the Rayleigh distance. The transverse Laplacian of the Gaussian amplitude gives, to the same accuracy,

$$W(s,\zeta,\omega) = \frac{4v^2}{\omega^2 b^4(s)}[\zeta^2 - b^2(s)], \tag{I10}$$

which is negative in the core of the beam, $|\zeta| < b(s)$, and positive on its flanks: the core is effectively faster than the medium and the flanks slower. The Eigenwaves are the streamlines of $\nabla_{\mathbf{x}}\varphi$, $d\zeta/ds = \zeta s/(s^2+\ell_0^2)$, i.e. the hyperbolae

$$\zeta(s) = \zeta_0\sqrt{1+\frac{s^2}{\ell_0^2}},\quad \frac{d^2\zeta}{ds^2} = \frac{\zeta\,\ell_0^2}{(s^2+\ell_0^2)^2} = \frac{1}{2}\frac{\partial W}{\partial\zeta}, \tag{I11}$$

the second equality being the paraxial form of (I6). The trajectories of a beam therefore diverge - this is diffraction, expressed as the bending of the Eigenwaves by the potential - with the asymptotic angle $\zeta_0/\ell_0 = 2v\zeta_0/(\omega b_0^2)$, inversely proportional to the frequency; as $\omega\to\infty$, $\ell_0\to\infty$, $W\to 0$ and the parallel straight rays of the eikonal beam are recovered. Along the axis, $W = -4v^2/[\omega^2 b^2(s)]$ and the weak-coupling phase-time correction (39) integrates to the Gouy phase,

$$\delta\tau(s) = \frac{1}{2v}\int_0^s W\,(s',0,\omega)ds' = -\frac{1}{\omega}\arctan\frac{s}{\ell_0}, \tag{I12}$$

in agreement with the exact phase in (I9). Two features of this example are worth retaining. The magnitude of the potential on the axis, $|W| \le 4v^2/(\omega^2 b_0^2)$, is small precisely when the waist is wide compared with the wavelength - the perturbative regime of Section 6 - and is largest at the waist, where the amplitude varies most rapidly across the trajectories. And the corrections (I11)- (I12) do not scale as $\omega^{-2}$: the divergence angle scales as $\omega^{-1}$, and the Gouy phase time as $\omega^{-2}$ within the Rayleigh distance but as $\omega^{-1}$ beyond it. The $\omega^{-2}$ signature of Section 6 presupposes an amplitude field whose *shape* is frequency independent, as is the case for the ray-theoretical amplitude (40); when the shape of the amplitude is itself set by diffraction, the frequency dependence of the coupling scalar $\kappa$ must be retained.

*First-order Eigenwaves.* For the two-point problem between $\mathbf{x}_S$ and $\mathbf{x}_R$, at distance $\ell = |\mathbf{x}_R - \mathbf{x}_S|$, the weak-coupling equation (38) takes a particularly simple form in a homogeneous medium, where $L_{\mathbf{xx}} = L_{\mathbf{xr}} = \mathbf{0}$ and $L_{\mathbf{rr}} = (\mathbf{I} - \mathbf{r}_0 \otimes \mathbf{r}_0)/v$ by (G6):

$$\frac{d^2\delta\mathbf{x}}{ds^2} = \frac{1}{2}\nabla_\perp W, \quad \delta\mathbf{x}(0) = \delta\mathbf{x}(\ell) = \mathbf{0} \quad \Rightarrow$$
$$\delta\mathbf{x}(s) = -\frac{1}{2}\int_0^\ell g\,(s,s')\nabla_\perp W(s')ds', \quad g(s,s') = \frac{\min(s,s')[\ell - \max(s,s')]}{\ell}, \tag{I13}$$

with $\nabla_\perp = (\mathbf{I} - \mathbf{r}_0 \otimes \mathbf{r}_0)\nabla_\mathbf{x}$ evaluated along the straight ray $\mathbf{x}_0(s) = \mathbf{x}_S + s\,\mathbf{r}_0$. Equation (I13) gives, in quadratures, the frequency-dependent path for any prescribed potential and, together with (39), a closed-form benchmark pair $(\delta\mathbf{x}, \delta\tau)$ for the finite-element Eigenwave solver.

### I.3 The banana-doughnut pattern

Consider now the homogeneous medium of Appendix I.2 perturbed by a small, smooth velocity anomaly $\delta v(\mathbf{x})$, and ask how the finite-frequency phase time from $\mathbf{x}_S$ to $\mathbf{x}_R$ depends on the location of the anomaly relative to the straight ray. In ray theory the answer is trivial: only the part of the anomaly crossed by the ray matters, $\delta t_0 = -\int_{\mathbf{x}_0}\delta\, v\, ds/v^2$. The finite-frequency answer, obtained by Marquering et al. (1999) and Dahlen et al. (2000) from the Born approximation, is the banana-doughnut kernel, which is nonzero in a volume surrounding the ray and vanishes on the ray itself. We now show how this pattern arises in the Eigenwave framework.

To first order in $\delta v$, the field is $u = u_0(1+\eta)$, where $u_0$ is the unperturbed spherical wave (I7) and $\eta = u_1/u_0$ is the relative Born-scattered field,

$$\eta(\mathbf{x}_R, \omega) = -\frac{\omega^2}{2\pi v^3}\int \delta v(\mathbf{x}')\frac{\ell}{r_S(\mathbf{x}')r_R(\mathbf{x}')}e^{i\omega\,\Delta T(\mathbf{x}')}dV',$$
$$\Delta T(\mathbf{x}') = \frac{r_S(\mathbf{x}') + r_R(\mathbf{x}') - \ell}{v} \geq 0, \tag{I14}$$

with $r_S, r_R$ the distances of the scattering point from the source and from the receiver, and $\Delta T$ the detour time of the scattered wave; for a general field point $\mathbf{x}$ the same expression holds with $\mathbf{x}_R$ replaced by $\mathbf{x}$, $\ell$ by $r_S(\mathbf{x})$ and $r_R(\mathbf{x}')$ by $|\mathbf{x} - \mathbf{x}'|$. The amplitude and phase of the perturbed field are $a = a_0(1 + \operatorname{Re}\eta)$ and $\varphi = \varphi_0 + \operatorname{Im}\eta$, so that the phase-time perturbation at the receiver is

$$\delta\tau(\mathbf{x}_R,\omega) = \frac{\operatorname{Im}\eta(\mathbf{x}_R,\omega)}{\omega} = \int K_\omega(\mathbf{x}')\delta v(\mathbf{x}')dV',$$
$$K_\omega(\mathbf{x}') = -\frac{\omega}{2\pi v^3}\frac{\ell}{r_S(\mathbf{x}')r_R(\mathbf{x}')}\sin[\omega\,\varDelta T(\mathbf{x}')]. \tag{I15}$$

$K_\omega$ is the monochromatic phase-delay kernel, the "wavepath" of Woodward (1992). The traveltime measured by cross-correlating band-limited waveforms is the average of the monochromatic phase delays weighted by $\omega^2|m(\omega)|^2$, $m(\omega)$ being the spectrum of the unperturbed pulse (Dahlen et al., 2000), and the same average of $K_\omega$ is the homogeneous-medium banana-doughnut kernel of Dahlen et al. (2000) and Hung et al. (2000). Its pattern is read off (I15) directly. (i) On the ray, $\varDelta T = 0$ and $K_\omega = 0$: a scatterer placed exactly on the geometrical ray produces a wave that arrives in phase with the direct wave and alters only its amplitude, $\eta$ being real; this is the hole of the doughnut. (ii) Off the ray, at distance $\zeta$ from the ray at position $s$ along it, the paraxial detour time is $\varDelta T \simeq \zeta^2\ell/[2vs(\ell-s)]$; within the first Fresnel zone, $0 < \omega\,\varDelta T < \pi$, the kernel is negative - a fast anomaly beside the ray advances the arrival - with its extremum at $\omega\,\varDelta T = \pi/2$, and it changes sign in each successive Fresnel zone. The first zero,

$$\zeta_F(s) = \sqrt{\frac{2\pi v\,s(\ell-s)}{\omega\,\ell}}, \qquad \zeta_F(\ell/2) = \sqrt{\frac{\pi v\ell}{2\omega}}, \tag{I16}$$

is the radius of the first Fresnel zone, which gives the "banana" its shape: widest at the midpoint, pinched at the endpoints, with a width proportional to $\omega^{-1/2}$. (iii) The integral of the kernel over the plane normal to the ray is, to paraxial accuracy, $\int K_\omega\,dA = -1/v^2$ at every $s$ and every $\omega$: the kernel amplitude grows as $\omega$ while the Fresnel area shrinks as $\omega^{-1}$, so that as $\omega \to \infty$ the banana collapses onto the ray and reproduces the ray-theoretical sensitivity $\delta t_0$ for anomalies smooth on the Fresnel scale, whereas for anomalies narrower than $\zeta_F$ the ray-theoretical and finite-frequency sensitivities differ qualitatively.

The Eigenwave interpretation follows by evaluating the potential of the perturbed field. Outside the support of $\delta v$, $u_0(1+\eta)$ satisfies (I1) with the unperturbed velocity, whence $\nabla^2\eta + 2\nabla_\mathbf{x}\ln u_0 \cdot \nabla_\mathbf{x}\eta = 0$; taking the real part of this relation and using $\nabla^2 a_0 = 0$ in (I3),

$$W = \frac{v^2}{\omega^2}\left(\nabla^2\operatorname{Re}\eta + 2\frac{\nabla_\mathbf{x}a_0}{a_0}\cdot\nabla_\mathbf{x}\operatorname{Re}\eta\right) = \frac{2v}{\omega}\,\hat{\mathbf{r}}_S\cdot\nabla_\mathbf{x}\operatorname{Im}\eta, \tag{I17}$$

where $\hat{\mathbf{r}}_S = (\mathbf{x}-\mathbf{x}_S)/r_S$ is the direction of the unperturbed ray. The potential is generated entirely by the interference of the scattered wave with the direct wave - it vanishes with $\delta v$ - and it oscillates transversely on the Fresnel scale (I16). Its ray integral in the weak-coupling formula (39), with $\hat{\mathbf{r}}_S \cdot \nabla_\mathbf{x} = d/ds$ along the ray, is

$$\frac{1}{2}\int_{\mathbf{x}_0} \frac{W}{v}\,ds = \frac{1}{\omega}\int_0^{\ell} \frac{d\,\mathrm{Im}\,\eta}{ds}\,ds = \frac{\mathrm{Im}\,\eta(\mathbf{x}_R,\omega)}{\omega} = \delta\tau, \tag{I18}$$

so that, for an anomaly off the ray, the weak-coupling phase-time correction of Section 6 is the banana-doughnut integral (I15). More generally, for an anomaly located anywhere, the first-order Eigenwave phase time $\tau = \int nL\,ds$ with $L = 1/(v + \delta v)$ contains two contributions, the explicit one of the velocity and that of the index,

$$\delta\tau = -\int_{\mathbf{x}_0} \frac{\delta v}{v^2}\,ds + \frac{1}{2}\int_{\mathbf{x}_0} \frac{W}{v}\,ds = \int K_\omega\,\delta v\,dV', \tag{I19}$$

the path correction (I13) not contributing by Fermat's principle, and the identity holds because the exact phase satisfies (I3) with the local velocity. Equation (I19) is the trajectory-mechanics reading of the banana-doughnut pattern. Ray theory retains only the first term, which concentrates the whole sensitivity on the ray; the second term, the ray integral of the interference potential, redistributes it over the Fresnel volume: for an anomaly off the ray it supplies the entire (doughnut) sensitivity, for a sub-wavelength anomaly on the ray it cancels the ray-theoretical term, and for an anomaly smooth on the Fresnel scale it is negligible and ray theory is recovered. The "paradox" of a kernel that vanishes where the ray lies is thus not a failure of the stationary-path description but a manifestation of the very term neglected in the eikonal approximation - the coupling of the phase to the second spatial derivatives of the amplitude - which trajectory mechanics retains as the potential $W$.

Figure 11 illustrates the pattern and its decomposition: the kernel, its doughnut cross-section, and the two terms of (I19) for anomalies narrower than, and comparable with, the Fresnel radius.

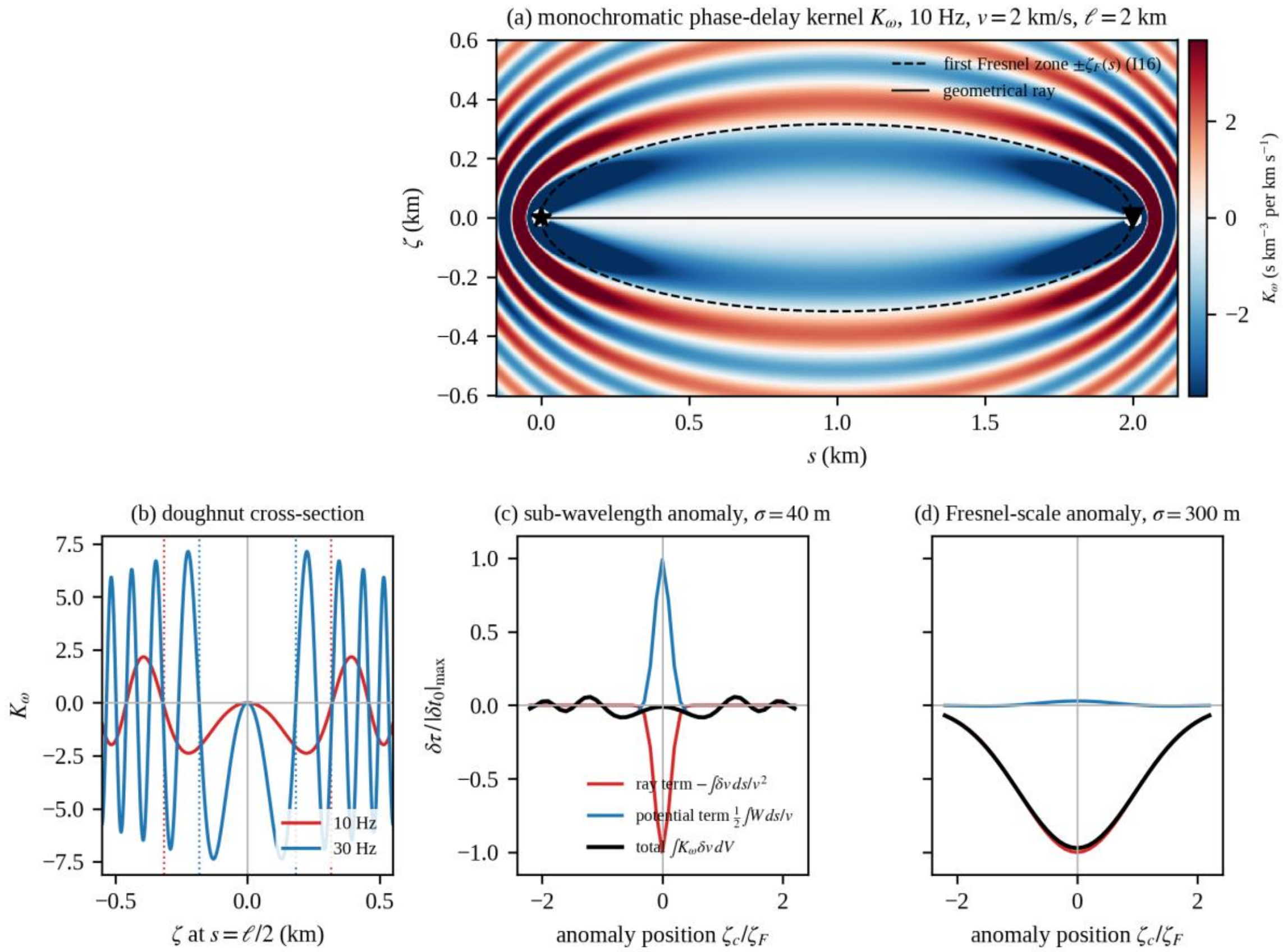

**Figure 11.** The banana-doughnut pattern in the Eigenwave framework (Appendix I.3): homogeneous medium, $v = 2$ km/s, $\ell = 2$ km. (a) Monochromatic phase-delay kernel $K_\omega$ of (I15) at 10 Hz in the plane of the ray; dashed: the first Fresnel zone $\pm\zeta_F(s)$ of (I16). The kernel vanishes on the ray (the hole of the doughnut), is negative in the first Fresnel zone and alternates in sign beyond it. (b) Cross-section at the midpoint of the ray at 10 and 30 Hz (dotted: $\pm\zeta_F$); the sensitivity region narrows as $\omega^{-1/2}$. (c), (d) The decomposition (I19) of the phase-time perturbation into the ray-theoretical term and the ray integral of the interference potential, for a Gaussian velocity anomaly moved across the ray at $s = \ell/2$, normalized by the peak of the ray-theoretical term; the total is the Born integral of $K_\omega$. For a sub-wavelength anomaly (c, $\sigma = 40$ m) the potential term cancels the ray term on the ray and the total is small everywhere; for an anomaly on the Fresnel scale (d, $\sigma = 300$ m $\approx \zeta_F$) the potential term is negligible and ray theory is recovered.

Finally, the trajectories themselves become frequency dependent in the perturbed medium: substituting (I17) into (I13) gives the first-order Eigenwave, which is deflected by the transverse gradient of the interference potential and therefore, like $K_\omega$, responds to the anomaly through the Fresnel structure $\sin(\omega\,\Delta T)$ rather than through the ray-theoretical criterion of whether the ray crosses it. As in the beam example, the potential (I17) does not carry the $\omega^{-2}$ signature of Section 6 - the Born factor $\omega^2$ in (I14) and the phase $\omega\,\Delta T$ make the frequency dependence of $\eta$, and hence of $W$, structural - so that the Eigenwaves of a scattering medium should be computed frequency by frequency, as in the continuation strategy of Section 8.4, rather than extrapolated from a single frequency with the $\omega^{-2}$ law.

### Appendix J. Relation to pilot-waves mechanics in quantum physics

Written in amplitude–phase form, $\psi = Re^{iS/\hbar}$ and $u = Ae^{i\omega T}$, the Schrödinger and Helmholtz equations yield formally identical pairs: a continuity (transport) equation, and a Hamilton–Jacobi equation modified by an amplitude-curvature term. That term is the quantum potential $Q = -\frac{\hbar^2}{2m}\frac{\nabla^2 R}{R}$ of pilot-wave (Bohmian) mechanics (Bohm, 1952; Wyatt, 2005) and the wave potential $W$ of (9) here. The correspondence $\hbar \leftrightarrow 1/\omega$ maps the classical limit onto the ray limit; both corrections are of second order, $Q = O(\hbar^2)$ and $W = O(\omega^{-2})$; and in both theories the trajectories are driven by the classical force plus a wave-induced force, which couples them to the global interference structure of the field. Two differences are essential. First, Bohmian trajectories are integral curves of a single-valued phase and do not cross, whereas multipathing and caustics require several phase branches or a phase-space representation; this is why the Eigenwave closes each ray branch separately and uses a Maslov representation near caustics (Sections 8.1 and 9.3). Second, a Bohmian trajectory is postulated to be a material particle path, whereas an Eigenwave is a computational construct that organizes the phase of a classical wavefield. Gaussian-beam theory differs from both: there, finite frequency modifies the field around classical central rays rather than the trajectories themselves.

### Appendix K. Maslov (phase-space) asymptotic theory

This appendix summarizes the elements of Maslov's asymptotic theory (Maslov and Fedoriuk, 1981; Chapman and Drummond, 1982; Thomson and Chapman, 1985; Chapman, 2004) used to build the closure field of Section 9.3. We state it in two dimensions, for a family of rays crossing horizontal levels $z = \text{const}$, which is the form used in the numerical example; the extension to three dimensions is outlined at the end.

#### K.1 The ray family as a Lagrangian manifold

A family of rays parametrized by the take-off angle $\gamma$ defines, on each level $z$, a curve in the phase plane $(x, p_x)$: the points $\left(X(\gamma), p_x(\gamma)\right)$, where $X$ is the horizontal position at which the ray crosses the level and $p_x$ its horizontal slowness. Together with the phase time $\tau(\gamma)$, which obeys $\partial_\gamma \tau = p_x\, \partial_\gamma X$ along the level, this curve is the section of the Lagrangian manifold of the family. The ray field is its projection onto the $x$ axis:

$$u(x, z, \omega) \approx \sum_k \sqrt{\frac{v}{8\pi\omega|J_k|}} \exp\left[i\left(\omega\tau_k + \frac{\pi}{4} - m_k\frac{\pi}{2}\right)\right], \quad J_k = \partial_\gamma X \cos\theta, \tag{K1}$$

where the sum runs over the rays with $X(\gamma) = x$, $m_k$ is their KMAH index and $\theta$ the ray angle from the vertical. Caustics are the points where the projection is singular, $\partial_\gamma X = 0$: there the Jacobian vanishes, (K1) diverges, and two or more rays coalesce. The manifold itself is regular at a caustic; only its projection onto $x$ is not. This is the central observation of the theory: a caustic in the $x$ chart is generally not a caustic in the conjugate chart $p_x$, where the relevant Jacobian is $\partial_\gamma p_x$.

#### K.2 Mixed representation and the Maslov integral

In the slowness chart the field is represented through its horizontal Fourier transform,

$$u(x,z) = \frac{\omega}{2\pi}\int \tilde{u}(p_x,z)\, e^{i\omega p_x x}\mathrm{d}p_x, \quad \tilde{u} \approx \tilde{A}\, e^{i\omega\tilde{\tau}}, \quad \tilde{\tau} = \tau - p_x X, \tag{K2}$$

where the phase $\tilde{\tau}$ is the Legendre transform of the phase time with respect to $x$, so that its derivative along the family is $\partial_\gamma\tilde{\tau} = -X\,\partial_\gamma p_x$. The spectral amplitude is finite wherever $\partial_\gamma p_x \neq 0$. Changing the integration variable from $p_x$ to $\gamma$ gives the Maslov integral

$$u(x,z,\omega) = \int \Phi(\gamma)B(\gamma,z)e^{i\omega\psi(x,\gamma)}\mathrm{d}\gamma, \quad \psi = \tau + p_x(x - X), \tag{K3}$$

with a smooth aperture taper $\Phi(\gamma)$. The amplitude $B$ is fixed by requiring that stationary phase reproduce (K1). Since $\partial_\gamma\psi = \partial_\gamma p_x(x - X)$, the stationary points are the rays through $x$, and there

$$\partial_\gamma^2\psi = -\,\partial_\gamma p_x\ \partial_\gamma X, \qquad |B|^2 = \frac{v|\partial_\gamma p_x|}{16\pi^2\cos\theta}, \qquad \arg B = \frac{\pi}{4} - \frac{\pi}{4}\mathrm{sgn}\big(-\,\partial_\gamma p_x\big), \tag{K4}$$

which is (53). The factor $\partial_\gamma X$ of the Jacobian cancels between the ray amplitude and the stationary-phase factor, and $B$ contains only $\partial_\gamma p_x$: the integrand is regular on the caustics in $x$. Because the sign of $\partial_\gamma^2\psi$ changes with that of $\partial_\gamma X$, each crossing of a caustic of the projection automatically produces the phase shift $-\pi/2$ of the KMAH index; the index need not be counted along the rays. Near a fold, where two stationary points coalesce, the phase $\psi$ becomes cubic in $\gamma$ and (K3) reduces to an Airy function; near a cusp it becomes quartic and (K3) reduces to the Pearcey integral (Pearcey, 1946; Berry and Upstill, 1980). Evaluated numerically, the integral reproduces these uniform forms without any special function, and it is accurate to leading order in $\omega^{-1}$, like the ray field it replaces away from caustics.

**K.3 Chart changes and the complex shear**

The slowness chart fails where $\partial_\gamma p_x = 0$, i.e. at the caustics of the projection onto $p_x$. There one must return to the $x$ chart or use another one. The general linear chart is a shear, $p_x - \kappa x$, whose Maslov integral carries the additional phase $\frac{\kappa}{2}(x - X)^2$ and whose amplitude contains $D = \partial_\gamma p_x - \kappa\,\partial_\gamma X$ instead of $\partial_\gamma p_x$. A real shear removes the singularity only if the tangent directions $(\partial_\gamma X, \partial_\gamma p_x)$ of the curve span less than a half-plane; otherwise several charts must be patched with a partition of unity, and their stationary-phase contributions agree only asymptotically. An imaginary shear, $\kappa = i\kappa_i$, avoids the choice altogether: $Im\,D = -\kappa_i\,\partial_\gamma X$ and $Re\,D = \partial_\gamma p_x$ cannot vanish together, so the amplitude never vanishes or diverges. This is the uniform closure (54). The imaginary part of the phase is a Gaussian window of width $w_b = \sqrt{2/(\omega\kappa_i)}$ around each ray, which relates (54) to Gaussian-beam summation (Červený et al., 1982; Popov, 1982). The difference is that the window is set on each level rather than at the source, so it is not bound to exceed the Fresnel scale of beams launched from a point source.

**K.4 Numerical implementation**

In Section 9 the integral is evaluated level by level, for the $x$ positions where the potential is needed. The ray quantities $X, \tau, p_x, \theta$ on each level are interpolated from rays traced by the Eigenray solver, and $\partial_\gamma X$ and $\partial_\gamma p_x$ are obtained by spline differentiation along the family. Three practical conditions apply. First, the take-off angles must resolve the oscillation of the integrand, $\omega\big|\partial_\gamma p_x\big||x - X|\Delta\gamma < \pi$. Second, the aperture taper $\Phi$ must be smooth, because a truncated family produces aperture diffraction; the result at the receiver was insensitive to the taper for the amplitude, but not for the potential derived from it (Section 9.5). Third, the chart must be checked on every level, which in the example is done by monitoring the sign of $\partial_\gamma p_x$. In three dimensions the level curve becomes a two-dimensional section of the manifold, and the mixed chart becomes a choice among partial Legendre transforms in one or both horizontal slowness components; the complex shear generalizes to a complex symmetric matrix. The cost is proportional to the number of rays times the number of evaluation points and frequencies, and the construction uses only quantities that the Eigenray solver already provides.

### Appendix L. The closure field and the Eigenwave compared

The closure field and the Eigenwave are complementary constructions. The closure field is an approximate finite-frequency wavefield built from the ray family (Section 9.3 and Appendix K). The Eigenwave is a single stationary path of the phase time in the effective medium $v/n$, with $n = \sqrt{1+W}$, whose potential is itself derived from the amplitude of a closure field. The Eigenwave therefore depends on the closure field, but the two answer different questions. Table 4 lists their main differences, with the evidence of Section 9.

**Table 4.** Main differences between the closure field and the Eigenwave.

| | **Closure field (phase-space / Maslov)** | **Eigenwave** |
|---|---|---|
| **What it is** | An approximate finite-frequency wavefield $u(\mathbf{x},\omega)$, built as an integral over a family of rays in a mixed chart, eqs (53)–(54) | A two-point stationary path of the phase-time functional in the effective medium $v/n$, one per arrival and frequency |
| **Inputs** | Kinematic and dynamic quantities of the Eigenray family: $X, \tau, p_x, \partial_\gamma p_x$, KMAH index | The potential $W$ of (9), which requires the amplitude of a closure field, and the Eigenray as initial guess |
| **Output** | Amplitude and phase of the field wherever it is evaluated | Frequency-dependent path $\mathbf{x}(s;\omega)$, phase time $\tau(\omega)$, dynamic quantities, and $\lvert W\rvert$ along the path |
| **Amplitude** | Its own modulus, finite at caustics, including the cusp | None of its own near caustics, where the transport amplitude (41) remains singular; it is taken from the closure field |
| **Phase** | Its own phase, the interference of all rays of the family | Recomputed by integrating $n/v$ along the path; the phase of the closure field is discarded |
| **Caustics** | Handled by construction; the KMAH shift emerges from stationary phase | Handled only through $W$: on the axis of a symmetric focus it carries the focal phase; off the axis, or for split caustics, only part of it, and the KMAH bookkeeping has no reliable rule yet |
| **Multipathing** | One integral represents the whole family, and the arrivals interfere automatically | One path per arrival; multipathing is retained by treating branches separately |
| **Numerical sensitivity** | Integrates amplitudes; well conditioned and robust to the aperture | The phase depends on second derivatives of the amplitude over many |
| | | wavelengths, with error $\approx \pi N_\lambda \delta \overline{W}$; sensitive to smoothing, stencil and aperture |
| **Accuracy in Section 9** | Full gather: median misfit 0.20, phase errors at most 0.13 rad, including the cusp | 0.02–0.17 rad on the axis, depending on the aperture; median misfit 0.36 and up to 0.6–0.9 rad near the axis; no gain far from caustics |
| **Relation to exact theory** | Asymptotic; not an exact solution of the Helmholtz equation | Exact trajectory mechanics for the exact $W$; with the exact potential it matched finite differences to 0.19 rad |
| **Cost and locality** | A fan of rays and one integral per evaluation point; inexpensive in two dimensions | One warm-started Newton solve per frequency, but $W$ requires a closure field around the path |
| **Best use** | Synthetic seismograms and amplitudes, especially near caustics | Frequency-dependent paths, phase times away from caustics, sensitivity kernels, and $\lvert W \rvert$ as an indicator of the validity of ray theory |

In short, the closure field is the better tool for predicting the wavefield, and in the example it reproduces the finite-difference seismograms along the whole receiver line. The Eigenwave is the better tool for describing how a finite-frequency arrival samples the medium, provided it has a closure field to rely on and stays out of the immediate neighbourhood of caustics. A formulation in which the members of the Maslov integral themselves carry the finite-frequency phase would combine the two, and remains to be developed.

**References**

Berry, M.V. and Upstill, C. (1980) Catastrophe optics: morphologies of caustics and their diffraction patterns. *Progress in Optics*, 18, 257-346.

Bliss, G. (1916) Jacobi's condition for problems of the calculus of variations in parametric form. *Transactions of the American Mathematical Society*, 17, 195-206.

Bohm, D. (1952) A suggested interpretation of the quantum theory in terms of 'hidden variables' I, II. *Physical Review*, 85, 166-193.

Bona, A. and Slawinski, M. (2003) Fermat's principle for seismic rays in elastic media. *Journal of Applied Geophysics*, 54, 445-451.

Cerjan, C., Kosloff, D., Kosloff, R. and Reshef, M. (1985) A nonreflecting boundary condition for discrete acoustic and elastic wave equations. *Geophysics*, 50, 705-708.

Červený, V. (2000) *Seismic Ray Theory*. Cambridge University Press.

Červený, V., Popov, M. and Pšenčík, I. (1982) Computation of wave fields in inhomogeneous media - Gaussian beam approach. *Geophysical Journal International*, 70, 109-128.

Chapman, C. (2004) *Fundamentals of Seismic Wave Propagation*. Cambridge University Press.

Chapman, C.H. and Drummond, R. (1982) Body-wave seismograms in inhomogeneous media using Maslov asymptotic theory. *Bulletin of the Seismological Society of America*, 72, S277-S317.

Dahlen, F.A., Hung, S.-H. and Nolet, G. (2000) Fréchet kernels for finite-frequency traveltimes - I. Theory. *Geophysical Journal International*, 141, 157-174.

Farra, V. (1992) Bending method revisited: a Hamiltonian approach. *Geophysical Journal International*, 109, 138-150.

Foreman, T.L. (1989) An exact ray theoretical formulation of the Helmholtz equation. *Journal of the Acoustical Society of America*, 86, 234-246.

Hung, S.-H., Dahlen, F.A. and Nolet, G. (2000) Fréchet kernels for finite-frequency traveltimes - II. Examples. *Geophysical Journal International*, 141, 175-203.

Koren, Z. and Ravve, I. (2020a) Eigenrays in 3D heterogeneous anisotropic media, Part I - Kinematics, variational formulation. arXiv:2003.09406.

Koren, Z. and Ravve, I. (2020b) Eigenrays in 3D heterogeneous anisotropic media, Part II - Kinematics, validation of the Lagrangian. arXiv:2003.09407.

Koren, Z. and Ravve, I. (2020c) Eigenrays in 3D heterogeneous anisotropic media, Part III - Kinematics, finite-element implementation. arXiv:2003.09408.

Koren, Z. and Ravve, I. (2021) Eigenrays in 3D heterogeneous anisotropic media, Part I: Kinematics. *Geophysical Prospecting*, 69, 3-27.

Kravtsov, Yu.A. (1964) A modification of the geometrical optics method. *Radiofizika*, 7, 664-673.

Ludwig, D. (1966) Uniform asymptotic expansions at a caustic. *Communications on Pure and Applied Mathematics*, 19, 215-250.

Maslov, V.P. and Fedoriuk, M.V. (1981) *Semi-Classical Approximation in Quantum Mechanics*. Reidel, Dordrecht.

Marquering, H., Dahlen, F.A. and Nolet, G. (1999) Three-dimensional sensitivity kernels for finite-frequency traveltimes: the banana-doughnut paradox. *Geophysical Journal International*, 137, 805-815.

Pearcey, T. (1946) The structure of an electromagnetic field in the neighbourhood of a cusp of a caustic. *Philosophical Magazine*, 37, 311-317.

Popov, M. (1982) A new method of computation of wave fields using Gaussian beams. *Wave Motion*, 4, 85-97.

Protasov, M. and Gadylshin, K. (2017) Computational method for exact frequency-dependent rays on the basis of the solution of the Helmholtz equation. *Geophysical Journal International*, 210, 525-533.

Ravve, I. and Koren, Z. (2019) Directional derivatives of ray velocity in anisotropic elastic media. *Geophysical Journal International*, 216, 859-895.

Ravve, I. and Koren, Z. (2020a) Eigenrays in 3D heterogeneous anisotropic media, Part IV - Geometric spreading from traveltime Hessian. arXiv:2003.10242.

Ravve, I. and Koren, Z. (2020b) Eigenrays in 3D heterogeneous anisotropic media, Part V - Dynamics, variational formulation. arXiv:2003.10243.

Ravve, I. and Koren, Z. (2020c) Eigenrays in 3D heterogeneous anisotropic media, Part VI - Dynamics, Lagrangian vs. Hamiltonian approaches. arXiv:2003.10244.

Ravve, I. and Koren, Z. (2020d) Eigenrays in 3D heterogeneous anisotropic media, Part VII - Dynamics, finite-element implementation. arXiv:2003.11418.

Ravve, I. and Koren, Z. (2021) Eigenrays in 3D heterogeneous anisotropic media, Part II: Dynamics. *Geophysical Prospecting*, 69, 28-52.

Snieder, R. and Spencer, C. (1993) A unified approach to ray bending, ray perturbation and paraxial ray theories. *Geophysical Journal International*, 115, 456-470.

Thomsen, L. (1986) Weak elastic anisotropy. *Geophysics*, 51, 1954-1966.

Thomson, C.J. and Chapman, C.H. (1985) An introduction to Maslov's asymptotic method. *Geophysical Journal of the Royal Astronomical Society*, 83, 143-168.

Vasco, D.W. (2018) An extended trajectory mechanics approach for calculating the path of a transient pressure: derivation and illustration. *Water Resources Research*, 54, doi:10.1002/2017WR021360.

Vasco, D.W. and Nihei, K.T. (2019a) A trajectory mechanics approach for the study of wave propagation in an anisotropic elastic medium. *Geophysical Journal International*, 219, 1885-1899.

Vasco, D.W. and Nihei, K.T. (2019b) Broad-band trajectory mechanics. *Geophysical Journal International*, 216, 745-759.

Woodward, M.J. (1992) Wave-equation tomography. *Geophysics*, 57, 15-26.

Wyatt, R.E. (2005) *Quantum Dynamics with Trajectories*. Springer.